\documentclass[aps,amsfonts,amsmath,nofootinbib,byrevtex,%
  preprintnumbers,showkeys,preprint,prd]{revtex4}
\usepackage{graphicx}
\usepackage{amssymb}
\usepackage{hyperref}
\usepackage{bm} 
\usepackage{subfigure} 
\usepackage{rotating} 
\usepackage{color}

\allowdisplaybreaks[1]
\def\be#1\ee{\begin{equation}#1\end{equation}}
\def\bal#1\eal{\begin{align}#1\end{align}}
\def\bat#1\eat{\begin{alignat}{2}#1\end{alignat}}
\def\bmu#1\emu{\begin{multline}#1\end{multline}}
\def\bga#1\ega{\begin{gather}#1\end{gather}}
\newcommand{\ba}{\begin{array}}
\newcommand{\ea}{\end{array}}
\newcommand{\n}{\notag}

\renewcommand{\d}{\partial}

\renewcommand{\cal}{\mathcal}

\newcommand{\ds}{\displaystyle}
\newcommand{\ts}{\textstyle}
\newcommand{\imineq}[2]{\vcenter{\hbox{\includegraphics[height=#2ex]{#1}}}}

\begin{document}

\title{Exploiting the scheme dependence of the renormalization group
improvement in infrared Yang-Mills theory}
\date{\today}

\author{Pietro Dall'Olio$^{\dag}$, Axel Weber$^\ast$}
\affiliation{$\dag$Centro de Ciencias Matem\'aticas, UNAM - Campus Morelia,
  Antigua Carretera a P\'atzcuaro 8701,
Col.\ Ex Hacienda San Jos\'e de la Huerta,
58089 Morelia, Michoac\'an, Mexico \\
$\ast$Instituto de F\'isica y Matem\'aticas, 
Universidad Michoacana de San Nicol\'as de Hidalgo, 
Edificio C-3, Ciudad Universitaria, A. Postal 2-82, 58040 Morelia, 
Michoac\'an, Mexico}
%

\begin{abstract}
Within the refined Gribov-Zwanziger scenario
for four-dimensional Yang-Mills theory in the Landau gauge, a
gluon mass term is generated from the restriction of the
gauge field configurations to the first Gribov region. Tissier and
Wschebor have pointed out that simply adding a gluon mass term to the
usual Faddeev-Popov action yields one-loop renormalization group
improved gluon and ghost propagators which are
in good agreement with the lattice data even in the infrared
regime. In this work, we extend their analysis to
several alternative renormalization schemes and show how the
renormalization scheme dependence can be used to achieve
an almost perfect matching to the lattice data for the gluon and 
ghost propagators.
\end{abstract}

\maketitle

\section{Introduction}

Yang-Mills theory, the pure gauge sector of QCD, ows its nontrivial dynamics 
to the non-Abelian nature of the gauge symmetry, here taken as SU($N$). 
It is presumably responsible for the peculiar features of the strong 
interaction such as color confinement and chiral symmetry breaking, which are
not present in its Abelian counterpart QED. These phenomena, which are 
related to the infrared (IR) behavior of the theory and its highly nontrivial 
vacuum state, are usually considered to be inaccessible to a perturbative 
approach due to the presence of a Landau pole for the running coupling
at the mid-momentum scale $\Lambda_{\text{QCD}}$. 

In his seminal paper \cite{Gri78}, Gribov showed that the Faddeev-Popov 
procedure, aimed at fixing the gauge in a covariant way in the 
non-Abelian case, is valid only perturbatively since for large gauge fields 
the gauge orbits intersect the gauge-fixing hypersurface in several points 
(Gribov copies), thus invalidating the Faddeev-Popov construction on a 
nonperturbative level. In an attempt to overcome the gauge copy ambiguity, 
Gribov proposed to restrict the functional integration to gauge field 
configurations for which the Faddeev-Popov operator $(-\d_\mu D_\mu)$ 
is positive definite in Euclidean space-time, or, equivalently
(in Landau gauge), to local minima of the functional 
$\int_x A^a_\mu(x) A^a_\mu(x)$ along the respective 
gauge orbits \cite{Zwa82, SF86}, thus defining the 
(first) Gribov region. Moreover, he was able to show that this restriction 
of the functional integral avoids the generation of a Landau pole 
at a nonvanishing scale and changes the behavior of the n-point functions 
in the IR, which are dominated in this regime by gauge field configurations
close to the boundary of the Gribov region \cite{footn3}. 

For the rest of this paper, we restrict our attention
to Yang-Mills theory in the Landau gauge.
In the \emph{refined} Gribov-Zwanziger scenario \cite{DGS08}, the possibility 
of the formation of condensates of the gluon field and various auxiliary 
fields introduced in order to realize Gribov's restriction in a local fashion 
\cite{Zwa89, Zwa93}, is taken into account. The gluon propagator then
develops a massive behavior in the IR, approaching a non-zero value at 
vanishing momentum. Such an IR finite gluon propagator, together with an IR 
finite ghost dressing function, corresponds to the \emph{decoupling} solutions 
found within the framework of the nonperturbative Dyson-Schwinger equations
\cite{AN04, BBL06}, and also with a mapping to $\lambda \phi^4$ theory
\cite{Fras08}. On the other hand, the \emph{scaling} solutions of the
Dyson-Schwinger equations \cite{SHA97, SHA98, FA02, Zwa02, LS02}, and also
  of the functional renormalization group equations \cite{PLN04}, are realized
when the \emph{horizon condition} is imposed in the absence of condensates,
in which case it amounts to forcing the inverse ghost dressing function to
vanish at zero momentum \cite{Zwa93, Zwa02}.

Perhaps more importantly, the decoupling behavior of the gluon propagator in
the IR is in agreement, in three and four space-time dimensions, with Monte 
Carlo simulations in the minimal Landau gauge on the largest lattices to date
\cite{BIMS07, CM07, SSL07}.
In the minimal Landau gauge, the restriction
to the Gribov region is implemented numerically by selecting, for each gauge 
orbit, a local minimum of the (appropriately discretized) quadratic 
functional $\int_x A^a_\mu(x) A^a_\mu(x)$.

A large body of work has been put together over the years, aiming at an ever
more complete and consistent determination of the n-point function of Landau
gauge Yang-Mills theory with functional methods.
Refs.\ \cite{AP06, BLY08, HAFS08, FMP09, AHS10, IP13, AIP13, HS13, BHM14, EWAV14, CHS15, ABF16, CFM16, Hub20}
should give an impression of the variety of the approaches and
of the sophistication achieved, although our selection of works is
inevitably subjective.

In 2010,
Tissier and Wschebor \cite{TW10} have shown that the lattice results for 
the gluon and ghost propagators in the IR can be reproduced rather well by 
plain one-loop calculations with a transverse massive gluon propagator. 
This massive extension of the Faddeev-Popov action corresponds to a 
particular case (in the Landau gauge) of a class of massive models originally 
studied by Curci and Ferrari \cite{CF76} with the aim of regularizing
the infrared divergences of Yang-Mills theory in a renormalizable way, where 
the mass parameter was thought to be eventually sent to zero. In our present
context, however, the mass parameter is supposed to have a definite non-zero
value which is ultimately related to the Gribov parameter that emerges from 
the restriction to the Gribov region. 

One way to understand the appearance of a gluon mass term is from the fact
that the restriction of the gauge fields to the Gribov region breaks the 
nilpotent BRST symmetry of the Faddeev-Popov action \cite{Zwa93}. In the 
absence of this symmetry, an effective massive behavior of the gluon 
propagator can be shown to correspond (in three and four space-time 
dimensions) to an infrared attractive, hence physically relevant, fixed point 
\cite{Web12}. Such a fixed point is of the 
same nature as the high-temperature fixed point of the Ginzburg-Landau model 
\cite{Bel91}, see also Ref.\ \cite{RSTW17} for an accurate study of the 
renormalization group flow in the space of the mass and coupling parameters.
In the latter two-parameter space, families of IR divergent and IR safe 
solutions (with and without a Landau pole, respectively) are separated by a 
critical trajectory corresponding to a nontrivial IR fixed point. 

The triviality of the IR fixed point for the IR safe family of flow functions
was already pointed out in Ref.\ \cite{TW011}, where it was shown that a 
particular renormalization scheme leads to a change of the sign of the beta 
function at low momenta, thus avoiding the generation of a Landau pole in 
agreement with the Gribov-Zwanziger scenario, and driving the running coupling 
towards vanishing values in the extreme IR (note that, due to the 
presence of a mass parameter, the beta function for the coupling constant is 
\emph{not} universal even at the one-loop level). 
This state of affairs \textit{a posteriori} justifies 
the straightforward perturbative approach of Ref.\ \cite{TW10}, which has 
produced results for the gluon and ghost propagators in the IR in reasonable
agreement with the lattice data. A renormalization group improvement
was then applied to extend these results to
the entire range of momenta for the propagators \cite{TW011}, to the vertex
functions \cite{TW13}, and to the fermionic sector \cite{PTW14, PTW15}.
The renormalization group improvement of the n-point functions is required
in order to accomplish an accurate description at all scales. In
renormalization group jargon, what is needed is a quantitative description
of the crossover from the Gaussian fixed point in the ultraviolet (UV)
to the IR high-temperature 
fixed point in the IR safe schemes, the crossover being driven by the flow of 
the running mass parameter $m(\mu)$ with the renormalization scale $\mu$,
or rather by the flow of the dimensionless combination $m(\mu)/\mu$.  

In this paper, we scrutinize the relevant elements for a renormalization 
scheme to be both IR safe and to provide the correct behavior for large 
momenta, i.e., in the ultraviolet regime, after renormalization 
group improvement. In the perturbative UV regime where the relevant gauge
field configurations are far from the boundary of the 
Gribov region, the nilpotent BRST symmetry should be 
restored and, in particular, the gluon mass term should become negligible.
Among the renormalization schemes which fulfill these requirements, we will
determine the one that yields renormalization group improved propagators which 
best match the lattice results.

Just as in the work of Tissier and Wschebor, we employ
Callan-Symanzik equations to implement the renormalization group
improvement. To be clear, we use the term ``Callan-Symanzik equations'',
in accord with modern textbooks, to refer to the renormalization group 
equations that describe the variation of the renormalized n-point 
functions and parameters with an off-shell renormalization scale \cite{GP76}, 
whereas in the original version of the equations the physical mass was used 
as a scale instead \cite{Cal70, Sym70}. We should also like to stress that, 
contrary to the renormalization of the theory at a fixed scale,
the resummation effected by the renormalization group equations carries
a nontrivial dependence on the renormalization scheme since it is the
renormalization scheme dependent finite contributions that get resummed.

It is noteworthy that a massive behavior of the gluon propagator also emerges 
within a recent approach aimed to bypass the Gribov ambiguity from first 
principles, by taking a proper weighted average over all Gribov copies 
\cite{ST12, NRS20} instead of trying to select a unique representative for 
each gauge orbit as in the Gribov-Zwanziger scenario. However, our
present goal is merely to determine a formulation of the theory in the
continuum that reproduces the lattice data in the 
minimal Landau gauge, and the simple extension to the Curci-Ferrari model 
improved by the renormalization group may be all that is needed for an 
accurate description of the full propagators of Yang-Mills theory 
in this sense. On the other hand, considering that the horizon function that 
is added to the Faddeev-Popov action in the Gribov-Zwanziger 
formulation \cite{Zwa89} in order to implement the restriction to the Gribov 
region (where the Gribov parameter in the horizon function has to 
be adjusted to fulfill the so-called horizon condition)
is a complicated nonlocal functional of the gauge field, one might expect 
the sole addition of a gluon mass term to be insufficient to reproduce all 
the (proper) n-point functions of the theory correctly.

Despite its importance for the consistency of any theory, we will not address 
the issue of unitarity here. In the case at hand, unitarity is seriously 
compromised by the breaking of the nilpotent BRST symmetry whose 
cohomology is traditionally used to construct the physical Hilbert space 
equipped with a positive norm \cite{KO79}. While this fact has been
employed as an argument against the physical relevance of the 
Curci-Ferrari model \cite{Oji82, BSN96}, it cannot at
present be excluded that a proper definition of the Hilbert space exists 
that would restore the unitarity of the theory, perhaps by exploiting the 
cohomology of an extended BRST transformation \cite{CD15}.

This work is organized as follows. In Section II we examine different 
renormalization schemes within the massive extension of Yang-Mills theory 
in the Landau gauge, for which the one-loop counterterms are fixed 
by imposing 
normalization conditions on the proper two-point functions, and by the 
definition of the renormalized coupling constant. We elaborate on the liberty 
of shifting among renormalization schemes while maintaining the IR safety 
property and the correct UV  behavior, elucidating in particular the role 
of the longitudinal part of the proper gluonic two-point function in the 
renormalization process. In Section III we integrate the 
Callan-Symanzik equations for the proper ghost and gluon two-point functions 
in the different renormalization schemes and compare the numerical results 
for the renormalization group resummed propagators to the lattice 
data. We also present a detailed analysis of the analytic 
behavior of the resummed two-point functions in the deep UV and in the 
extreme IR, confirming, for instance, the increase of the gluon propagator 
with momentum in the latter regime (a feature which is not unambiguously 
exhibited by the lattice data in four dimensions). Furthermore we show, both 
numerically and analytically, that the renormalization group resummation 
produces, for some of the renormalization schemes considered, a nontrivial 
violation of a Slavnov-Taylor identity that is derived from the non-nilpotent
extension of the BRST symmetry to the massive case. 
Finally, in Section IV we present our conclusions. 
We relegate to the appendix the details of the derivation of the 
Slavnov-Taylor identity just mentioned, and the explicit expressions
for the one-loop diagrams.

A preliminary account of some of the results presented here was
given in Ref.\ \cite{Web16}.

\section{Renormalization schemes}

\subsection{Tissier-Wschebor scheme and a Slavnov-Taylor identity}

As discussed in the Introduction, we shall add a gluonic mass term to the
standard Faddeev-Popov action for SU($N$) Yang-Mills theory in the Landau 
gauge and thus consider the following action in $D$-dimensional Euclidean 
space-time (a Curci-Ferrari model \cite{CF76}):
\be
S=\int d^D x\left[\frac{1}{4} F^a_{\mu \nu} F^a_{\mu \nu}
+\partial_\mu \bar{c}^a(D_\mu c)^a+iB^a\partial_\mu A^a_\mu
+\frac{m^2}{2} A^a_\mu A^a_\mu \right],
\label{CF}
\ee
where we have implemented the Landau gauge condition via a Nakanishi-Lautrup
auxiliary field $B^a$ \cite{Oji82}. The action
\eqref{CF} is invariant under the extended BRST transformation
\bat
s A_\mu^a &= (D_\mu c)^a, &\qquad s c^a &= - \frac{g}{2}f^{abc}c^b c^c, \n \\
s\bar{c}^a &= iB^a, &\qquad s B^a &= -im^2 c^a.
\label{BRST}
\eat
Note the Grassmann nature of the operator $s$ which anticommutes
with all Grassmann fields in the generalized Leibniz rule. The
massive extension of the BRST transformation is not nilpotent, i.e.,
$s^2 \neq 0$ as long as $m^2 \neq 0$.

The most important consequence of the invariance of the action under the
transformation \eqref{BRST} for our purposes is the identity \cite{TW011}
\be
\Gamma^B_{c\bar c}(p^2) \, \Gamma^{\parallel\,B}_{AA}(p^2)=p^2 m_B^2 
\label{STI}
\ee
for the bare two-point functions (we have made explicit that this is
a relation between the bare quantities because, as we shall see in the 
following, depending on the renormalization scheme the corresponding relation 
for the renormalized quantites may or may not be fulfilled). Our notation 
for the proper (1PI) two-point functions is
\be
\left. (2\pi)^{2D} \frac{\delta^2 \Gamma}{\delta c^b(-q) \delta \bar{c}^a(p)}
\, \right|_{(A,c,\bar{c},B)=0}
= \Gamma_{c\bar{c}}(p^2) \delta^{ab} (2\pi)^D \delta^D(p-q) \label{notateGamma}
\ee
(putting all fields equal to zero after taking the derivatives on the 
left-hand side), and similarly for $\Gamma_{A_\mu A_\nu}(p)$ which we decompose
into a transverse and a longitudinal part:
\be
\Gamma_{A_\mu A_\nu}(p) = \Gamma^{\perp}_{AA}(p^2) \left( \delta_{\mu\nu} 
- \frac{p_\mu p_\nu}{p^2} \right)
+ \Gamma^{\parallel}_{AA}(p^2) \frac{p_\mu p_\nu}{p^2} \,. \label{decomp}
\ee
For the convenience of the reader, we have reproduced in Appendix \ref{derSTI} 
the derivation of the identity \eqref{STI} from a Slavnov-Taylor 
identity resulting from the invariance of the action 
under the extended BRST transformation, the 
antighost equation, and the Dyson-Schwinger equation for the Nakanishi-Lautrup 
field. According to Eq.\ \eqref{STI}, the longitudinal part of the 
proper gluonic two-point function vanishes only in the massless case. 

For later use we remark that the
(connected) two-point functions or propagators
are given by the inverse of the proper two-point functions,
\bal
\langle c^a (-p) \bar{c}^b (q) \rangle
&= \big( \Gamma_{c\bar{c}}(p^2) \big)^{-1} \delta^{ab} 
(2\pi)^D \delta^D(p-q) \,, \n \\
\langle A^a_\mu (-p) A^b_\nu (q) \rangle
&= \big( \Gamma^{\perp}_{AA}(p^2) \big)^{-1} \left( \delta_{\mu\nu} 
- \frac{p_\mu p_\nu}{p^2} \right) \delta^{ab} 
(2\pi)^D \delta^D(p-q) \,. \label{propagators}
\eal
Contrary to the proper gluonic two-point function, the 
gluon propagator is always transverse in the Landau gauge, and it is 
determined by the transverse part of the proper two-point function alone,
see Appendix \ref{transverse}.

In the IR safe renormalization scheme proposed by Tissier and Wschebor in 
Ref.\ \cite{TW011}, the following normalization conditions are imposed on 
the renormalized proper two-point functions at the renormalization 
scale $\mu$:
\bal
\Gamma^\perp_{AA} (\mu^2) &= m^2 + \mu^2 \,, \label{TWs1} \\
\Gamma^\parallel_{AA} (\mu^2) &= m^2 \,, \label{TWs2} \\
\Gamma_{c\bar{c}} (\mu^2) &= \mu^2 \,. \label{TWs3}
\eal
As usual, the renormalized fields are related to the bare fields
(which receive an index $B$) via $A^B_\mu = Z_A^{1/2} A_\mu$, 
$c^B = Z_c^{1/2} c$, etc., so that the normalization
conditions \eqref{TWs1}--\eqref{TWs3} implicitly define the field 
renormalization constants $Z_A$ and $Z_c$ as functions of $\mu^2$.

Defining, furthermore, $Z_{m^2}$ through $m_B^2 = Z_{m^2} \, m^2$, the 
Slavnov-Taylor identity \eqref{STI} can be written
in terms of the renormalized quantities as
\be
\Gamma_{c\bar{c}} (p^2) \, \Gamma^\parallel_{AA} (p^2) = (Z_A Z_c Z_{m^2})
p^2 m^2 \,. \label{STIz}
\ee
If we specialize to $p^2 = \mu^2$ and apply the normalization condition
\eqref{TWs3}, the Slavnov-Taylor identity implies that
\be
\Gamma^\parallel_{AA} (\mu^2) = (Z_A Z_c Z_{m^2}) m^2 \,. \label{STIGl}
\ee
It is hence apparent that the normalization condition \eqref{TWs2} is
equivalent to $Z_A Z_c Z_{m^2} = 1$ [provided that $\Gamma_{c\bar{c}} (p^2)$
is normalized as in Eq.\ \eqref{TWs3}], the form in which 
it was originally
proposed in Ref.\ \cite{TW011}. We furthermore conclude from Eq.\ 
\eqref{STIz} that the normalization conditions \eqref{TWs2}
and \eqref{TWs3} imply the
renormalized counterpart of the Slavnov-Taylor identity,
\be
\Gamma_{c\bar{c}} (p^2) \, \Gamma^\parallel_{AA} (p^2) = p^2 m^2 \,.
\label{STIr}
\ee
Since one of the principles in renormalizing a quantum field theory is to
preserve the symmetries of the classical theory (whenever possible, i.e.,
in the absence of anomalies), the fact that the renormalization scheme 
\eqref{TWs1}--\eqref{TWs3} implies the identity \eqref{STIr} for the
renormalized quantities is certainly satisfactory. We shall have
more to say on this issue in the next subsection.

A second renormalization scheme presented in Ref.\ \cite{TW011}
replaces the normalization condition \eqref{TWs2} with
\be
\Gamma^\parallel_{AA} (0) = m^2 \,. \label{TWu2}
\ee
In fact, in Ref.\ \cite{TW011} this normalization condition was 
imposed on $\Gamma^\perp_{AA} (0)$, however, in three and four space-time
dimensions we find that
\be
\Gamma^{\perp \, B}_{AA} (0) = \Gamma^{\parallel \, B}_{AA} (0)
\label{local}
\ee
to any order in perturbation theory, meaning that the proper gluonic 
two-point function $\Gamma^B_{A_\mu A_\nu} (p)$ is local [cf.\ Eq.\
\eqref{decomp}], a property that carries over to the renormalized two-point 
function. Since $\Gamma^\parallel_{AA} (0) \neq \Gamma^\parallel_{AA} (\mu^2)$ 
in general, and following the argument above up to Eq.\ \eqref{STIGl},
the normalization condition \eqref{TWu2} implies that $Z_A Z_c Z_{m^2} \neq 1$,
hence the Slavnov-Taylor identity
\eqref{STIr} for the renormalized quantities is not fulfilled in
the renormalization scheme with condition \eqref{TWu2}. It was also shown
in Ref.\ \cite{TW011} that this scheme is not IR safe, i.e., the integration
of the renormalization group equations leads to a Landau pole 
(note, in particular, that
the flow functions are not universal to one-loop order in the
presence of a mass parameter). We shall come back to the property of IR 
safeness below.

\subsection{Alternative renormalization schemes for the two-point functions
\label{altRG}}

We shall now explore the possibility of formulating alternative IR safe 
renormalization schemes. To this end, we need to take a closer look
at the IR and, as it turns out, also the UV behavior of the flow functions. 
We start
by presenting the one-loop corrections to the two-point functions in
the IR limit $p^2 \ll m^2$ and the UV limit $p^2 \gg m^2$. The complete
one-loop expressions, separated into the contributions of the 
different diagrams, can be found in Appendix \ref{twop}. For $p^2 \ll m^2$, 
we have in four space-time dimensions using dimensional regularization
with $D = 4 - \epsilon$,
\bal
\Gamma^\perp_{AA} (p^2) &= m^2 + \frac{3}{4} \, m^2 \frac{N g^2}{(4\pi)^2}
\left( \overline{\frac{2}{\epsilon}} - \ln \frac{m^2}{\kappa^2} 
+ \frac{5}{6} \right) + m^2 (\delta Z_{m^2} + \delta Z_A) \n \\
&\phantom{=} {}+ p^2 - \frac{13}{6} \, p^2 \frac{N g^2}{(4\pi)^2}
\left( \overline{\frac{2}{\epsilon}} - \ln \frac{m^2}{\kappa^2} 
- \frac{1}{26} \ln \frac{p^2}{m^2} - \frac{3}{52} \right) + p^2 \delta Z_A 
\,, \label{GAtir} \\
\Gamma^\parallel_{AA} (p^2) &= m^2 + \frac{3}{4} \, m^2 \frac{N g^2}{(4\pi)^2}
\left( \overline{\frac{2}{\epsilon}} - \ln \frac{m^2}{\kappa^2} 
+ \frac{5}{6} \right) + m^2 (\delta Z_{m^2} + \delta Z_A) \n \\
&\phantom{=} {}- \frac{1}{4} \, p^2 \frac{N g^2}{(4\pi)^2}
\left( -\ln \frac{p^2}{m^2} + \frac{11}{6} \right) \,, \label{GAlir} \\
\Gamma_{c\bar{c}} (p^2) &= p^2 - \frac{3}{4} \, p^2 \frac{N g^2}{(4\pi)^2}
\left( \overline{\frac{2}{\epsilon}} - \ln \frac{m^2}{\kappa^2} 
+ \frac{5}{6} \right) + p^2 \delta Z_c \n \\
&\phantom{=} {}+ \frac{1}{4} \, p^2 \frac{p^2}{m^2}
\frac{N g^2}{(4\pi)^2} \left( -\ln \frac{p^2}{m^2} + \frac{11}{6} \right)
\,, \label{Gcir}
\eal
where we have introduced the abbreviation
\be
\overline{\frac{2}{\epsilon}} = \frac{2}{\epsilon} - \gamma_E + \ln (4 \pi) \,,
\label{overeps}
\ee
and $\kappa$ is an arbitrary unit of mass. Note that we have written 
the renormalized two-point functions in terms of the renormalized parameters.
The counterterms $\delta Z_A$, $\delta Z_c$, $\delta Z_{m^2}$ with
$Z_A = 1 + \delta Z_A$, etc., will be determined through
the normalization conditions. The advantage of this representation
is that the identification of the flow functions in the different
renormalization schemes to be considered 
will be particularly simple in this way.
The locality condition \eqref{local} is obviously fulfilled for the bare and 
the renormalized gluonic two-point function. Also notice that the 
coefficients of $\ln p^2$ do not coincide with the (negative of
the) coefficients of 
$2/\epsilon$ in the one-loop corrections, which is a consequence of the 
appearance of the dimensionful parameter $m^2$. In the opposite limit 
$p^2 \gg m^2$, we find
\bal
\Gamma^\perp_{AA} (p^2) &= m^2 + \frac{3}{4} \, m^2 \frac{N g^2}{(4\pi)^2}
\left( \overline{\frac{2}{\epsilon}} - \ln \frac{m^2}{\kappa^2} 
+ \frac{79}{6} \right) + m^2 (\delta Z_{m^2} + \delta Z_A) \n \\
&\phantom{=} {}+ p^2 - \frac{13}{6} \, p^2 \frac{N g^2}{(4\pi)^2}
\left( \overline{\frac{2}{\epsilon}} - \ln \frac{p^2}{\kappa^2} 
+ \frac{97}{78} \right) + p^2 \delta Z_A \,, \label{GAtuv} \\
\Gamma^\parallel_{AA} (p^2) &= m^2 + \frac{3}{4} \, m^2 \frac{N g^2}{(4\pi)^2}
\left( \overline{\frac{2}{\epsilon}} - \ln \frac{p^2}{\kappa^2} 
+ \frac{4}{3} \right) + m^2 (\delta Z_{m^2} + \delta Z_A) \n \\
&\phantom{=} {}+ \frac{3}{4} \, m^2 \, \frac{m^2}{p^2}
\frac{N g^2}{(4\pi)^2} \left( -\ln \frac{p^2}{m^2} - \frac{1}{2} \right)
\,, \label{GAluv} \\
\Gamma_{c\bar{c}} (p^2) &= p^2 - \frac{3}{4} \, p^2 \frac{N g^2}{(4\pi)^2}
\left( \overline{\frac{2}{\epsilon}} - \ln \frac{p^2}{\kappa^2} 
+ \frac{4}{3} \right) + p^2 \delta Z_c \n \\
&\phantom{=} {}- \frac{3}{4} \, p^2 \, \frac{m^2}{p^2}
\frac{N g^2}{(4\pi)^2} \left( -\ln \frac{p^2}{m^2} - \frac{1}{2} \right)
\,. \label{Gcuv}
\eal

Once the counterterms are determined from the normalization conditions,
the corresponding flow functions can be extracted as follows:
\bal
\gamma_A &= \mu^2 \frac{d}{d \mu^2} \ln Z_A 
= \mu^2 \frac{d}{d \mu^2} \, \delta Z_A \,, \label{gammaA} \\
\gamma_c &= \mu^2 \frac{d}{d \mu^2} \ln Z_c 
= \mu^2 \frac{d}{d \mu^2} \, \delta Z_c \,, \label{gammac} \\
\beta_{m^2} &= \mu^2 \frac{d}{d \mu^2} \, m^2
= -m^2 \left( \mu^2 \frac{d}{d \mu^2} \right) \delta Z_{m^2} \,. \label{betam2}
\eal
The last expressions in each line are only correct to one-loop order. The
$\mu^2$-derivatives are to be taken with the bare parameters held fixed,
which are, however, equal to the renormalized parameters to the order
considered. The Slavnov-Taylor identity \eqref{STIr} for the renormalized
quantities is fulfilled to one-loop order if and only if
\be
\delta Z_A + \delta Z_c + \delta Z_{m^2} = 0 \,, \label{STIc}
\ee
see Eq.\ \eqref{STIz}, in which case
\be
\beta_{m^2} = m^2 (\gamma_A + \gamma_c) \,. \label{betam}
\ee

In Ref.\ \cite{TW011}, the renormalized coupling constant $g$ was defined
from the renormalized proper ghost-gluon vertex in the Taylor limit of
vanishing ghost momentum. There are no quantum corrections to the vertex
in this limit \cite{Tay71}, so that $g$ is simply given by
\be
g = Z_A^{1/2} Z_c \, g_B \,, \label{defg}
\ee
and the beta function is
\be
\beta_g = \mu^2 \frac{d}{d \mu^2} \, g 
= \frac{g}{2} \, (\gamma_A + 2 \gamma_c) \,. \label{betag}
\ee
In the next subsection, we shall consider, as an alternative, the definition
of $g$ from the ghost-gluon vertex at the symmetry point. The corresponding
one-loop corrections depend on the momentum scale, however, they turn out
to be suppressed in the IR limit $p^2 \ll m^2$ and become constant in the
UV limit $p^2 \gg m^2$, so that Eq.\ \eqref{betag} is approximately
fulfilled in both limits even for this alternative definition of the 
renormalized coupling constant. We refer the reader to the following 
subsection for details.

Let us now look at the two renormalization schemes considered in Ref.\
\cite{TW011} again. In both schemes, $Z_c$ is determined from condition
\eqref{TWs3}. A look at Eq.\ \eqref{Gcir} shows that, in the IR limit
$\mu^2 \ll m^2$, the contribution to $\delta Z_c$ of the order 
$(\mu^2/m^2)^0$ is $\mu^2$-independent, so that the anomalous dimension
$\gamma_c$ in Eq.\ \eqref{gammac} is of the order $(\mu^2/m^2)$ and
hence suppressed relative to $\gamma_A$ (see below). As a result, the
beta function \eqref{betag} for the coupling constant, and particularly its 
sign, is given by $\gamma_A$ alone in the IR limit.

In the renormalization scheme with the condition \eqref{TWu2}, Eq.\
\eqref{GAlir} shows that $(\delta Z_{m^2} + \delta Z_A)$ is 
$\mu^2$-independent, hence in the IR limit $\mu^2 \ll m^2$ all the 
$\mu^2$-dependence of $\delta Z_A$ (and thus of $\delta Z_{m^2}$)
comes from the second line of Eq.\ \eqref{GAtir} for 
$\Gamma_{AA}^\perp (p^2)$. In particular, $\delta Z_A$ contains a
$\mu^2$-dependent term of the order $(\mu^2/m^2)^0$ which dominates over the
$\mu^2$-dependence of $\delta Z_c$ in the IR limit.
We read off that $\gamma_A$ and hence $\beta_g$ are negative: the integration 
of the renormalization group equations will lead to a Landau pole for 
the running coupling constant in the IR. The explicit evaluation of
$(\delta Z_{m^2} + \delta Z_A)$ and $\delta Z_c$ from the expressions
\eqref{GAlir} and \eqref{Gcir} [with the normalization condition \eqref{TWs3}]
shows that the Slavnov-Taylor identity \eqref{STIc} for the one-loop 
counterterms is violated by terms of the order of $(\mu^2/m^2)$ [times
$N g^2/(4\pi)^2$] in the IR limit, and, using Eq.\ \eqref{Gcuv}, by terms 
of the order of $(m^2/\mu^2)^0$ in the 
UV limit, where we formally consider the logarithmic terms
$\ln (\mu^2/m^2)$ to be of the order one.

For the IR safe scheme with condition \eqref{TWs2},
$\delta Z_A$ is conveniently extracted from the difference
\be
\Gamma^\perp_{AA} (\mu^2) - \Gamma^\parallel_{AA} (\mu^2) = \mu^2 \,.
\label{GAdiff}
\ee
Taking the difference of Eqs.\ \eqref{GAtir} and \eqref{GAlir} 
again shows that the $\mu^2$-dependence of $\delta Z_A$ is of the
order $(\mu^2/m^2)^0$ and dominates over the $\mu^2$-dependence of
$\delta Z_c$ in the IR limit. Then the beta function for the coupling
constant is
\be
\beta_g = \frac{g}{2} \left( -\frac{1}{12} + \frac{1}{4} \right)
\frac{N g^2}{(4\pi)^2}
= \frac{1}{12} \frac{N g^2}{(4\pi)^2} \, g \label{TWbetag}
\ee
in the IR limit. The beta function is positive and the corresponding IR stable
fixed point of the coupling constant is trivial. We have separated
the contributions of $\Gamma_{AA}^\perp (p^2)$ and $\Gamma_{AA}^\parallel (p^2)$
to $\gamma_A$, and hence to $\beta_g$, in Eq.\ \eqref{TWbetag}. Note that the
positive contribution to the beta function stems from the
$p^2$-dependence of $\Gamma^\parallel_{AA} (p^2)$. The triviality of
the IR fixed point of the coupling constant is the reason why one-loop
perturbation theory works so well in the IR regime when a gluon mass term
is added to the action \cite{TW10}.

Since the Slavnov-Taylor identity \eqref{STIr} 
for the renormalized quantities is exactly fulfilled
in this renormalization scheme, the beta function for the mass parameter
can directly be determined from Eq.\ \eqref{betam}. The result is
\be
\beta_{m^2} = \frac{1}{6} \frac{N g^2}{(4 \pi)^2} \, m^2 \label{TWbetam}
\ee
in the IR limit, which implies that $m^2$, just like $g^2$, tends
logarithmically to zero with $\mu^2$. In fact, $m^2$ becomes proportional to
$g^2$ in the IR limit. Even though $m^2$ vanishes in the limit
$\mu^2 \to 0$, $\Gamma_{AA}^\perp (p^2)$ and $\Gamma_{AA}^\parallel (p^2)$
tend towards a finite constant for $p^2 \to 0$. The reason is the behavior
of $\delta Z_A$ (and hence of $\gamma_A$) for small $\mu^2$, 
cf.\ Eqs.\ \eqref{2pARG}, \eqref{2pAparRG}, \eqref{runningmir} and 
\eqref{intgamAirLO} in the next section.
Note that, since the fall-off of the mass parameter in the IR limit is only
logarithmic, our characterization of the IR regime as the region where
$\mu^2 \ll m^2$ holds, is still adequate.

For the UV limit in the same renormalization scheme, we use the expressions
\eqref{GAtuv}--\eqref{Gcuv} and apply the conditions \eqref{TWs3} and
\eqref{GAdiff} to extract $\delta Z_A$ and $\delta Z_c$. To the 
leading order $(m^2/\mu^2)^0$, the well-known
perturbative values for the anomalous dimensions $\gamma_A$ and $\gamma_c$
in the Landau gauge are recovered, and, 
in particular, with the help of Eq.\ \eqref{betag} the familiar result
\be
\beta_g = -\frac{11}{6} \frac{N g^2}{(4\pi)^2} \, g \label{UVbetag}
\ee
for the beta function of the coupling constant in the usual
perturbative Yang-Mills theory is obtained in the UV limit. For the 
beta function of the mass, Eq.\ \eqref{betam} gives
\be
\beta_{m^2} = -\frac{35}{12} \frac{N g^2}{(4\pi)^2} \, m^2 \,. 
\label{UVbetam}
\ee
Consequently, the mass parameter goes logarithmically to zero also 
in the UV limit $\mu^2 \to \infty$, and the usual (not extended) BRST
symmetry is recovered for $p^2 \gg m^2$. For example, the 
proper gluonic two-point function becomes transverse for 
$p^2 \gg m^2$, which is most clearly seen after 
integrating the renormalization group equations, cf.\ the next section.

We conclude from the foregoing discussion 
of the IR limit in the two different renormalization 
schemes that extracting the $\mu^2$-dependence of
$\delta Z_A$ from $\Gamma^\perp_{AA} (p^2)$
alone will not lead to a positive beta function for the coupling constant
in the IR, and hence cannot avoid the appearance of a Landau pole. Indeed,
the \emph{a priori} appealing possibility to determine both $Z_A$ and
$Z_{m^2}$, and thus $\gamma_A$ and $\beta_{m^2}$, from the transverse part
$\Gamma^\perp_{AA} (p^2)$ alone fails in the sense that it necessarily
leads to a negative beta function for the coupling constant. Such
a renormalization scheme would seem to be appealing at first because,
in the Landau gauge, all quantum corrections to any proper
n-point function (with the exception of those n-point functions that
involve the Nakanishi-Lautrup field) are determined by the transverse part
$\Gamma^\perp_{AA} (p^2)$ alone given that the gluon propagator function is 
the inverse of $\Gamma^\perp_{AA} (p^2)$, see Eq.\ \eqref{propagators}.
One would then expect normalization 
conditions like
\bal
\frac{\d}{\d p^2} \, \Gamma^\perp_{AA} (p^2) \Big|_{p^2 = \mu^2}
&= 1 \,, \label{tt1} \\
\Gamma^\perp_{AA} (\mu^2) &= m^2 + \mu^2 \,, \label{tt2}
\eal
that involve only the transverse part,
to lead to an optimal approximation of the full gluon propagator at momentum
scales of the order of $\mu^2$, and thus include as many higher-loop
contributions as possible in the one-loop renormalization group-improved
proper n-point functions. However, it is easy to see from Eq.\ 
\eqref{GAtir} that such normalization conditions lead to a negative beta 
function, and hence to a Landau pole, for the coupling constant in the IR.

In the example given in Eqs.\ \eqref{tt1} and \eqref{tt2} above,
it was necessary to involve the $p^2$-derivative of the 
proper two-point function 
in the normalization conditions to be able to determine the two 
counterterms $\delta Z_A$ and $\delta Z_{m^2}$ from the single function 
$\Gamma^\perp_{AA} (p^2)$ (at the same scale $p^2 = \mu^2$, in order to 
obtain a good approximation of the one-loop propagator at that scale). Even 
in renormalization schemes that involve, in addition, the longitudinal part
$\Gamma^\parallel_{AA} (p^2)$, however, one may employ the $p^2$-derivatives 
of the proper n-point functions in the normalization conditions. Thus, for an
alternative to the renormalization scheme of Eqs.\ 
\eqref{TWs1}--\eqref{TWs3}, one may replace Eq.\ \eqref{GAdiff} with
\be
\frac{\d}{\d p^2} \left( \Gamma^\perp_{AA} (p^2) 
- \Gamma^\parallel_{AA} (p^2) \right) \! \bigg|_{p^2 = \mu^2} = 1 \label{TWd1}
\ee
and, by analogy,
\be
\frac{\d}{\d p^2} \, \Gamma_{c\bar{c}} (p^2) \Big|_{p^2 = \mu^2} = 1 \,.
\label{TWd3}
\ee
The condition \eqref{TWd1} appears more natural when one 
considers the decomposition of the gluonic two-point function not 
in its transverse and longitudinal parts, but rather in
\be
\Gamma_{A_\mu A_\nu}(p) = \left( \Gamma^{\perp}_{AA}(p^2) 
- \Gamma^\parallel_{AA} (p^2) \right) \left( \delta_{\mu\nu} 
- \frac{p_\mu p_\nu}{p^2} \right)
+ \Gamma^{\parallel}_{AA}(p^2) \delta_{\mu \nu} \,, \label{decomp1}
\ee
which mimics the grouping of terms in the classical action \eqref{CF}
[see also Eq.\ \eqref{local} in this context].

A third normalization condition is required to complete the renormalization
scheme given by Eqs.\ \eqref{TWd1} and \eqref{TWd3} so far. One might
be inclined to use the condition \eqref{tt2} to this end, however, as we
shall now argue, such a scheme runs into trouble in the UV limit
$\mu^2 \gg m^2$: applying condition \eqref{TWd1} to the approximate results
\eqref{GAtuv} and \eqref{GAluv} and concentrating on the leading contributions
in the UV limit which come exclusively from the transverse part, 
one obtains
\be
\delta Z_A = \frac{13}{6} \frac{N g^2}{(4\pi)^2}
\left( \overline{\frac{2}{\epsilon}} - \ln \frac{\mu^2}{\kappa^2} 
+ \frac{19}{78} \right) \label{dZAUVtrouble}
\ee
[plus terms of the order of $(m^2/\mu^2)$]. Substituting this result back
into Eq.\ \eqref{GAtuv} leaves one with
\be
\Gamma^\perp_{AA} (\mu^2) = \mu^2 - \frac{13}{6} \, \mu^2
\frac{N g^2}{(4\pi)^2} + m^2 \delta Z_{m^2} \,,
\ee
neglecting terms of the order of $\mu^2 (m^2/\mu^2)$ [i.e., terms
which are supressed by one power of $(m^2/\mu^2)$ relative to the leading
order $\mu^2$]. Applying the 
normalization condition \eqref{tt2} to the latter expression, 
$\delta Z_{m^2}$ would have to be of the order of $\mu^2/m^2$ and, e.g.,
\be
\beta_{m^2} = -\frac{13}{6} \, \mu^2 \frac{N g^2}{(4\pi)^2}
\label{betamincons}
\ee
to leading order. Then the renormalized Slavnov-Taylor identity is 
flagrantly violated, see Eq.\ \eqref{STIc}, and we shall show in the 
next section through the approximate integration of the renormalization group 
equations that in this scheme the 
longitudinal part $\Gamma^\parallel_{AA} (p^2)$
of the proper gluonic two-point function grows without bound for large $p^2$,
so that one does not recover the usual, not extended, BRST symmetry in the UV
limit.

Instead of using Eq.\ \eqref{tt2}, we hence combine the normalization 
conditions \eqref{TWd1} and \eqref{TWd3} with the condition \eqref{TWs2},
\be
\Gamma^\parallel_{AA} (\mu^2) = m^2 \,, \label{TWd2}
\ee
in order to complete the renormalization scheme. Note that it turns
out to be necessary to involve the longitudinal part of the proper gluonic
two-point function in the normalization conditions, yet again, this time in
order to obtain the correct UV behavior.

In the next section, we will compare the ghost and gluon propagators 
obtained from the renormalization group improvement in the different
renormalization schemes to the lattice data. As far as this comparison is
concerned, there will be no compelling reason to prefer the scheme defined
by the normalization conditions \eqref{TWd1}, \eqref{TWd3} and \eqref{TWd2}
over Tissier-Wschebor's IR safe scheme [defined by Eqs.\ 
\eqref{TWs1}--\eqref{TWs3}, or with Eq.\ \eqref{GAdiff} replacing Eq.\ 
\eqref{TWs1}] or vice versa, however, there is one interesting aspect
to the new scheme: the renormalized Slavnov-Taylor identity \eqref{STIr}
is violated. This happens because we have changed the normalization condition
\eqref{TWs3} to \eqref{TWd3}, so that the condition \eqref{TWs2} [or
\eqref{TWd2}] does not guarantee the Slavnov-Taylor identity for the
renormalized quantities any more. The violation 
of the Slavnov-Taylor identity
is of the same order as in the (not IR safe) renormalization scheme
defined by the conditions \eqref{TWs1}, \eqref{TWs3} and \eqref{TWu2},
i.e., the corrections to Eq.\ \eqref{STIc} are of the order of
$(\mu^2/m^2)$ in the IR limit, and of the order of $(m^2/\mu^2)^0$ 
in the UV limit \cite{footn2}. 

One may argue that the
normalization conditions \eqref{TWd1}, \eqref{TWd3} and \eqref{TWd2} 
do not define a proper renormalization of the theory since they do not
respect the extended BRST symmetry of the classical action. Despite the fact
that the Slavnov-Taylor identity \eqref{STIr} is violated, the 
resulting flow functions coincide with the ones of the Tissier-Wschebor scheme
\eqref{TWs1}--\eqref{TWs3} in both the IR and the UV limit, in particular,
the new scheme is IR safe and the usual, not extended, BRST symmetry
is recovered in the UV limit. It then appears that the exact symmetry 
of the renormalized theory under the extended BRST transformation \eqref{BRST} 
is not essential to the quantitative results for the 
two-point functions (see also the next section). To our purpose
of describing, in the simplest possible manner, Yang-Mills theory including 
the restriction of the functional integration to the Gribov region, 
the transformation \eqref{BRST} and the Slavnov-Taylor identity \eqref{STIr}
by themselves are not of primordial interest, because they are special to 
the Curci-Ferrari model. As briefly mentioned in the Introduction,
for a better or more complete description of the properties of Yang-Mills 
theory, in the future we may consider extensions of the 
Faddeev-Popov action that go beyond a gluon mass term.

The use of the $p^2$-derivative actually allows for a greater flexibility in 
the normalization conditions: one may interpolate between the conditions
\eqref{tt1} and \eqref{TWd1} by introducing a parameter $\zeta$ and
replacing the condition \eqref{TWd1} with
\be
\frac{\d}{\d p^2} \left( \Gamma^\perp_{AA} (p^2) 
- \zeta \, \Gamma^\parallel_{AA} (p^2) \right) \! \bigg|_{p^2 = \mu^2} = 1 
\,, \label{TWz1}
\ee
while keeping the conditions \eqref{TWd3} and \eqref{TWd2} [the condition
\eqref{tt2} cannot be used, for the same reason as before]. The
normalization condition \eqref{TWz1} is consistent with the classical 
action \eqref{CF} for any value of $\zeta$ since the longitudinal part of the 
second derivative of the classical action with respect to the gluon field (the 
mass squared) is $p^2$-independent. We shall refer
to this class of normalization conditions as (general) derivative schemes. Note that a
condition analogous to \eqref{TWz1} but without the $p^2$-derivative, like
\be
\Gamma^\perp_{AA} (\mu^2) - \zeta \, \Gamma^\parallel_{AA} (\mu^2)
= \mu^2 + (1 - \zeta) m^2 \,, \label{TWsalt}
\ee
would be completely equivalent to the condition \eqref{TWs1} in the
original IR safe scheme of Tissier and Wschebor as can explicitly be seen
by adding
\be
\zeta \, \Gamma^\parallel_{AA} (\mu^2) = \zeta m^2
\ee
to Eq.\ \eqref{TWsalt}. The normalization condition \eqref{TWz1} corresponds 
to the decomposition
\be
\Gamma_{A_\mu A_\nu}(p) = \left( \Gamma^{\perp}_{AA}(p^2) 
- \zeta \, \Gamma^\parallel_{AA} (p^2) \right) \left( \delta_{\mu\nu} 
- \frac{p_\mu p_\nu}{p^2} \right)
+ \Gamma^{\parallel}_{AA}(p^2) \left( \zeta \, \delta_{\mu \nu} 
+ (1 - \zeta) \frac{p_\mu p_\nu}{p^2} \right) \label{decompz}
\ee
of the gluonic two-point function which interpolates linearly between
the standard decomposition \eqref{decomp} for $\zeta = 0$ and the
decomposition \eqref{decomp1} analogous to the classical action for 
$\zeta = 1$.

With the help of the approximate expressions \eqref{GAtir} and \eqref{GAlir},
we find for the beta function in the IR limit [cf.\ Eq.\ \eqref{TWbetag}]
\be
\beta_g = \frac{g}{2} \left( -\frac{1}{12} + \frac{\zeta}{4} \right)
\frac{N g^2}{(4\pi)^2} \,,
\ee
so that the renormalization scheme is IR safe, with a trivial IR fixed point
of the coupling constant, for $\zeta > 1/3$. For $\zeta < 1/3$, on the other
hand, the integration of the renormalization group equations generates a
Landau pole. The case $\zeta = 1/3$ is particularly interesting, and the
renormalization group equations will be numerically integrated for this
case in the next section. We will find that the renormalized coupling
constant $g$ converges to a nontrivial constant in the limit 
$\mu^2 \to 0$, however, the actual value of $g$ in this limit depends on the 
initial conditions. The beta function for the mass parameter is
\be
\beta_{m^2} = m^2 \left( -\frac{1}{12} + \frac{\zeta}{4} \right)
\frac{N g^2}{(4 \pi)^2}
\ee
in the IR limit, so that $m^2$ tends to zero for $\mu^2 \to 0$ along with $g^2$
in the case $\zeta > 1/3$, and it diverges like $g^2$ at a finite scale
$\mu^2$ for $\zeta < 1/3$. For the case $\zeta = 1/3$, we shall find in
the next section that $m^2$ has a nonzero limit for $\mu^2 \to 0$.

In the UV limit, on the other hand, one again finds the beta functions
\eqref{UVbetag} and \eqref{UVbetam} of Tissier-Wschebor's IR safe scheme,
and also of the simple derivative scheme with parameter $\zeta = 1$. In
particular, the usual (not extended) BRST symmetry obtains in this 
limit. The Slavnov-Taylor identity in the form \eqref{STIc} is violated both 
in the IR and the UV limit, with corrections of the same order as in the case
$\zeta = 1$.

The last renormalization scheme we shall consider is motivated by 
the observation that the standard decomposition \eqref{decomp} 
of the proper gluonic two-point function is particularly natural in the UV 
limit where its longitudinal part has to vanish as a consequence 
of the reinstatement of the
proper (not extended) BRST symmetry, while the decomposition
\eqref{decomp1} is rather adequate in the IR limit where the difference
$(\Gamma^{\perp}_{AA}(p^2) - \Gamma^\parallel_{AA} (p^2))$ goes to zero, 
see Eq.\ \eqref{local}. It might then seem to
be a good idea to allow the parameter $\zeta$ that interpolates between the
two decompositions in Eq.\ \eqref{decompz} to depend on the renormalization
scale $\mu$. The simplest choice is to replace $\zeta$ in the normalization
condition \eqref{TWz1} with
\be
\zeta = \frac{1}{1 + (\mu^2/m^2)} \,, \label{zetadyn}
\ee
so that $\zeta \to 1$ for $\mu^2 \ll m^2$ and $\zeta \to 0$ for 
$\mu^2 \gg m^2$ [the other normalization conditions \eqref{TWd3} and
\eqref{TWd2} go unchanged]. Note that $\zeta$ is defined in Eq.\
\eqref{zetadyn} in terms of the renormalized mass (at the renormalization
scale $\mu$). We shall refer to this renormalization scheme
as the scale-dependent derivative scheme.

The flow functions in this scale-dependent derivative scheme show the same
behavior in the IR and UV limits as in Tissier-Wschebor's IR safe scheme
and the $\zeta=1$ derivative scheme, and the Slavnov-Taylor identity
\eqref{STIc} is violated in the two limits to the same order as in the
latter derivative scheme.

\subsection{Renormalization of the coupling constant}

Another way in which a quantitative difference can be obtained 
for the resummed two-point functions arises from a different choice of the 
momentum configuration at which 
the renormalized coupling constant is defined from the renormalized proper 
ghost-gluon vertex. In Ref.\ \cite{TW011} the Taylor scheme was used, i.e., the
limit of vanishing momentum for the external ghost. In this configuration, the 
quantum corrections to the vertex vanish to any perturbative order in the 
Landau gauge as a result of Taylor's nonrenormalization theorem \cite{Tay71}, 
hence the renormalized coupling constant 
is given by the bare one multiplied with 
the field renormalization constants, see Eq.\ \eqref{defg}. Consequently,
the beta function for the coupling constant can be expressed in terms of 
the anomalous dimensions $\gamma_A$ and $\gamma_c$, Eq.\ \eqref{betag}.

We shall also consider, as an alternative, a scheme where the renormalized 
coupling is defined from the ghost-gluon vertex at the symmetry point 
$k^2=p^2=(p-k)^2=\mu^2$, where the momenta $k, -p, p-k$ (taken as
\emph{incoming}, cf.\ the diagrams in Appendix \ref{threep}) are those of the
external ghost, antighost, and gluon, in this order. To compare with, in the
Taylor limit we have $k^2=0$ and $p^2=(p-k)^2=\mu^2$. The global symmetries of
the action (disregarding the BRST symmetry, for the time being) restrict the
proper bare ghost-gluon vertex to be of the general form
\be
\Gamma^B_{c^b \bar c^a A^c_\mu}(k,-p,p-k)=-ig_Bf^{abc}\big[ p_\mu A \big(k^2,p^2,(p-k)^2
  \big)+(p_\mu-k_\mu) B\big(k^2,p^2,(p-k)^2\big) \big] \,,
\ee
where $A$ and $B$ are Lorentz scalars, and thus are functions of the
invariants $k^2$, $p^2$ and $(p-k)^2$. The dressing functions $A$ and $B$ take
the values one and zero, respectively, at tree level.

As is common in Landau gauge, we shall concentrate on the transverse part
of the vertex, with respect to the gluon momentum, and define the
renormalized coupling at the symmetry point as
\be
g =  Z_A^{1/2}Z_c\, g_B\, A\big(k^2=p^2=(p-k)^2=\mu^2\big) \,. \label{symmg}
\ee
The reason for considering the transverse part is that, in the 
Landau gauge, all (improper) n-point correlation functions, and the
transverse parts of all proper n-point functions, are determined from the
transverse part of this vertex (and the transverse parts of all other
vertices) alone. Notice, however, that the longitudinal part of the proper 
gluonic two-point function has turned out to be important in the 
renormalization of the theory, as we have emphasized in the previous 
subsection.

Up to the one-loop level, we find for the dressing function at the symmetry
point in the IR limit $\mu^2 \ll m^2$,
\be
A(\mu^2) = 1 + \frac{1}{288} \frac{N g^2}{(4 \pi)^2} \,
\frac{\mu^2}{m^2} \left( -30 \ln \frac{\mu^2}{m^2} + 235 - 12 \tilde{J}_0
\right) \,,
\ee
where
\be
\tilde{J}_0 = -2 \int_0^1 d x\, \frac{\ln x}{1 - x + x^2} \,.
\ee
In the UV limit $\mu^2 \gg m^2$, on the other hand,
\be
A(\mu^2) = 1 + \frac{1}{24} \frac{N g^2}{(4 \pi)^2}
\big( 18 + 5 \tilde{J}_0 \big) \,,
\ee
where we have suppressed corrections of the order $(m^2/\mu^2)$ to the
term in parenthesis. The complete expressions, separated into the 
contributions of the two one-loop diagrams, are given in Appendix \ref{threep},
with a brief description of how they are obtained from the Feynman
parameterization and the Cheng-Wu trick \cite{CW87}. Note, in particular, 
that the dressing function $A$ does vary with the scale $\mu^2$, while the
corresponding expression calculated with a massless gluon propagator would
be $\mu^2$-independent (for dimensional reasons).

It is clear from the definition \eqref{symmg} at the symmetry point that
the beta function for the coupling constant inherits an extra term as
compared to the Taylor scheme:
\be
\beta_g = \mu^2 \frac{d}{d\mu^2} \, g
= \frac{g}{2} \left( \gamma_A + 2 \gamma_c + 2 \mu^2 \frac{d}{d \mu^2}
A(\mu^2) \right) \,. \label{betagsymm}
\ee
We read off from the explicit expressions above that the contribution of
the dressing function $A$ to the beta function
is of the order $(\mu^2/m^2)$ in the IR limit and
of the order $(m^2/\mu^2)$ in the UV limit, and is thus suppressed with
respect to the leading contributions from the anomalous dimensions in both
limits. Therefore, the beta function in the renormalization scheme at the 
symmetry point quantitatively changes, with respect to the one in the Taylor 
scheme, only in the intermediate region of momenta.

In the context of the scale-dependent derivative scheme introduced at the end 
of the previous section, it is natural to think of an interpolation between the
two definitions of the renormalized coupling constant as the renormalization
scale varies between the UV and IR limits. For small external momenta, the most
important contributions to the one-loop diagrams are expected to come from 
the region in internal momentum space where the momentum of the (massless)
internal ghost propagator is small, i.e., close to the Taylor limit of the
ghost-gluon vertices, while for large momenta all the propagators become
effectively massless and the symmetry point, which is 
also the most common choice for the 
definition of a renormalized coupling constant, seems more adequate. We thus 
propose a third definition of the renormalized coupling as
\be
g =  Z_A^{1/2}Z_c\, g_B \big[ \zeta + (1 - \zeta) A(\mu^2) \big] \,, 
\label{dyng}
\ee
with a $\mu^2$-dependent parameter $\zeta$ that tends to one in the IR limit
and to zero in the UV limit, as in Eq.\ \eqref{zetadyn}. The beta function
is, correspondingly,
\be
\beta_g = \frac{g}{2} \left( \gamma_A + 2 \gamma_c + 2 \mu^2 \frac{d}{d \mu^2}
\big[ \zeta + (1 - \zeta) A(\mu^2) \big] \right) \,, \label{betagdyn}
\ee
where the $\mu^2$-dependence of $\zeta$ has to be taken into account in the
differentiation. As before, the beta function
changes with respect to the one in the Taylor scheme only in the 
intermediate momentum regime.

In principle, it would be possible to define the coupling constant using
other proper renormalized vertices than the ghost-gluon vertex, for
example, the three-gluon vertex. We shall explore these possibilities in
a future publication.

\section{Renormalization group improvement}

\subsection{Integration of the renormalization group equations \label{intRGE}}

Before we properly discuss the implementation of the renormalization group,
let us comment on the changes that arise when one goes from one
renormalization scheme to another (or from one value of the renormalization
scale to another). In different renormalization schemes defined by their 
corresponding normalization conditions, the
renormalized proper n-point functions 
can differ at most by finite constant factors (the ratios of the field
renormalization constants), provided they refer to the same bare theory. This 
property holds to any fixed perturbative order and, of course, it is true for 
the fully dressed (exact) proper n-point functions, and it also carries over 
to any physical quantities that are built from the n-point functions (except 
that the physical quantities are unaffected by the renormalization scheme 
dependent constant factors).

This statement can become rather subtle when it comes to the preservation
of symmetries in the renormalized theory [as an example, think of the case
of the axial U(1) anomaly]. Here we will focus on the action \eqref{CF} 
and the massive extension \eqref{BRST} of the BRST symmetry, where no 
such subtlety arises. As we have seen in the previous section, the 
Slavnov-Taylor identity \eqref{STI} for the bare quantities implies
the identity
\be
\Gamma_{c\bar{c}} (p^2) \, \Gamma^\parallel_{AA} (p^2) = (Z_A Z_c Z_{m^2})
p^2 m^2 \label{STIzrep}
\ee
for the renormalized quantities, in any renormalization scheme. In 
Tissier-Wschebor's original IR safe scheme one has $Z_A Z_c Z_{m^2} = 1$, so
that the Slavnov-Taylor identity properly holds in the renormalized theory
(to any perturbative order, and for any value of the renormalization scale), 
i.e.,
\be
\Gamma_{c\bar{c}} (p^2) \, \Gamma^\parallel_{AA} (p^2) = p^2 m^2 \,.
\label{STIrrep}
\ee
In the derivative schemes (including the simple scheme with $\zeta = 1$
and the scale-dependent derivative scheme), however, $Z_A Z_c Z_{m^2} \neq 1$, so that
the renormalized counterpart of the Slavnov-Taylor identity does not take
the form \eqref{STIrrep}. Writing Eq.\ \eqref{STIzrep} in the form
\be
\frac{\Gamma_{c\bar{c}} (p^2) \, \Gamma^\parallel_{AA} (p^2)}{p^2} 
= (Z_A Z_c Z_{m^2}) m^2 \,, \label{STIcomb}
\ee
we can still conclude that the combination 
$(\Gamma_{c\bar{c}} (p^2) \, \Gamma^\parallel_{AA} (p^2)/p^2)$ is a
$p^2$-independent (finite) constant, even if it is not equal to $m^2$.
Note that this conclusion holds to any perturbative order since it 
follows from the exact property \eqref{STIzrep} of the renormalized theory.

More generally, in the Tissier-Wschebor scheme the renormalized n-point 
functions satisfy all the Slavnov-Taylor identities which are derived from
the extended BRST symmetry, in their usual form. This is no longer true for 
the derivative schemes, however, the extended BRST symmetry of the bare action
is still present in the renormalized theory, if somewhat hidden. More
precisely, the Slavnov-Taylor identities for the n-point functions get 
modified by finite factors that involve the renormalization constants $Z_A$, 
$Z_c$, $Z_{m^2}$ and $Z_g = g_B/g$, just as in Eq.\ \eqref{STIzrep}. We shall
come back to this point later in this subsection.

We will now turn to the renormalization group improvement. Even if, in
four space-time dimensions, straightforward one-loop perturbation theory for 
the action \eqref{CF} reproduces the IR lattice data for the gluon and ghost 
propagators with unexpected precision \cite{TW10}, for a reliable description
of the large-momentum region it is necessary to improve the perturbative
results by applying the renormalization group. This is a well-known fact
in the perturbative regime of the usual massless formulation of Yang-Mills
theory, with which our present description coincides in the UV limit.

An intuitive understanding of the renormalization group improvement begins
with the observation that (renormalized) perturbation theory provides an 
excellent description when the relevant momentum scales are close to the 
renormalization scale (in a typical case, the logarithms that arise in the 
perturbative expansion are then small). What the renormalization group does
is, in simple terms, to continuously move the renormalization scale along
with the momentum scale considered.

Technically, the Callan-Symanzik renormalization group equations 
\cite{Cal70, Sym70}
exploit the trivial fact that the bare (unrenormalized) n-point functions do 
not depend on the renormalization scale. Concerning the name of
the equations, we remark that the original Callan-Symanzik equations were 
actually derived in an on-shell renormalization scheme. The version of the
equations where the renormalized n-point functions are defined at an off-shell
spacelike momentum, was formulated by Georgi and Politzer \cite{GP76}. 
As a simple example, one has for the transverse part of the proper gluonic
two-point function that
\bal
\mu^2 \frac{d}{d \mu^2} \, \Gamma^\perp_{AA} (p^2,\mu^2) 
&= \mu^2 \frac{d}{d \mu^2} \Big( Z_A (\mu^2) \, \Gamma^{\perp \, B}_{AA} 
(p^2) \Big) \n \\
&= \left( \mu^2 \frac{d}{d \mu^2} \, Z_A (\mu^2) \right)
\Gamma^{\perp \, B}_{AA} (p^2) \n \\
&= \gamma_A (\mu^2) \, \Gamma^\perp_{AA} (p^2,\mu^2) \,,
\eal
where we have used the general (all-order) definition of
the anomalous dimension $\gamma_A$ [first
equation in \eqref{gammaA}] in the last step. Integrating the latter
equation relates the two-point function at two different renormalization 
scales $\mu^2$ and $\bar \mu^2$,
\be
\Gamma^\perp_{AA}(p^2, \mu^2) = \Gamma^\perp_{AA}(p^2, \bar \mu^2) \,
\exp \left( \int_{\bar \mu^2}^{\mu^2} \frac{d \mu^{\prime \, 2}}{\mu^{\prime \, 2}} \, 
\gamma_A (\mu^{\prime \, 2}) \right) \,. \label{2pmudep}
\ee
Note that this is an exact relation that the full proper gluonic two-point
function must fulfill, provided that the full (all-order) anomalous
dimension $\gamma_A$ is inserted on the right-hand side. The renormalization 
scale dependence of the two-point function in itself is usually not of much 
interest, however, we may now move the reference renormalization scale 
$\bar \mu^2$ to $p^2$, so that we can rely on perturbation
theory to evaluate $\Gamma^\perp_{AA}(p^2, \bar \mu^2)$. In the case of 
Tissier-Wschebor's original renormalization scheme, we can directly use
the normalization condition \eqref{TWs1} (at $\bar \mu^2 = p^2$) to conclude 
that
\be
\Gamma^\perp_{AA}(p^2, \mu^2) = \big( m^2(p^2) + p^2 \big)
\exp \left( -\int_{\mu^2}^{p^2} \frac{d \mu^{\prime \, 2}}{\mu^{\prime \, 2}} 
\, \gamma_A (\mu^{\prime \, 2}) \right) \,. \label{2pARG}
\ee
We have arrived at an equation that describes the $p^2$-dependence of the
gluonic two-point function [exactly, if $\gamma_A$ and $m^2 (\mu^2)$
are evaluated exactly]. By the same
argument, one arrives at an exact equation for the proper ghost two-point
function in the Tissier-Wschebor scheme,
\be
\Gamma_{c\bar{c}}(p^2, \mu^2) = p^2 \,
\exp \left( -\int_{\mu^2}^{p^2} \frac{d \mu^{\prime \, 2}}{\mu^{\prime \, 2}} \, 
\gamma_c (\mu^{\prime \, 2}) \right) \, \label{2pcRG}
\ee
where we have used the normalization condition \eqref{TWs3}.

The main interest in these exact equations lies in the fact that they
can be used to improve the expressions for the two-point functions (in our
case) that are generated by straightforward perturbation 
theory to some finite order. If one inserted the perturbative expressions as 
such for $\gamma_A$, $\gamma_c$ and $m^2$ in Eqs.\ \eqref{2pARG} and 
\eqref{2pcRG}, however, there would actually be no improvement at all. The
improvement arises from expressing the perturbative results for 
$\gamma_A (\mu^2)$ and $\gamma_c (\mu^2)$, up to the order in $g_B^2$ to which 
one has evaluated $\Gamma^{\perp \, B}_{AA} (p^2)$ and $\Gamma^B_{c\bar{c}} (p^2)$ 
in the first place, in terms of the renormalized coupling constant $g (\mu^2)$
and the renormalized mass squared $m^2 (\mu^2)$ at the same value of 
the renormalization scale (we have done so implicitly in the previous 
section). In this way, one evaluates the anomalous dimensions with the values 
of the coupling constant and the mass parameter that best describe the theory 
at the corresponding scale. 

The renormalized coupling constant and mass themselves are
obtained through the integration of the differential equations
\be
\mu^2 \frac{d}{d\mu^2} \, g (\mu^2) = \beta_g (\mu^2) \,, \qquad
\mu^2 \frac{d}{d\mu^2} \, m^2 (\mu^2) = \beta_{m^2} (\mu^2) \,. 
\label{2coupdiffeq}
\ee
The beta functions on the right-hand sides are, again, calculated in 
perturbation theory to the same order in $g_B^2$ as before and, subsequently, 
the bare coupling constant and the bare mass parameter are expressed in terms 
of the renormalized quantities $g (\mu^2)$ and $m^2 (\mu^2)$, for the reasons 
explained above. Note that, for the determination of the anomalous 
dimensions $\gamma_A$ and $\gamma_c$ and the beta functions $\beta_{m^2}$ and 
$\beta_g$ in the first place, the $\mu^2$-derivatives have to be taken 
with all bare parameters held fixed, for which it is most convenient 
to write $Z_A$, $Z_c$ and $Z_{m^2}$ [and, possibly, $A(\mu^2)$ for the beta
function of the coupling constant, see Eq.\ \eqref{betagsymm}]
in terms of the bare parameters $g_B$ and $m_B^2$. In this way, one makes sure 
that the renormalized theories with different values of the renormalization 
scale $\mu^2$ all correspond to the same bare theory. Note that, on 
the other hand, in the scale-dependent derivative scheme the definition of the
parameter $\zeta$ involves the renormalized mass at the renormalization
scale $\mu^2$, and this implicit $\mu^2$-dependence has to be taken into
account in the $\mu^2$-derivatives for the determination of the flow functions.
To the one-loop order, however, this subtlety does not affect the results for
the anomalous dimensions nor the beta functions.

Calculating the proper two-point functions from the integrated form
\eqref{2pARG} and \eqref{2pcRG} of the Callan-Symanzik equations, with
the anomalous dimensions determined according to the procedure explained 
above, is tantamount to including the contribution of the leading 
logarithmic terms to all orders in the perturbative expansion. Now, different 
renormalization schemes (like the Tissier-Wschebor scheme or the
derivative schemes, with different values of $\zeta$) correspond to different 
choices of the finite parts of the counterterms, and, with
the application of the renormalization group equations, this difference
in the finite part of the counterterms is propagated to all perturbative
orders. When one uses perturbative approximations of the flow functions to 
some finite order, there is no guarantee that this difference between
renormalization schemes that is propagated to higher orders in
perturbation theory by the renormalization group, can still be absorbed
in overall constants multiplying the n-point functions: the result of
the renormalization group improvement for the n-point functions generally
depends nontrivially on the renormalization scheme.

A related aspect of the renormalization group improvement concerns
the BRST symmetry and, in particular, the Slavnov-Taylor identity
\eqref{STIzrep}. As we have shown before, in Tissier-Wschebor's IR safe 
scheme, the Slavnov-Taylor identity for the renormalized quantities is 
fulfilled in the form \eqref{STIrrep}. In 
Subsection \ref{analytical}, we 
shall explicitly demonstrate that the identity is still valid in the same form
after the renormalization group improvement. This is, however, not true for
the derivative schemes. One might already suspect that the $\mu^2$-dependent 
factor $(Z_A Z_c Z_{m^2})$  in Eq.\ \eqref{STIcomb} could cause a problem. In
effect, we shall see in the next subsections, both numerically and
analytically, that the combination $(\Gamma_{c\bar{c}} (p^2) \, 
\Gamma^\parallel_{AA} (p^2)/p^2)$ develops a $p^2$-dependence after the
renormalization group improvement (with the flow functions
determined to one-loop order), which indicates a serious violation
of the extended (massive) BRST symmetry. As we have mentioned
in the Introduction, the addition of a gluon mass term to the 
Faddeev-Popov action may be expected to be insufficient for a complete 
description of IR Yang-Mills theory (in the minimal Landau gauge). 
Consequently, we do not consider the extended BRST symmetry, which 
is special to the Curci-Ferrari model, to be an essential ingredient 
for such a description (while one has to insist on the recovery of 
the usual, not extended, BRST symmetry in the UV limit).

To end this subsection, we will discuss the formal integration of the
Callan-Symanzik equations for the derivative schemes. First of all, as a
consequence of the normalization condition \eqref{TWs2} which is used
in all renormalization schemes, we find for the longitudinal part of the
proper gluonic two-point function
\be
\Gamma^\parallel_{AA}(p^2, \mu^2) = m^2(p^2) \,
\exp \left( -\int_{\mu^2}^{p^2} \frac{d \mu^{\prime \, 2}}{\mu^{\prime \, 2}} 
\, \gamma_A (\mu^{\prime \, 2}) \right) \,, \label{2pAparRG}
\ee
in both the Tissier-Wschebor and (all) the derivative schemes.

Now consider the normalization condition \eqref{TWz1} in the derivative 
schemes. In analogy to Eq.\ \eqref{2pmudep}, one has
\be
\Gamma^\perp_{AA}(p^2, \mu^2) - \zeta \, \Gamma^\parallel_{AA}(p^2, \mu^2)
= \Big( \Gamma^\perp_{AA}(p^2, \bar \mu^2) 
- \zeta \, \Gamma^\parallel_{AA}(p^2, \bar \mu^2) \Big)
\exp \left( \int_{\bar \mu^2}^{\mu^2} 
\frac{d \mu^{\prime \, 2}}{\mu^{\prime \, 2}} \, 
\gamma_A (\mu^{\prime \, 2}) \right) \,. \label{2pmudepz}
\ee
Let us concentrate on the static (i.e., not scale-dependent) derivative 
schemes for a moment. We first differentiate Eq.\ \eqref{2pmudepz} with 
respect to $p^2$ and then set $\bar \mu^2$ to $p^2$. The normalization 
condition \eqref{TWz1} implies that
\be
\frac{\d}{\d p^2} \Big( \Gamma^\perp_{AA}(p^2, \mu^2) - 
\zeta \, \Gamma^\parallel_{AA}(p^2, \mu^2) \Big)
= \exp \left( - \int_{\mu^2}^{p^2} 
\frac{d \mu^{\prime \, 2}}{\mu^{\prime \, 2}} \, 
\gamma_A (\mu^{\prime \, 2}) \right) \,. \label{2pARGzdif}
\ee
Integrating this equation over $p^2$ yields $(\Gamma^\perp_{AA}(p^2, \mu^2) - 
\zeta \, \Gamma^\parallel_{AA}(p^2, \mu^2))$ in terms of its value at some
reference momentum scale. We choose zero as the reference momentum, where
locality implies that
\be
\Gamma^\perp_{AA} (0, \mu^2) = \Gamma^\parallel_{AA} (0, \mu^2) \label{local2}
\ee
[see Eq.\ \eqref{local}]. Then we find
\bal
\Gamma^\perp_{AA}(p^2, \mu^2) &= (1 - \zeta) \Gamma^\parallel_{AA}(0, \mu^2)
+ \zeta \, \Gamma^\parallel_{AA}(p^2, \mu^2) \n \\
&\phantom{=} {}+ \int_0^{p^2} dp^{\prime \, 2} \, \exp \left( 
-\int_{\mu^2}^{p^{\prime 2}} \frac{d \mu^{\prime \, 2}}{\mu^{\prime \, 2}} \, 
\gamma_A (\mu^{\prime \, 2}) \right) \,, \label{2pAperpRGz}
\eal
where $\Gamma^\parallel_{AA}(0, \mu^2)$ and $\Gamma^\parallel_{AA}(p^2, \mu^2)$
are given by Eq.\ \eqref{2pAparRG}. Equation \eqref{2pAperpRGz}
simplifies for the derivative scheme with $\zeta = 1$.

In an analogous way, the normalization condition \eqref{TWd3} leads to
\be
\Gamma_{c\bar{c}}(p^2, \mu^2) = \int_0^{p^2} dp^{\prime \, 2} \, \exp \left( 
-\int_{\mu^2}^{p^{\prime 2}} \frac{d \mu^{\prime \, 2}}{\mu^{\prime \, 2}} \, 
\gamma_c (\mu^{\prime \, 2}) \right) \,, \label{2pcRGz}
\ee
where we have used for the integration constant at $p^2 = 0$ that
$\Gamma_{c\bar{c}}(0, \mu^2) = 0$. The latter relation is valid to all orders 
in perturbation theory, and it can also be deduced from the identity 
\eqref{STI}, as long as $\Gamma^\parallel_{AA}(0, \mu^2) \neq 0$.

Finally, we comment on the scale-dependent derivative scheme. In this case, 
it is necessary to specify at what scale $\zeta$ is evaluated in Eq.\
\eqref{2pmudepz}. We take that scale to be $\bar \mu^2$ (on both sides of the
equation) in order to be able to take advantage of the normalization
condition \eqref{TWz1}. Then, after differentiating with respect to
$p^2$ and putting $\bar \mu^2 = p^2$, we have, instead of Eq.\ 
\eqref{2pARGzdif},
\be
\frac{\d}{\d p^2} \, \Gamma^\perp_{AA}(p^2, \mu^2) 
= \zeta (p^2) \, \frac{\d}{\d p^2} \, \Gamma^\parallel_{AA}(p^2, \mu^2)
+ \exp \left( - \int_{\mu^2}^{p^2} 
\frac{d \mu^{\prime \, 2}}{\mu^{\prime \, 2}} \, 
\gamma_A (\mu^{\prime \, 2}) \right) \,.
\ee
Use of the locality condition \eqref{local2} implies that
\bal
\Gamma^\perp_{AA}(p^2, \mu^2) &= \Gamma^\parallel_{AA}(0, \mu^2)
+ \int_0^{p^2} d p^{\prime \, 2} \, \zeta (p^{\prime \, 2}) \, \frac{\d}
{\d p^{\prime \, 2}} \, \Gamma^\parallel_{AA}(p^{\prime \, 2}, \mu^2) \n \\
&\phantom{=} {}+ \int_0^{p^2} dp^{\prime \, 2} \, \exp \left( 
-\int_{\mu^2}^{p^{\prime 2}} \frac{d \mu^{\prime \, 2}}{\mu^{\prime \, 2}} \, 
\gamma_A (\mu^{\prime \, 2}) \right) \,. \label{2pAperpRGdyn}
\eal
There are basically two ways to evaluate the first $p^{\prime \, 2}$-integral
in this equation. One is, by use of Eq.\ \eqref{2pAparRG}, to write
\be
\frac{\d}{\d p^{\prime \, 2}} \, \Gamma^\parallel_{AA}(p^{\prime \, 2}, \mu^2) 
= \frac{\beta_{m^2}(p^{\prime \, 2}) - m^2(p^{\prime \, 2}) \, 
\gamma_A(p^{\prime \, 2})}{p^{\prime \, 2}} \, 
\exp \left( -\int_{\mu^2}^{p^{\prime 2}} 
\frac{d \mu^{\prime \, 2}}{\mu^{\prime \, 2}} \, 
\gamma_A (\mu^{\prime \, 2}) \right) \,, 
\ee
the other would be to integrate by parts first and then use the known
$p^{\prime \, 2}$-dependence of $\zeta (p^{\prime \, 2})$ for the 
differentiation [taking into account the implicit $p^{\prime \, 2}$-dependence 
through $m^2(p^{\prime \, 2})$], and Eq.\ \eqref{2pAparRG} to replace 
$\Gamma^\parallel_{AA}(p^{\prime \, 2}, \mu^2)$. We have opted for the first
possibility in the numerical evaluation in Subsection \ref{numerical}.

\subsection{Numerical results \label{numerical}}

In this subsection we present the results of the numerical integration of 
the Callan-Symanzik equations, for the different renormalization schemes 
discussed in the previous section. We begin by briefly describing the
numerical procedure for the integration of the renormalization group
equations. The first step in this procedure is the solution of the pair
of coupled ordinary first-order differential equations \eqref{2coupdiffeq}
for $g (\mu^2)$ and $m^2 (\mu^2)$. The beta functions on the right-hand sides
of these equations are functions of $g (\mu^2)$, $m^2 (\mu^2)$, and also
explicitly of $\mu^2$, that are obtained from Eqs.\ 
\eqref{gammaA}--\eqref{betam2} and Eq.\ \eqref{betag} or, depending on the
renormalization scheme, Eq.\ \eqref{betagsymm} or \eqref{betagdyn},
by explicitly performing the differentiations of $\delta Z_A$, $\delta Z_c$, 
$\delta Z_{m^2}$ and (possibly) $A (\mu^2)$ with respect to $\mu^2$.
The functions $\delta Z_A$, $\delta Z_c$ and $\delta Z_{m^2}$ depend on the
renormalization scheme considered and result from applying the
corresponding normalization conditions to the one-loop expressions for
$\Gamma_{AA}^\perp (p^2)$, $\Gamma_{AA}^\parallel (p^2)$ and 
$\Gamma_{c\bar{c}} (p^2)$ that are presented in Appendix \ref{twop}. 
The function $A (\mu^2)$ is given by the sum of the 
tree-level vertex and the two diagrams evaluated in Appendix \ref{threep} at
the symmetry point.

One important issue for the numerical integration of the renormalization
group equations is the precise evaluation of the flow functions. It turns
out that, in the IR limit $\mu^2 \ll m^2$ as well as in the UV limit
$\mu^2 \gg m^2$, subtle cancellations of large terms take place in the
analytical expressions. Numerical precision is considerably improved by
replacing the exact expressions in both limits with the first 
two terms (at most) of the corresponding expansions in powers of 
$(\mu^2/m^2)$ or $(m^2/\mu^2)$. We have carefully checked that the 
expansions numerically coincide with the exact expressions
in a broad overlap region.

The system of differential equations \eqref{2coupdiffeq} is integrated
numerically with the Runge-Kutta method (up to fourth order) on a
logarithmic scale for $\mu^2$, for
given initial values $g (\mu_0^2)$ and $m^2 (\mu_0^2)$. In practice, we
use $\mu_0 = 3$ GeV (the physical scale is identified in the course of the
comparison to the lattice data, see below) and integrate the differential
equations towards smaller and towards larger values of $\mu^2$. The initial
values $g (\mu_0^2)$ and $m^2 (\mu_0^2)$ are chosen in such a way that they 
closely reproduce the lattice data for the gluon and ghost propagators, see 
below for the details of our fitting strategy.

The results for $g (\mu^2)$ and $m^2 (\mu^2)$, for any given renormalization
scheme, are then substituted in the expressions for the anomalous 
dimensions $\gamma_A (\mu^2)$ and $\gamma_c (\mu^2)$. Finally, the gluon and 
ghost two-point functions are obtained from the general formulas for the 
integrals of the Callan-Symanzik equations in Subsection \ref{intRGE}, Eqs.\
\eqref{2pARG}, \eqref{2pcRG} and \eqref{2pAparRG}, or \eqref{2pAperpRGz},
\eqref{2pcRGz} or \eqref{2pAperpRGdyn}, depending on the renormalization
scheme. The Riemann integrals in these formulas are evaluated 
numerically, using the Gauss-Legendre method to integrate over the logarithm 
of $\mu^{\prime \, 2}$ or $p^{\prime \, 2}$.

\begin{figure}
\begin{center}
\includegraphics[width=0.8\textwidth]{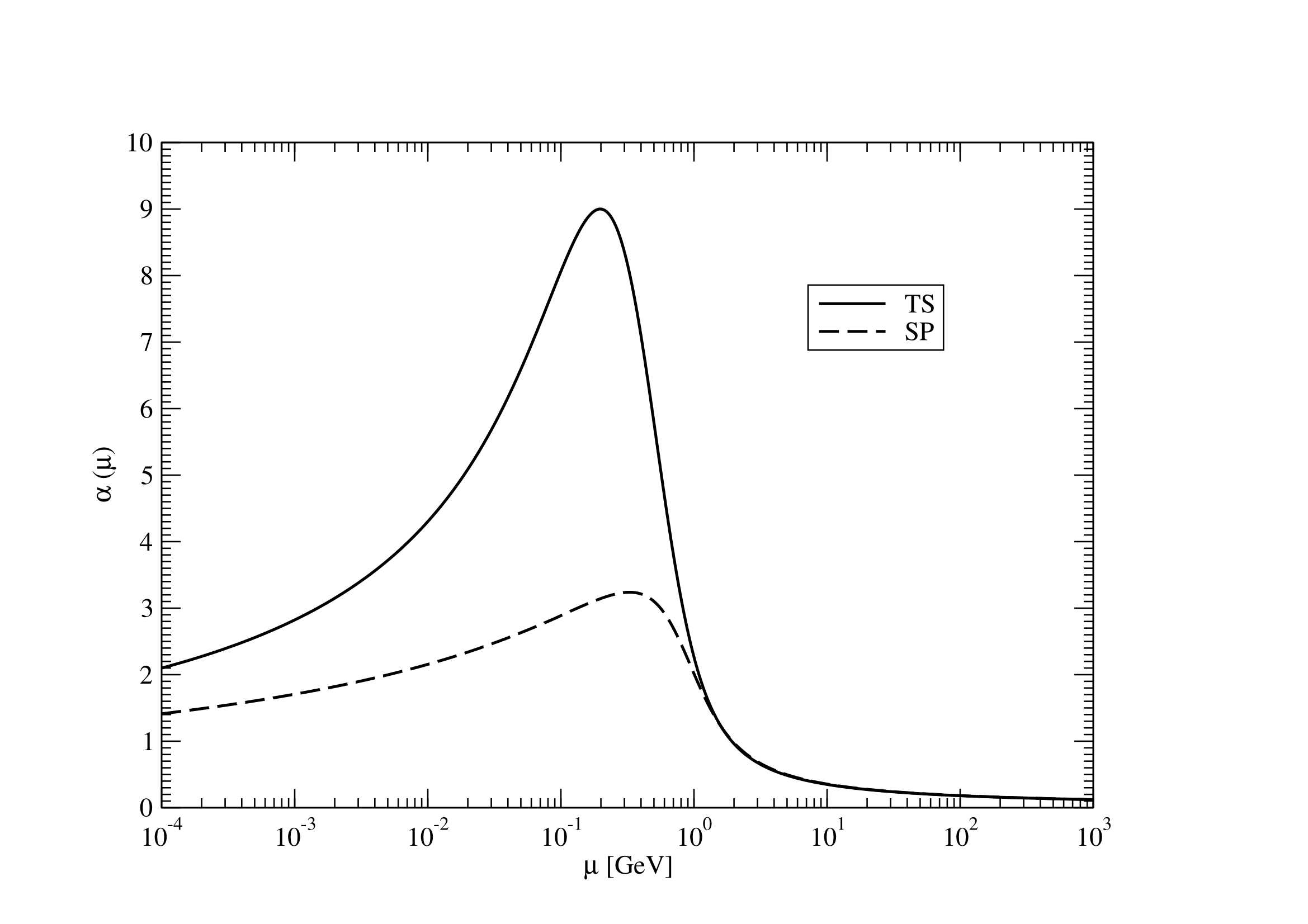}
\end{center}
\vspace{-1cm}
\caption{The renormalized strong fine structure constant $\alpha (\mu^2)
= g^2 (\mu^2)/4\pi$ as a function of the renormalization scale $\mu$
for the two renormalization schemes which define the coupling constant
from the ghost-gluon vertex in the Taylor limit (TS) or at
the symmetry point (SP). In both cases, the two-point functions are
normalized according to conditions \eqref{TWs1}--\eqref{TWs3}.} 
\label{figalpha}
\end{figure}

\begin{figure}
\begin{center}
\includegraphics[width=0.8\textwidth]{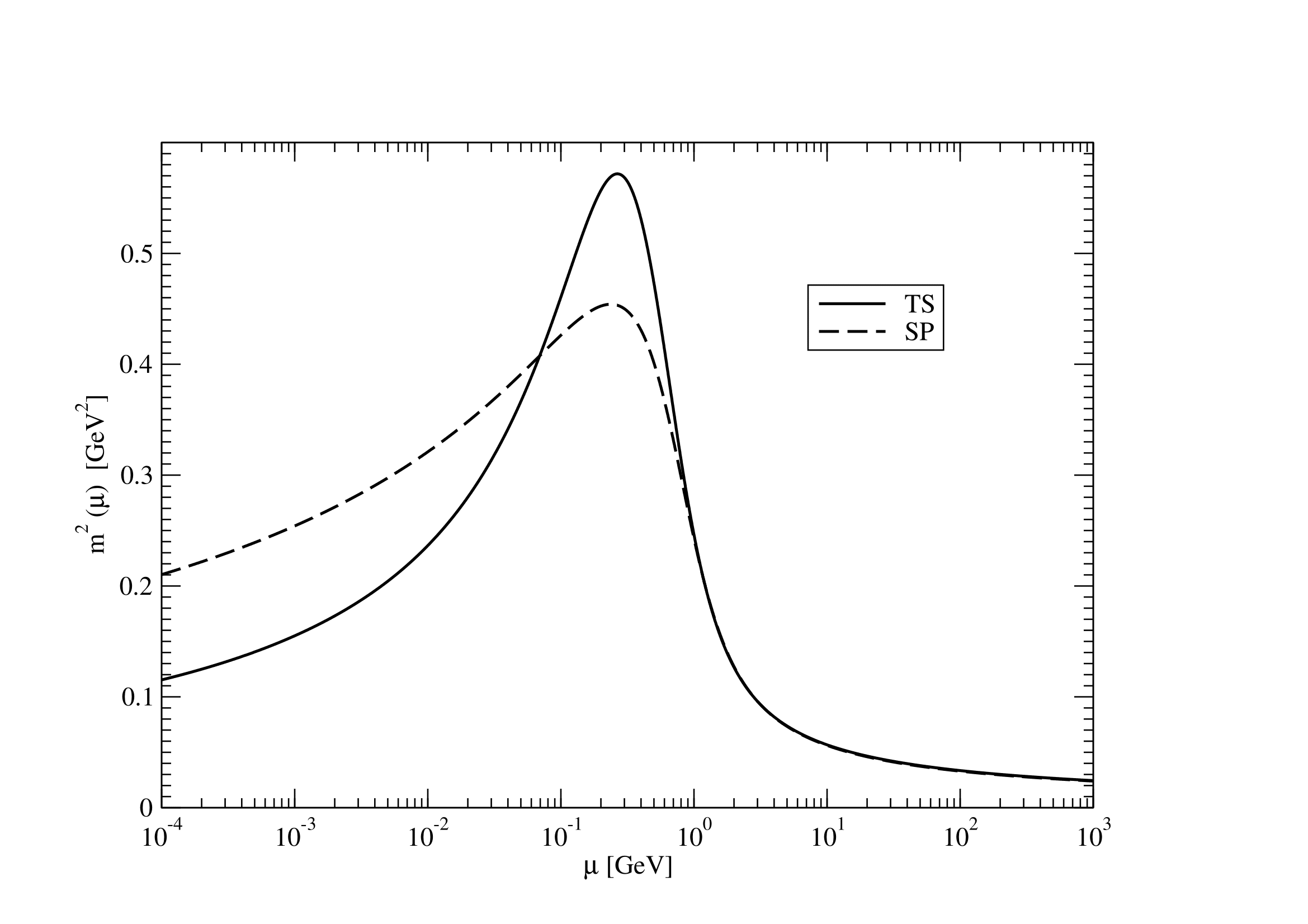}
\end{center}
\vspace{-1cm}
\caption{The square of the mass parameter as a function of the renormalization
scale $\mu$ for the same renormalization schemes as in Fig.\ \ref{figalpha}.} 
\label{figm2}
\end{figure}

In order to illustrate some of the general features of the solutions and
also the renormalization scheme dependence, we show in Figs.\ 
\ref{figalpha} and \ref{figm2}
the numerical results for $g (\mu^2)$ and $m^2 (\mu^2)$ in two different
renormalization schemes, the original Tissier-Wschebor scheme and a scheme
that uses the same normalization conditions \eqref{TWs1}--\eqref{TWs3}
for the two-point functions, but defines the coupling constant from
the ghost-gluon vertex at the symmetry point as in Eq.\ \eqref{symmg},
rather than in the Taylor limit as in Eq.\ \eqref{defg}.
The qualitative features are the same in both renormalization schemes: the
coupling constant $g (\mu^2)$ (or, rather, the strong fine structure constant
$\alpha = g^2/4 \pi$) tends logarithmically to zero in both the IR and the UV
limits, in accordance with the 
limiting behavior \eqref{TWbetag} and \eqref{UVbetag}
of the beta function. Hence, as expected, asymptotic freedom is recovered in
the UV limit, in both renormalization schemes. The smallness of the coupling
constant in the IR limit implies that perturbation theory (with a gluon mass
term) should give a precise description of the IR regime. This expectation 
was confirmed in Ref.\ \cite{TW10}. We remark that with
Dyson-Schwinger equations \cite{AN04}, a mapping to $\lambda \phi^4$ theory 
\cite{Fras08} and also with the epsilon expansion \cite{Web12}, 
the running coupling constant has been found to tend to zero with a power of 
the scale $\mu$ in the IR limit (for the decoupling solutions and the 
approach to the high-temperature fixed point, respectively), specifically 
$g^2 (\mu^2) \propto \mu^2$ for $\mu^2 \to 0$ in four dimensions. The apparent
discrepancy with the logarithmic dependency obtained in the present approach 
is simply due to a different definition of the coupling constant
(and of the gluon field renormalization constant).

The behavior of the running mass parameter $m^2 (\mu^2)$ is quite similar
to the coupling constant, with a logarithmic fall-off towards large $\mu^2$,
which is consistent with the recovery of the usual, not extended, BRST
symmetry in this limit. Rather unexpectedly, the mass parameter also
vanishes in the IR limit $\mu^2 \to 0$. The behavior of $m^2 (\mu^2)$ in
both limits is governed by the corresponding limits of the beta function
$\beta_{m^2}$, Eqs.\ \eqref{TWbetam} and \eqref{UVbetam}. As remarked before,
despite the vanishing of the mass parameter in the IR limit, both the 
transverse and longitudinal parts of the proper gluonic two-point function 
tend towards a finite value in this limit, see also below.

Even though the qualitative behavior is the same in both renormalization
schemes, in a quantitative sense the renormalization scheme 
dependence is rather strong. 
There is a considerable quantitative difference in the maximum 
values attained by $\alpha (\mu^2)$ and $m^2 (\mu^2)$, particularly 
in the case of $\alpha (\mu^2)$, and also in the fall-off
of $m^2 (\mu^2)$ towards smaller values of
$\mu^2$. In Figs.\ \ref{figalpha} and \ref{figm2},
we have integrated the renormalization group equations for $g (\mu^2)$
and $m^2 (\mu^2)$ in SU(2) Yang-Mills theory in the Tissier-Wschebor scheme
with the initial values $g (\mu_0^2) = 2{.}92$ and $m (\mu_0^2) = 0{.}31$ GeV
at $\mu_0 = 3$ GeV 
that we also use later  for the fits to the lattice data 
in this renormalization scheme.
In the symmetry point scheme, we use the same value $0{.}31$ GeV for
$m (\mu_0^2)$, but determine the value of $g (\mu_0^2)$
at $\mu_0 = 3$ GeV from a perturbative one-loop evaluation of the
ghost-gluon vertex at the symmetry point, with the same bare coupling
constant as in the Tissier-Wschebor scheme, compare Eqs.\ \eqref{defg}
and \eqref{symmg}. Note that taking the same value for $g (\mu_0^2)$ as
in the Tissier-Wschebor scheme would lead to slightly different results
for $g (\mu^2)$ and $m^2 (\mu^2)$. For the comparison to the lattice data 
below we have used the initial values that lead to the best fit to the data 
for each individual renormalization scheme. In particular, this procedure 
leads to other values for $g (\mu_0^2)$ and $m (\mu_0^2)$ in the symmetry
point scheme than the ones used in Figs.\ \ref{figalpha} and \ref{figm2},
and the difference between the fits to the gluon and ghost propagators
in the two renormalization schemes is much less than one would 
expect from the comparison of the schemes in Figs.\ \ref{figalpha} 
and \ref{figm2} (see below).

\begin{figure}
\begin{center}
\includegraphics[width=0.8\textwidth]{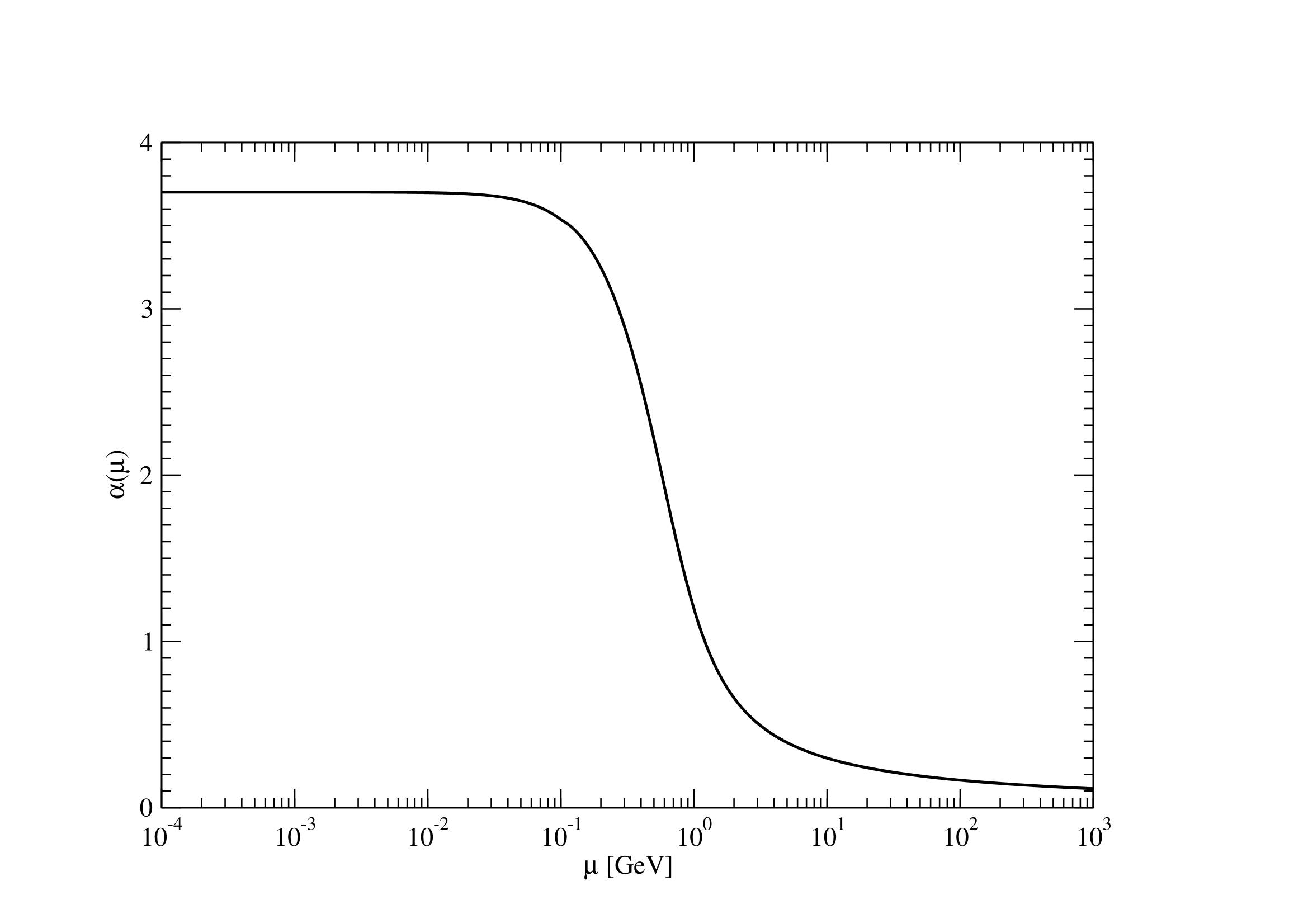}
\end{center}
\vspace{-1cm}
\caption{The strong fine structure constant $\alpha$ as a function of the 
  renormalization scale $\mu$ in the critical derivative scheme ($\zeta = 1/3$).}
\label{figalphacrit}
\end{figure}

\begin{figure}
\begin{center}
\includegraphics[width=0.8\textwidth]{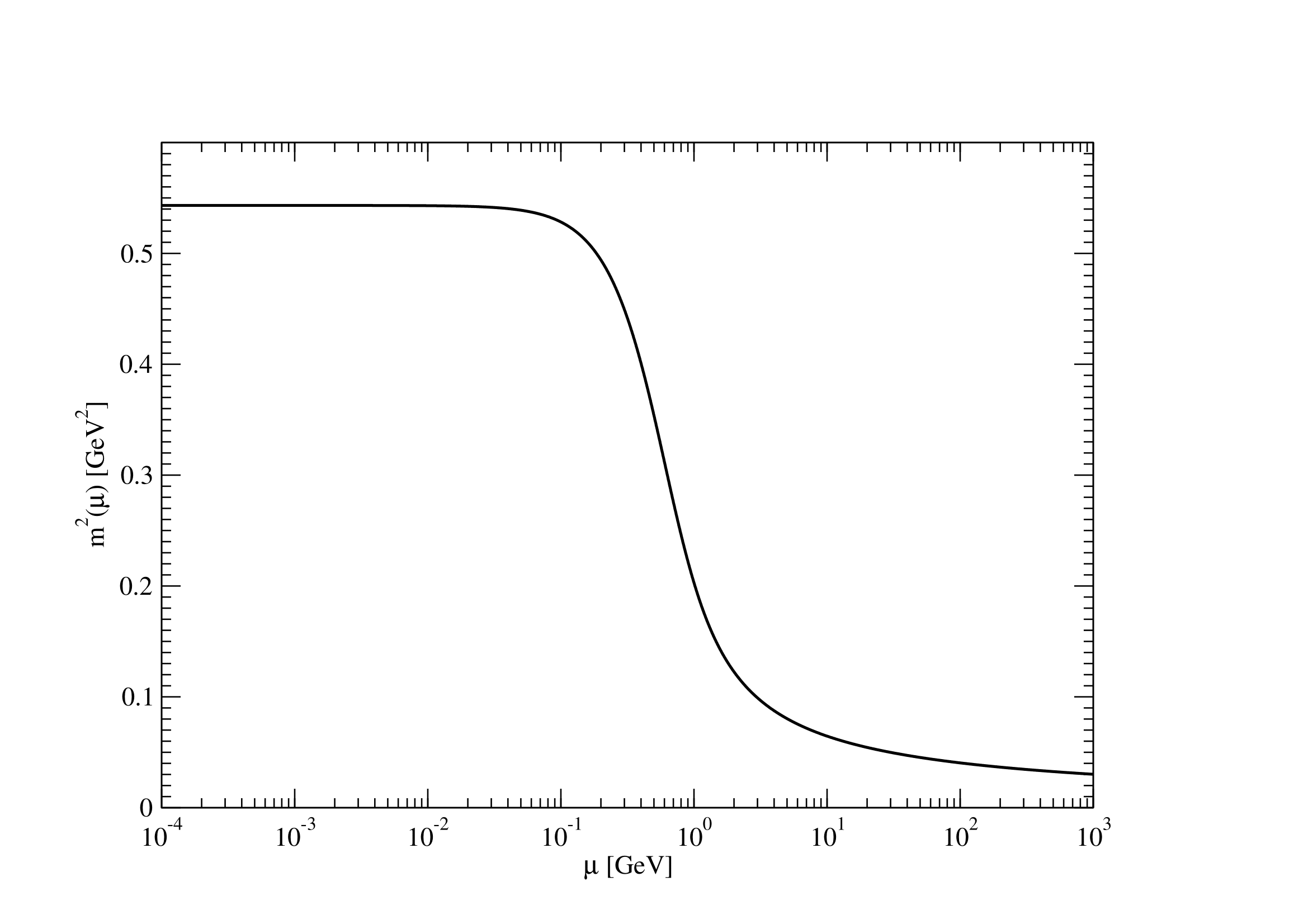}
\end{center}
\vspace{-1cm}
\caption{The square of the mass parameter as a function of the renormalization
  scale $\mu$ in the critical derivative scheme.}
\label{figm2crit}
\end{figure}

In all derivative schemes (including the simple scheme with $\zeta = 1$
and the scale-dependent scheme), the dependence of $\alpha$ and $m^2$
on the renormalization scale is qualitatively the same as in Figs.\
\ref{figalpha} and \ref{figm2}, 
with the exception of the critical $\zeta = 1/3$
scheme. In Figs.\ \ref{figalphacrit} and \ref{figm2crit}, we show the plots
of $\alpha (\mu^2)$ and $m^2 (\mu^2)$ in the critical derivative scheme, with
the initial values $g = 2{.}53$ and $m = 0{.}315$ GeV at $\mu = 3$ GeV.
While the UV behavior is the same as in all the other cases, in the IR limit
$\mu^2 \to 0$ both $\alpha$ and $m^2$ tend towards finite (non-zero)
values. The actual limiting values are found to depend on the initial
values $g (\mu_0^2)$ and $m^2 (\mu_0^2)$ for the integration of the
differential equations. In this sense we find a line of fixed points
of the system in the IR limit rather than a single fixed point in this
particular renormalization scheme.

We will now present the comparison of our results to the lattice data for
the gluon and ghost propagators in SU(2) Yang-Mills theory in the Landau
gauge. We shall use the data of Cucchieri and Mendes in Refs.\ \cite{CM08a} 
and \cite{CM08b} for the comparison. The gauge is fixed on the lattice
by minimizing the lattice discretization of the functional
\be
\int d^4 x \, A_\mu^a (x) A_\mu^a (x) \label{minfunc}
\ee
along the gauge orbits, which is equivalent to imposing the Landau (Lorenz)
gauge condition and restricting the gauge fields to the Gribov region at the
same time [if one accepts any local minimum of the functional
\eqref{minfunc}; determining the global minimum of \eqref{minfunc} instead
would correspond to the restriction of the gauge fields to the fundamental
modular region, see also our brief comment \cite{footn3}].

To compare our results with the lattice data, we will
determine the initial values
$g (\mu_0^2)$ and $m^2 (\mu_0^2)$ for each renormalization scheme (we shall
be using $\mu_0 = 3$ GeV throughout) in such a way that the integration
of the Callan-Symanzik equations leads to gluon and ghost propagators that
reproduce the results of the lattice simulations for these propagators as
closely as possible. We have determined the optimal initial values ``by
hand'', by varying $g (\mu_0^2)$ and $m^2 (\mu_0^2)$ in small steps and
comparing the plots of the propagators obtained by integrating the
Callan-Symanzik equations to the corresponding plots of the lattice data.

The procedure is not as straightforward as it may sound. One of the
difficulties is that the lattice propagators are not normalized 
and hence, for the comparison to the results of the integration
of the renormalization group equations, may still be multiplied with
arbitrary overall constants (in this sense, there are four parameters to
be adjusted). Another difficulty is the
comparatively lower precision of the lattice data
in the UV regime, cf.\ the plots of the dressing functions below, where the
data begin to oscillate for momenta above approximately $0{.}7$ GeV (for
the ghost dressing function), and the results at a given momentum
$p = (p^2)^{1/2}$ depend on the angles the momentum forms with the lattice 
axes (breaking of continuous rotational invariance on the lattice; the
different values are still almost compatible within the statistical errors).
For the representation of the lattice data in the figures we identify $p^2$
with the improved lattice momentum squared \cite{Ma00}.

After experimenting with different possible strategies for quite some time,
we have settled on the following procedure for the fits: for any given
renormalization scheme, we have determined $g (\mu_0^2)$ and $m^2 (\mu_0^2)$
(and the multiplicative factor for the normalization of the ghost propagator)
in such a way that the best possible fits to the ghost propagator 
$G_c (p^2) = (\Gamma_{c\bar{c}} (p^2))^{-1}$ [see Eq.\ \eqref{propagators}]
and the ghost dressing function $F_c (p^2) = p^2 \, G_c (p^2)$ are obtained, 
guided by the eye and making sure that the
renormalization group improved results fall inside the statistical error
bars of the data for all momenta below $3{.}7$ GeV (for some direction of
the momentum with respect to the lattice axes, for the values of $p$ where
the breaking of rotational invariance is visible in the lattice data). 
The latter condition turned out to be impossible to fulfill in the symmetry 
point scheme, see below. In any case, this procedure fixes $g (\mu_0^2)$ 
and $m^2 (\mu_0^2)$ to a fair precision. We remark that a similar strategy,
but adjusting our results to the gluon propagator and the gluon dressing
function as determined on the lattice rather than using the ghost propagator
and the ghost dressing function, or employing the ghost and gluon dressing
functions instead, would leave more freedom in the determination of the 
initial values. 

Having thus fixed $g (\mu_0^2)$ and $m^2 (\mu_0^2)$
for a given renormalization scheme by a fit to the
ghost propagator and the ghost dressing function, we use these same initial
values to calculate the renormalization group improved gluon propagator
$G_A (p^2) = (\Gamma_{AA}^\perp (p^2))^{-1}$ [see Eq.\ \eqref{propagators}]
and the gluon dressing function
$F_A (p^2) = p^2 \, G_A (p^2)$ (which is the dressing function
relative to a massless tree-level propagator). A comparison to the lattice
data for the latter functions, after adjusting the respective overall
multiplicative factor, gives an indication of how well the renormalization
group improvement reproduces the lattice data, for the renormalization
scheme considered. We shall show the comparisons for both the gluon propagator
and the gluon dressing function in the figures below, since the plots of
the propagator give a clearer picture of the quality of the fits in
the extreme IR regime, while the plots of the dressing function emphasize the
intermediate and large momentum regimes.

\begin{figure}
\begin{center}
\includegraphics[width=0.8\textwidth]{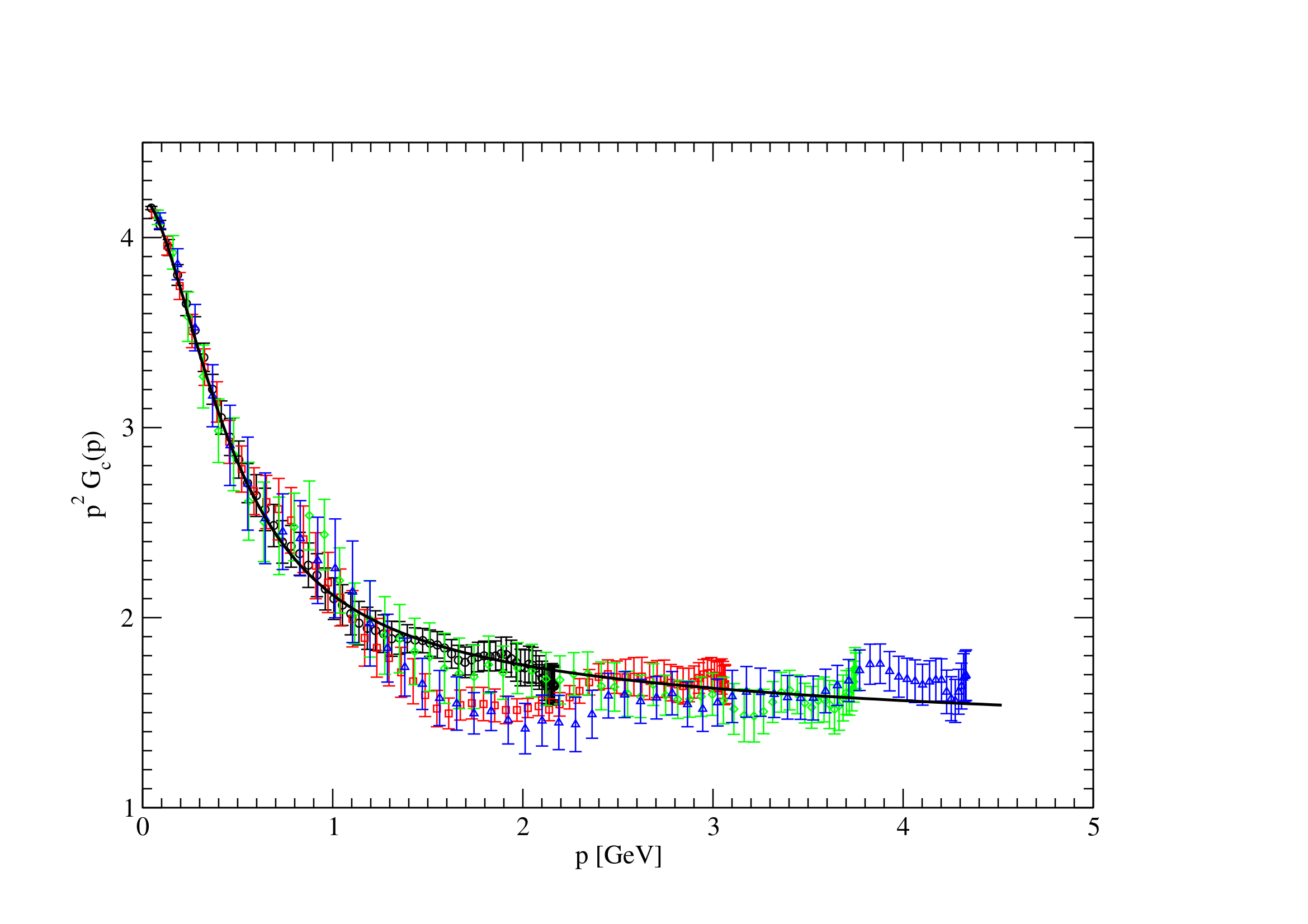} 
\end{center}
\vspace{-1cm}
\caption{The ghost dressing function $F_c (p^2) = p^2 \, G_c (p^2) =
  p^2/\Gamma_{c\bar{c}} (p^2)$ as a function of the momentum $p = (p^2)^{1/2}$
  in the original Tissier-Wschebor scheme defined by the normalization
  conditions \eqref{TWs1}--\eqref{TWs3} and the renormalized coupling
  constant determined from the Taylor limit of the ghost-gluon vertex.
  The initial conditions for the integration of the renormalization group
  equations have been 
  chosen in such a way that they produce the best possible fit 
  to the lattice data of Ref.\ \cite{CM08b}, which are also represented in the
  figure. Different colors of the data points (and error bars) in the online
  version correspond to different directions on the lattice.} 
\label{figghosttw}
\end{figure}

\begin{figure}
\begin{center}
\includegraphics[width=0.8\textwidth]{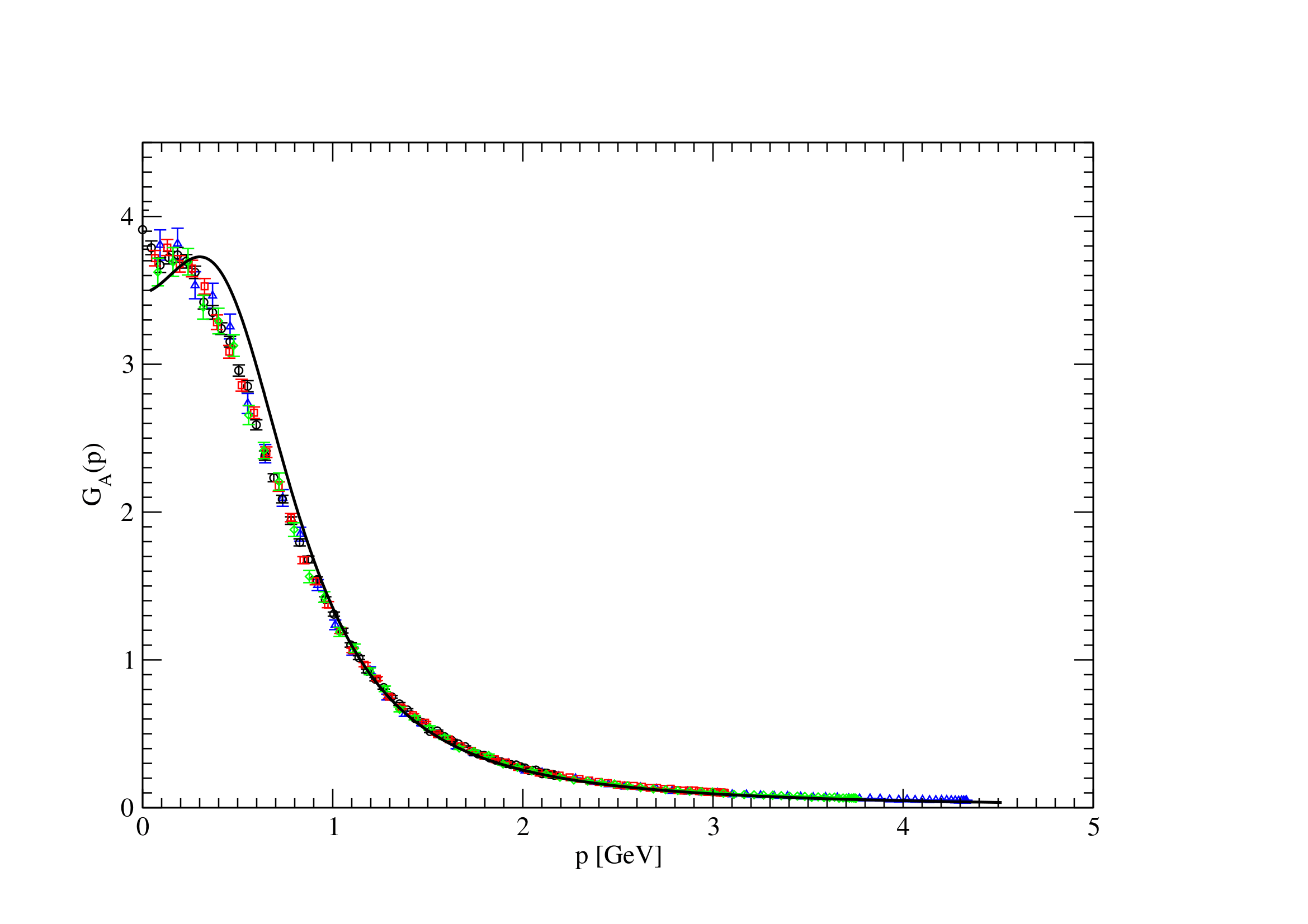} 
\end{center}
\vspace{-1cm}
\caption{The gluon propagator $G_A (p^2) = 1/\Gamma_{AA}^\perp (p^2)$ as a
  function of momentum in the original Tissier-Wschebor scheme with
  the same initial conditions for the integration of the
  renormalization group equations as in Fig.\ \ref{figghosttw}, compared to
  the lattice data of Ref.\ \cite{CM08a}.} 
\label{figgluontw}
\end{figure}

\begin{figure}
\begin{center}
\includegraphics[width=0.8\textwidth]{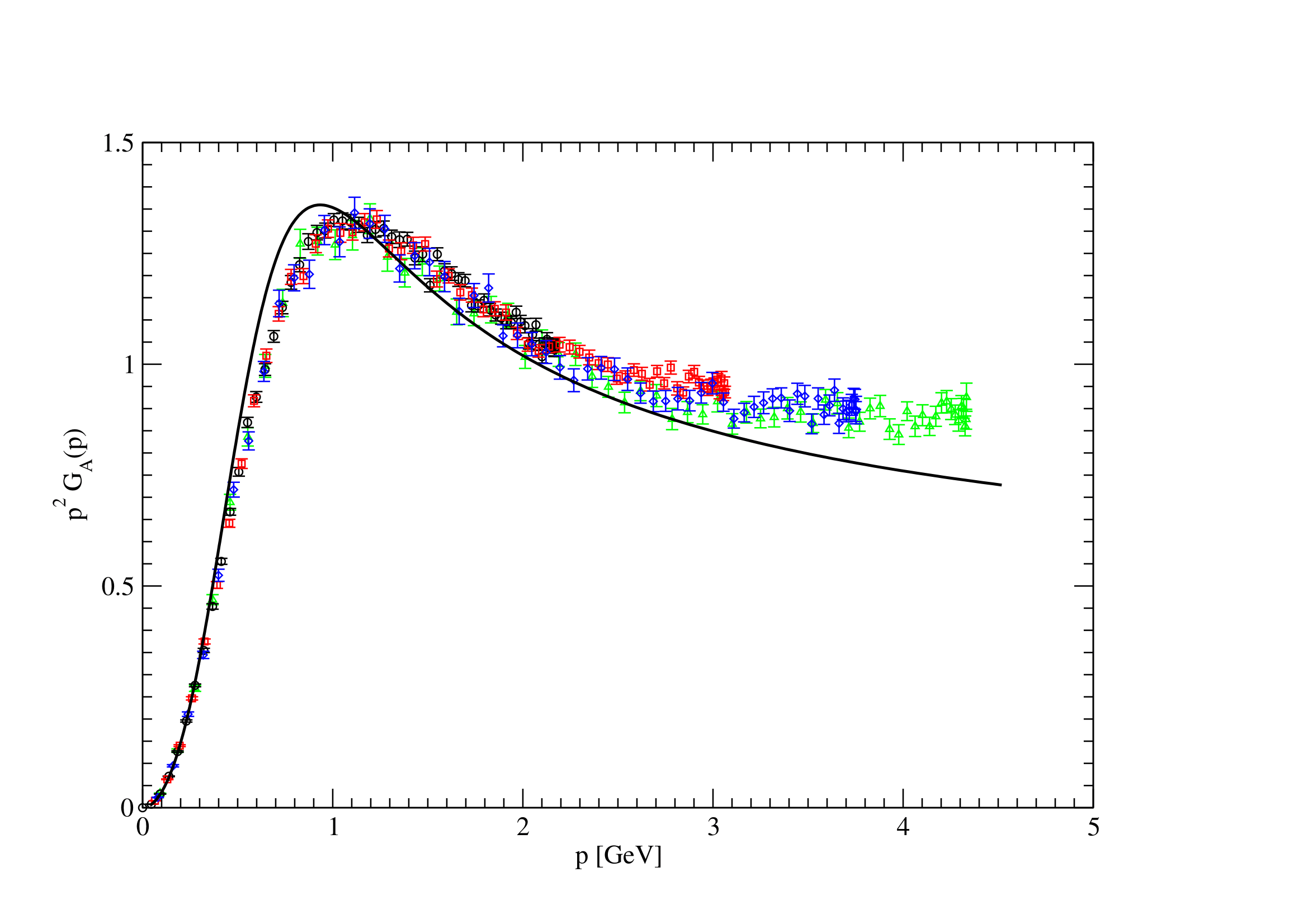} 
\end{center} 
\vspace{-1cm}
\caption{The gluon dressing function $F_A (p^2) = p^2 \, G_A (p^2) =
  p^2/\Gamma_{AA}^\perp (p^2)$ as a function of momentum in the original
  Tissier-Wschebor scheme with the same initial conditions as in Fig.\
  \ref{figghosttw}, compared to the lattice data of Ref.\ \cite{CM08a}.} 
\label{figgluondresstw}
\end{figure}

\begin{figure}
\begin{center}
\includegraphics[width=0.8\textwidth]{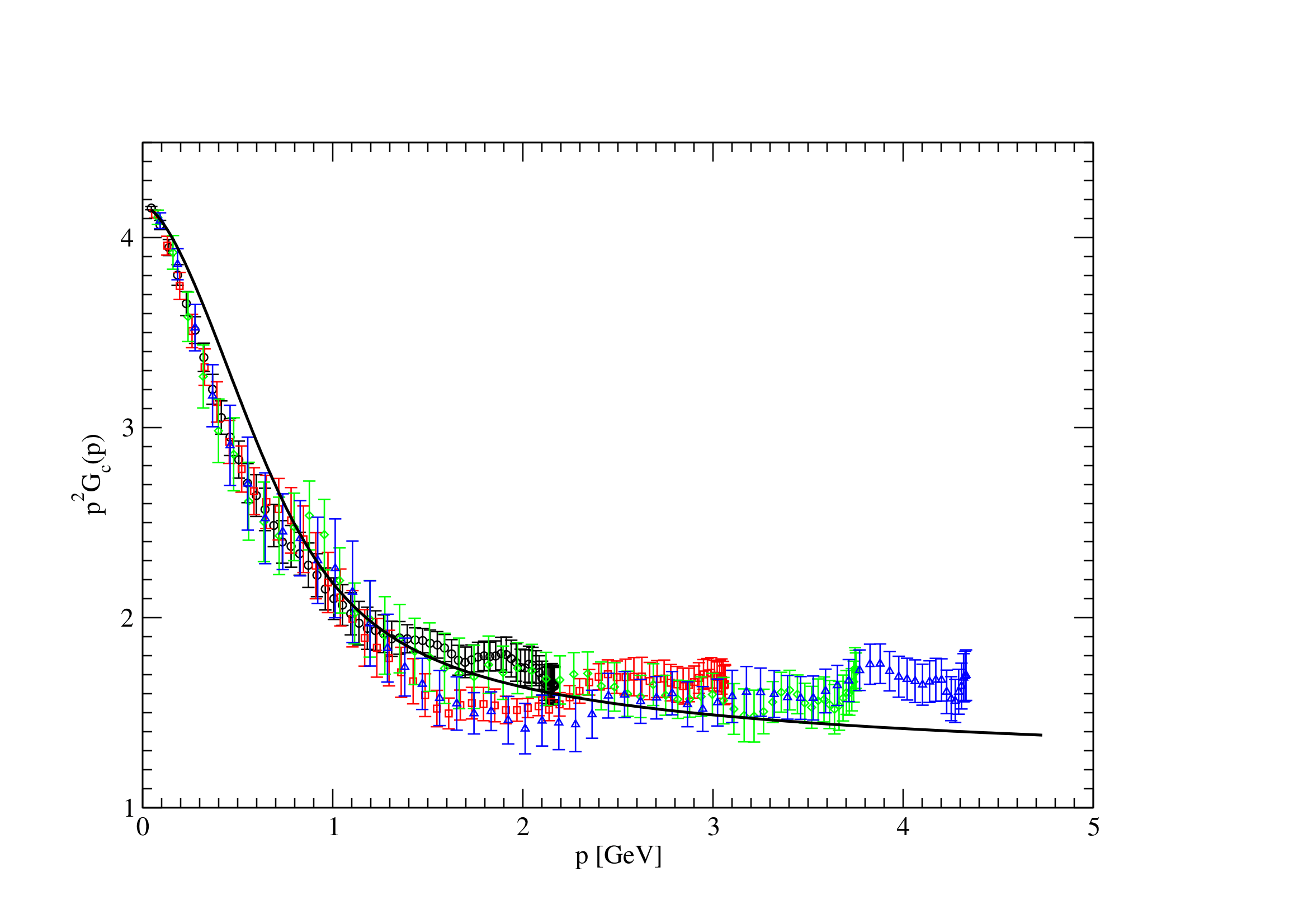} 
\end{center}
\vspace{-1cm}
\caption{The ghost dressing function $F_c (p^2) = p^2 \, G_c (p^2)$ as a 
  function of momentum in the renormalization scheme defined through the 
  conditions \eqref{TWs1}--\eqref{TWs3}, but with the renormalized coupling 
  constant determined from the ghost-gluon vertex at the symmetry point. The 
  initial conditions for the integration of the renormalization group 
  equations have been adjusted to produce the best possible fit 
  to the lattice data of Ref.\ \cite{CM08b},
  which are also represented in the figure.} 
\label{figghostsymm}
\end{figure}

\begin{figure}
\begin{center}
\includegraphics[width=0.8\textwidth]{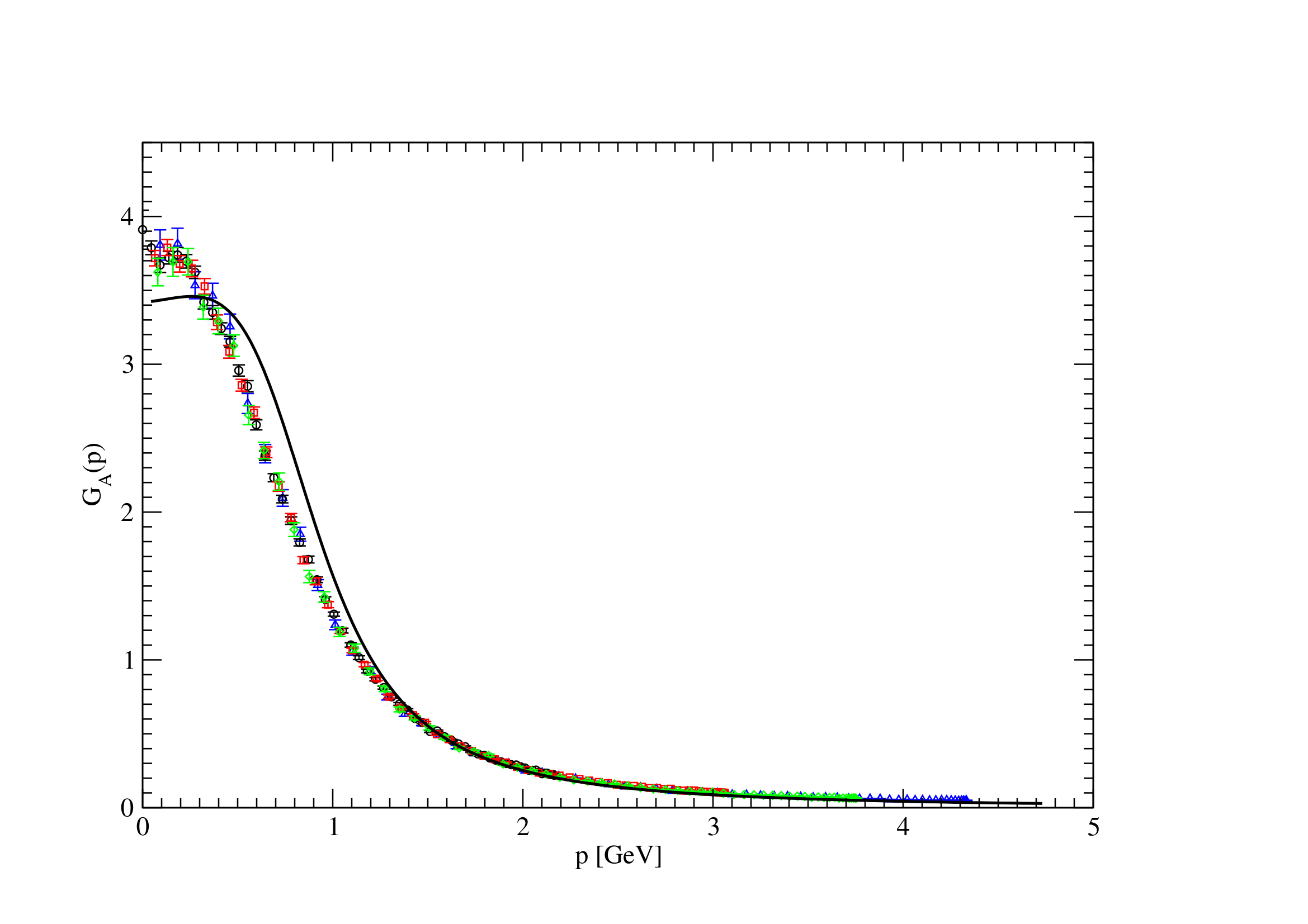} 
\end{center}
\vspace{-1cm}
\caption{The gluon propagator $G_A (p^2)$ as a function of momentum
  in the symmetry point scheme [with normalization conditions
  \eqref{TWs1}--\eqref{TWs3}]
  with the same initial conditions for the integration of the renormalization
  group equations as in Fig.\ \ref{figghostsymm}, compared to the lattice
  data from Ref.\ \cite{CM08a}.} 
\label{figgluonsymm}
\end{figure}

\begin{figure}
\begin{center}
\includegraphics[width=0.8\textwidth]{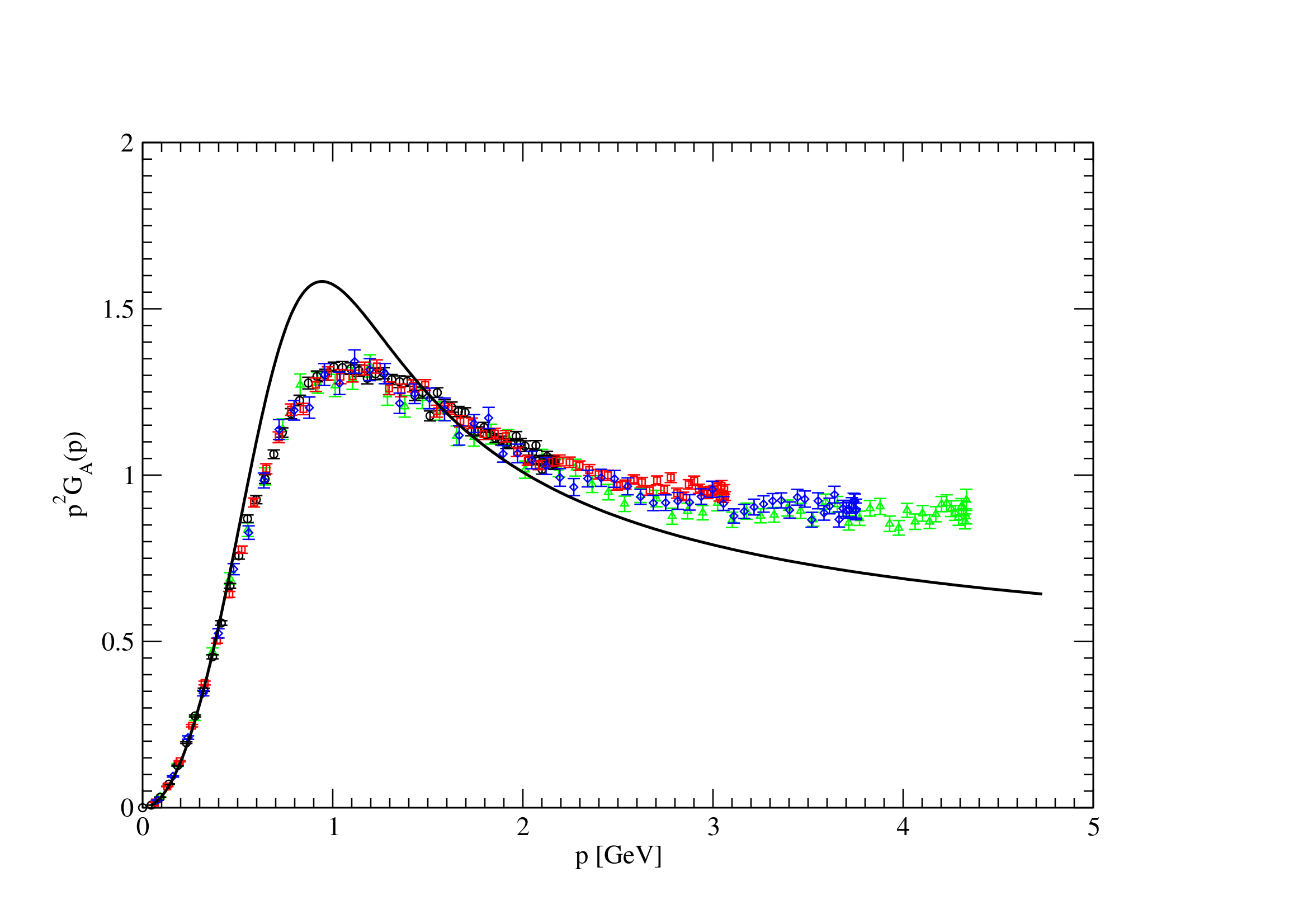} 
\end{center} 
\vspace{-1cm}
\caption{The gluon dressing function $F_A (p^2) = p^2 \, G_A (p^2)$ as a 
  function of momentum in the symmetry point scheme with the same initial 
  conditions as in Fig.\ \ref{figghostsymm}, compared to the lattice data 
  of Ref.\ \cite{CM08a}.} 
\label{figgluondresssymm}
\end{figure}

\begin{figure}
\begin{center}
\includegraphics[width=0.8\textwidth]{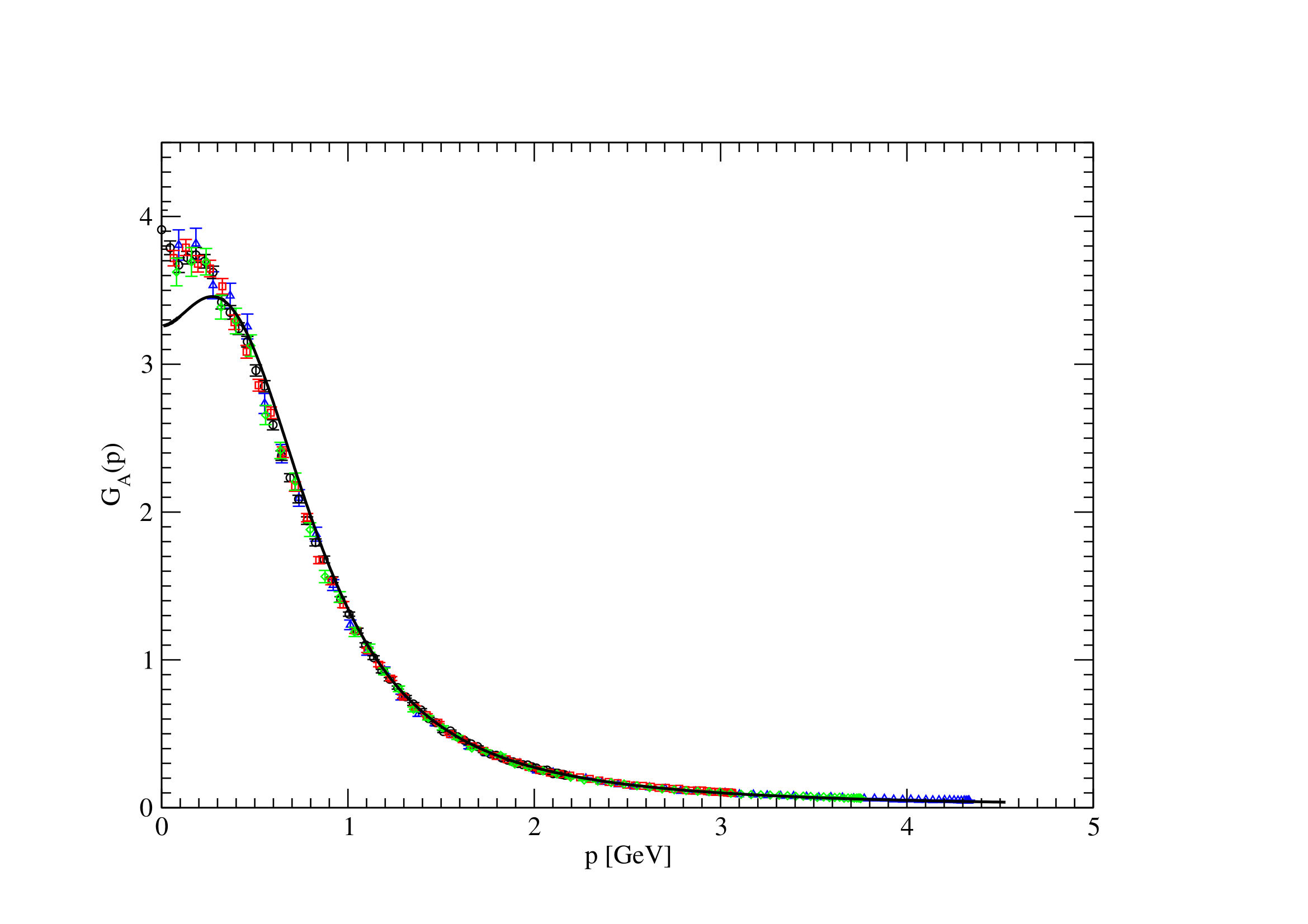} 
\end{center}
\vspace{-1cm}
\caption{The gluon propagator $G_A (p^2)$ as a function of momentum in the
  simple derivative scheme ($\zeta = 1$), where the renormalized coupling
  constant is defined from the ghost-gluon vertex in the Taylor limit.
  The initial conditions for the integration of the renormalization group
  equations are chosen in such a way that they produce the best possible fit 
  to the ghost propagator and the ghost dressing function 
  obtained in the lattice simulations of Ref.\
  \cite{CM08b}. The lattice data of Ref.\ \cite{CM08a} for the gluon
  propagator are also shown in the figure for comparison.} 
\label{figgluonder}
\end{figure}

\begin{figure}
\begin{center}
\includegraphics[width=0.8\textwidth]{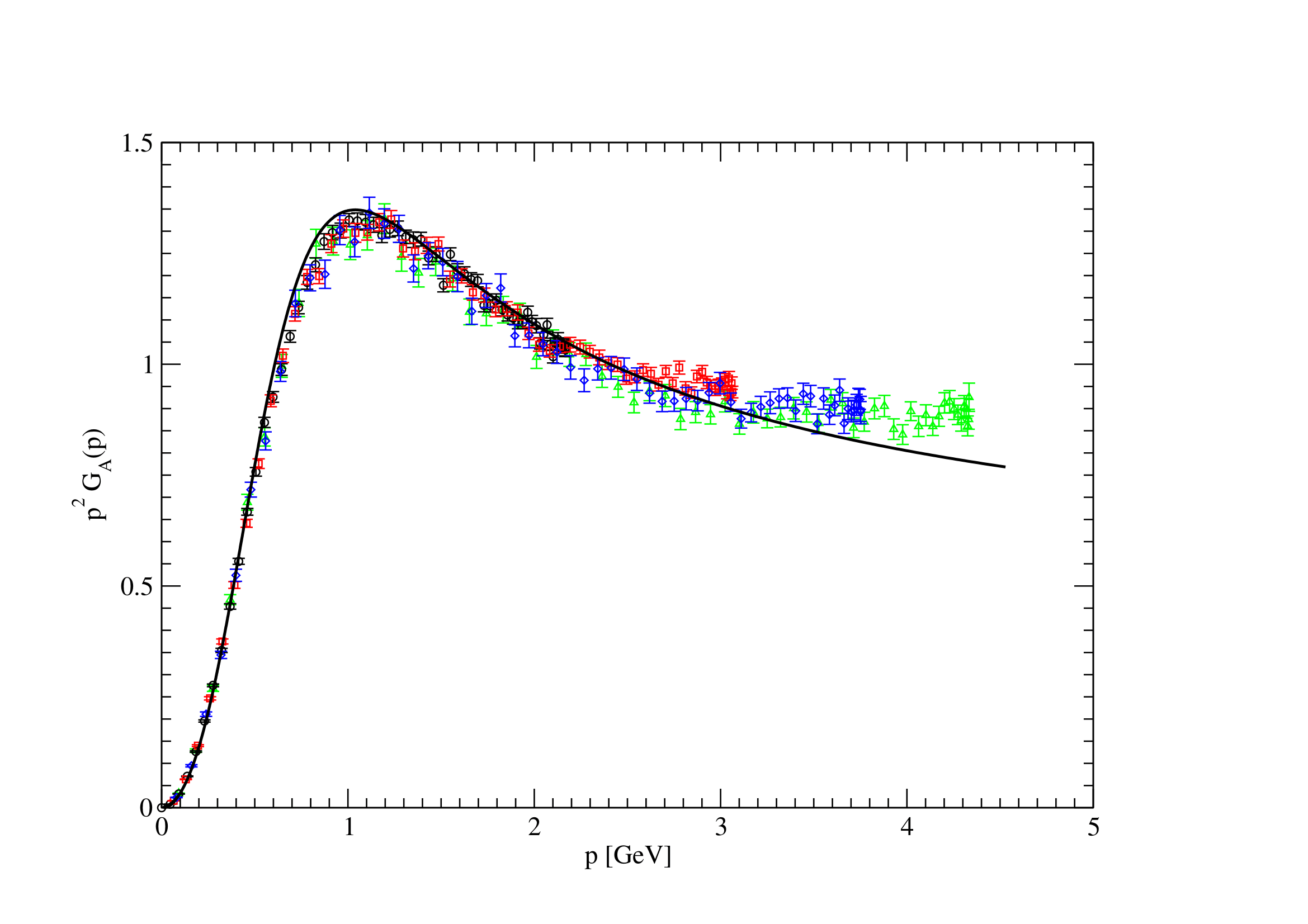} 
\end{center} 
\vspace{-1cm}
\caption{The gluon dressing function $F_A (p^2) = p^2 \, G_A (p^2)$ as a 
  function of momentum in the simple ($\zeta = 1$) derivative scheme with 
  the same 
  initial conditions as in Fig.\ \ref{figgluonder}, compared to the lattice 
  data of Ref.\ \cite{CM08a}.} 
\label{figgluondressder}
\end{figure}

\begin{figure}
\begin{center}
\includegraphics[width=0.8\textwidth]{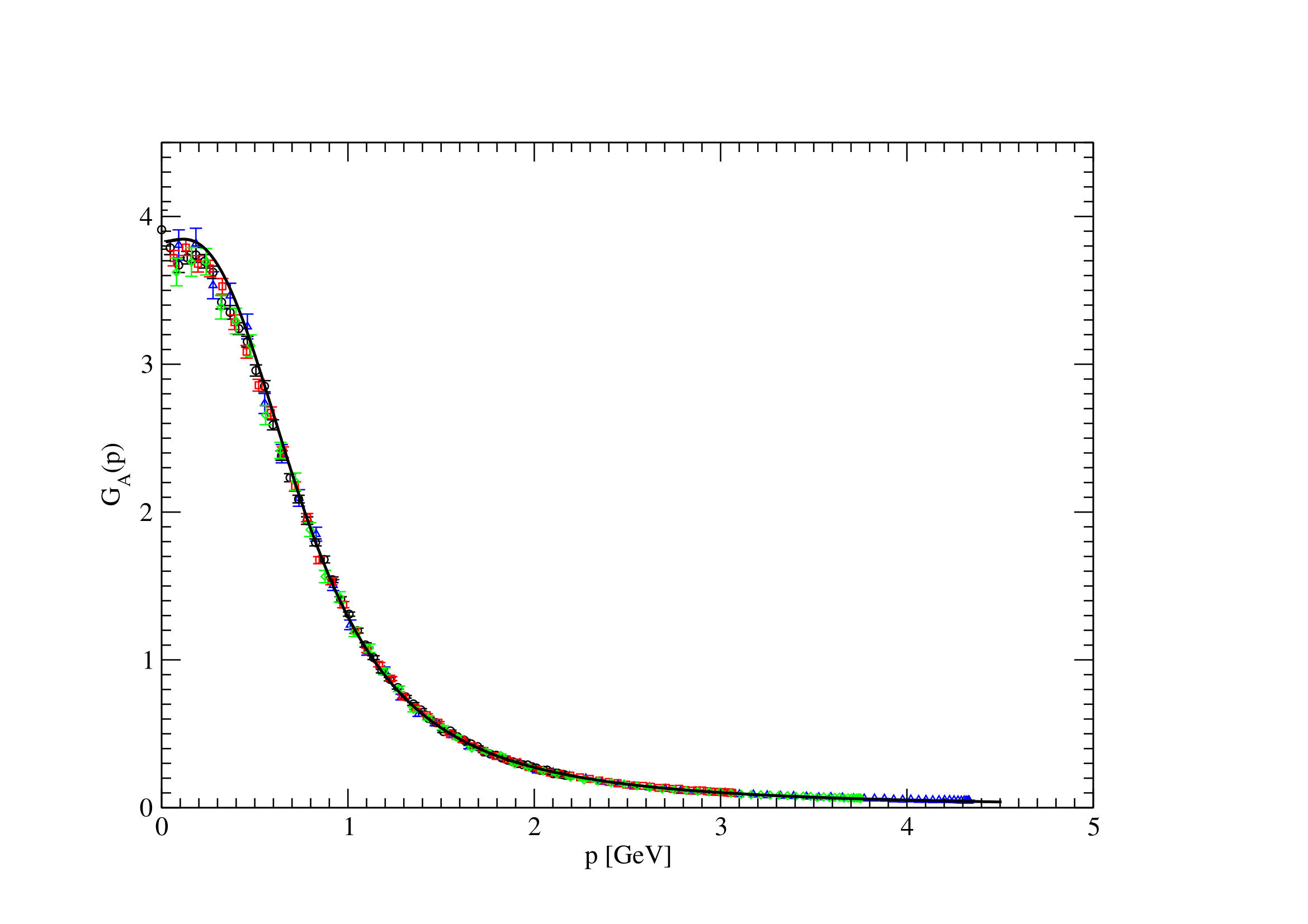} 
\end{center}
\vspace{-1cm}
\caption{The gluon propagator $G_A (p^2)$ as a function of momentum in the
  critical derivative scheme ($\zeta = 1/3$), where the renormalized coupling
  constant is defined from the ghost-gluon vertex in the Taylor limit.
  The initial conditions for the integration of the renormalization group
  equations are chosen in such a way that they produce the best possible fit
  to the ghost propagator and the ghost dressing function 
  obtained in the lattice simulations of Ref.\
  \cite{CM08b}. The lattice data of Ref.\ \cite{CM08a} for the gluon
  propagator are also shown in the figure for comparison.} 
\label{figgluoncrit}
\end{figure}

\begin{figure}
\begin{center}
\includegraphics[width=0.8\textwidth]{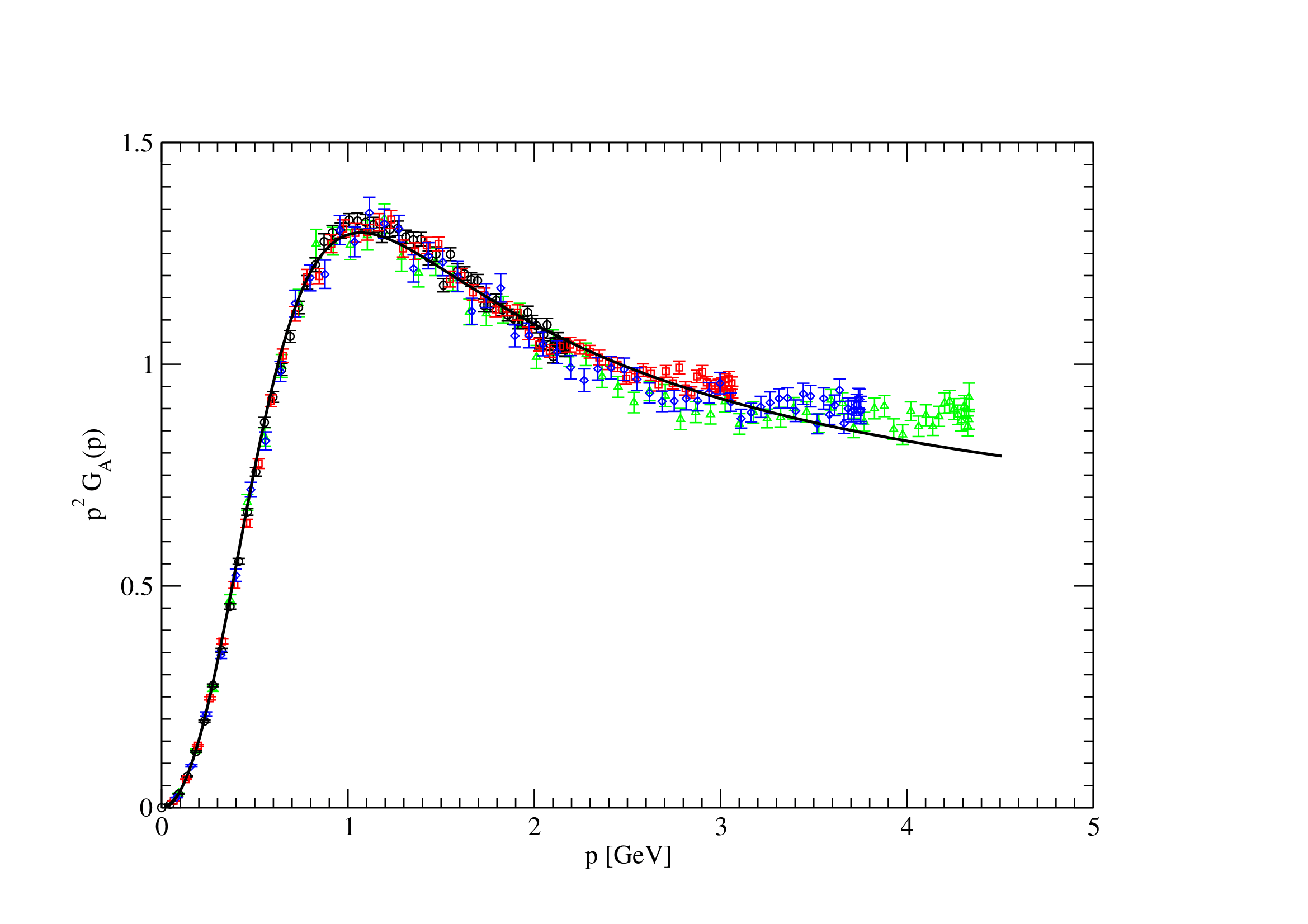} 
\end{center} 
\vspace{-1cm}
\caption{The gluon dressing function $F_A (p^2) = p^2 \, G_A (p^2)$ as a 
  function of momentum in the critical derivative scheme with the same initial
  conditions as in Fig.\ \ref{figgluoncrit}, compared to the lattice data of
  Ref.\ \cite{CM08a}.} 
\label{figgluondresscrit}
\end{figure}

\begin{figure}
\begin{center}
\includegraphics[width=0.8\textwidth]{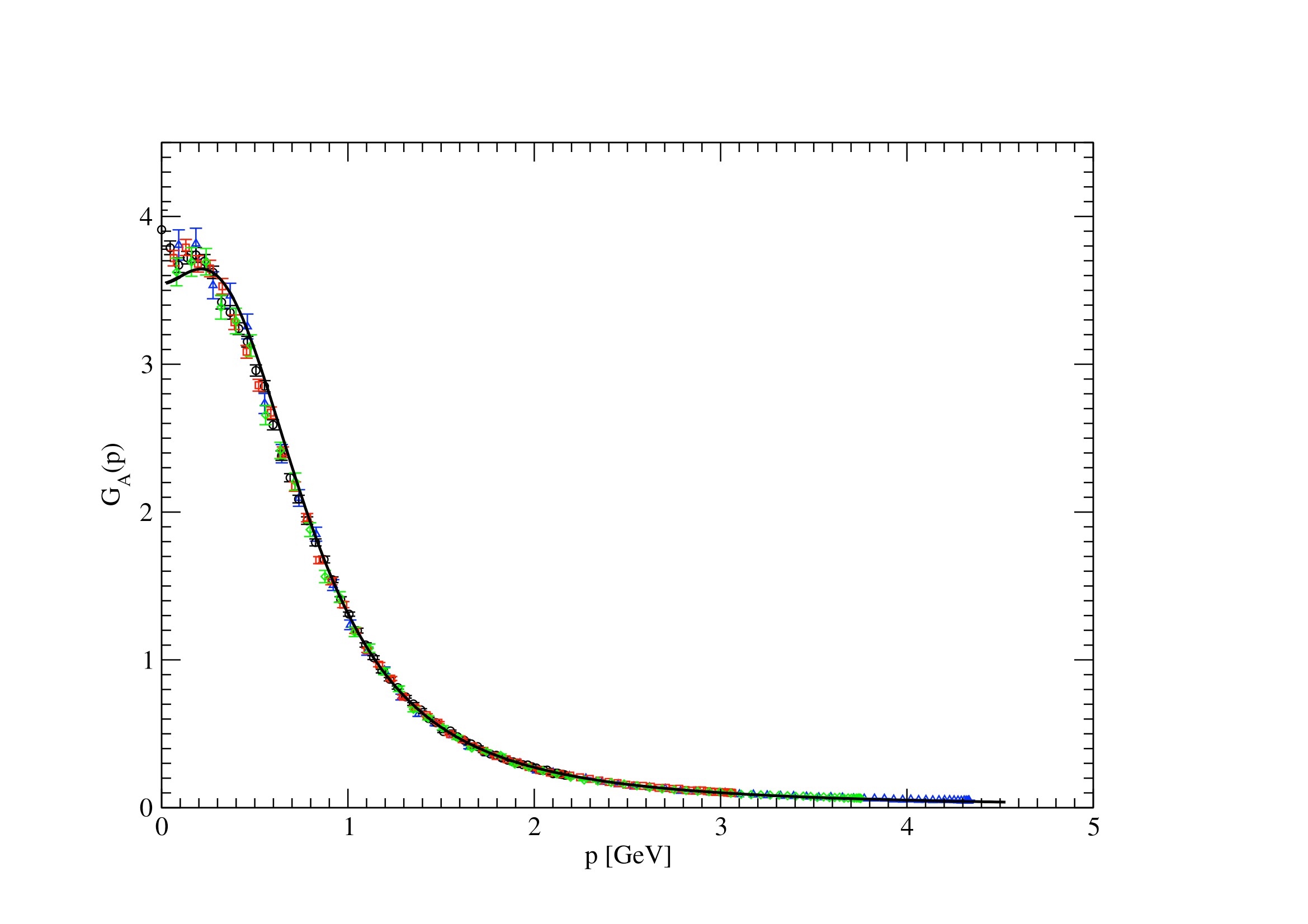} 
\end{center}
\vspace{-1cm}
\caption{The gluon propagator $G_A (p^2)$ as a function of momentum in the
  scale-dependent derivative scheme [with $\zeta$ from Eq.\ 
  \eqref{zetadyn}], where the renormalized coupling constant 
  is defined from the ghost-gluon vertex in the Taylor limit.
  The initial conditions for the integration of the renormalization group
  equations are chosen in such a way that they produce the best possible fit
  to the ghost propagator and the ghost dressing function 
  obtained in the lattice simulations of Ref.\
  \cite{CM08b}. The lattice data of Ref.\ \cite{CM08a} for the gluon
  propagator are also shown in the figure for comparison.} 
\label{figgluondyn}
\end{figure}

\begin{figure}
\begin{center}
\includegraphics[width=0.8\textwidth]{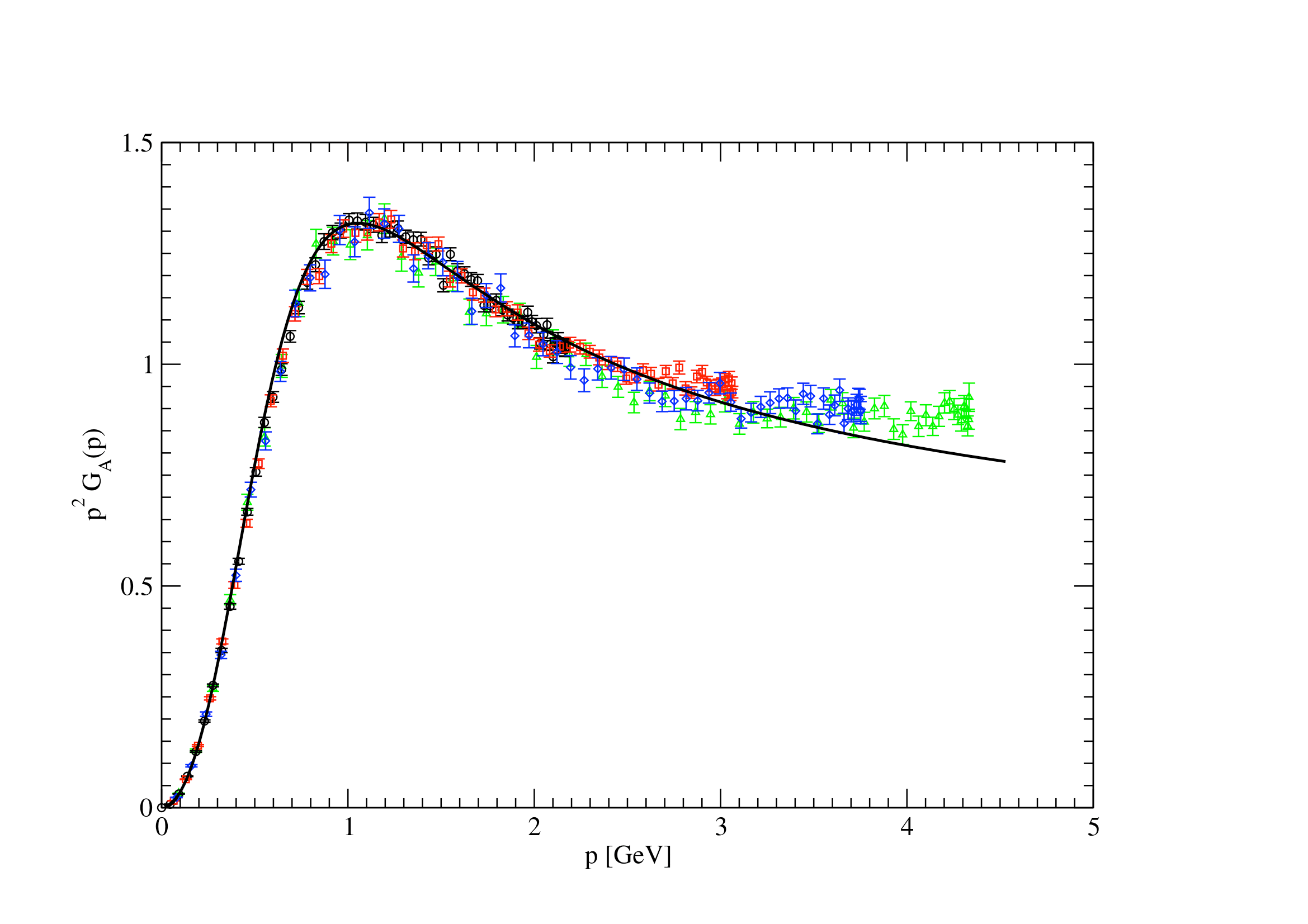} 
\end{center} 
\vspace{-1cm}
\caption{The gluon dressing function $F_A (p^2) = p^2 \, G_A (p^2)$ as a 
  function of momentum in the scale-dependent derivative scheme with the 
  coupling constant defined from the Taylor limit of the ghost-gluon vertex 
  and using the same initial conditions as in Fig.\ 
  \ref{figgluondyn}, compared to the lattice data of Ref.\ \cite{CM08a}.} 
\label{figgluondressdyn}
\end{figure}

\begin{figure}
\begin{center}
\includegraphics[width=0.8\textwidth]{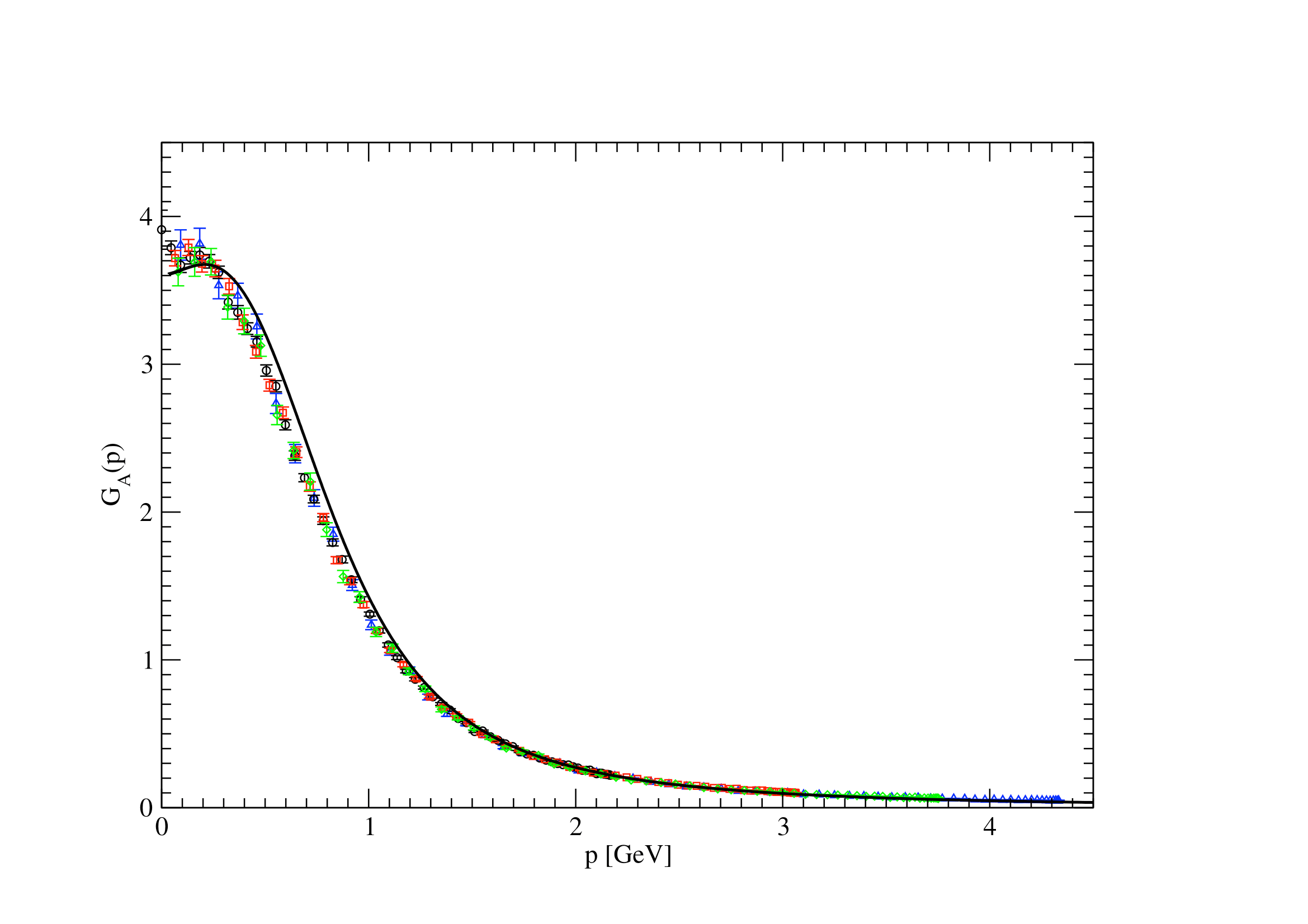} 
\end{center}
\vspace{-1cm}
\caption{The gluon propagator $G_A (p^2)$ as a function of momentum in the
  scale-dependent derivative scheme with the scale-dependent definition 
  \eqref{dyng} of the 
  renormalized coupling constant. The initial conditions for the integration 
  of the renormalization group equations are chosen in such a way 
  that they produce the best possible fit 
  to the ghost propagator and the ghost dressing function 
  obtained in the lattice simulations of Ref.\
  \cite{CM08b}. The lattice data of Ref.\ \cite{CM08a} for the gluon
  propagator are also shown in the figure for comparison.} 
\label{figgluongdyn}
\end{figure}

\begin{figure}
\begin{center}
\includegraphics[width=0.8\textwidth]{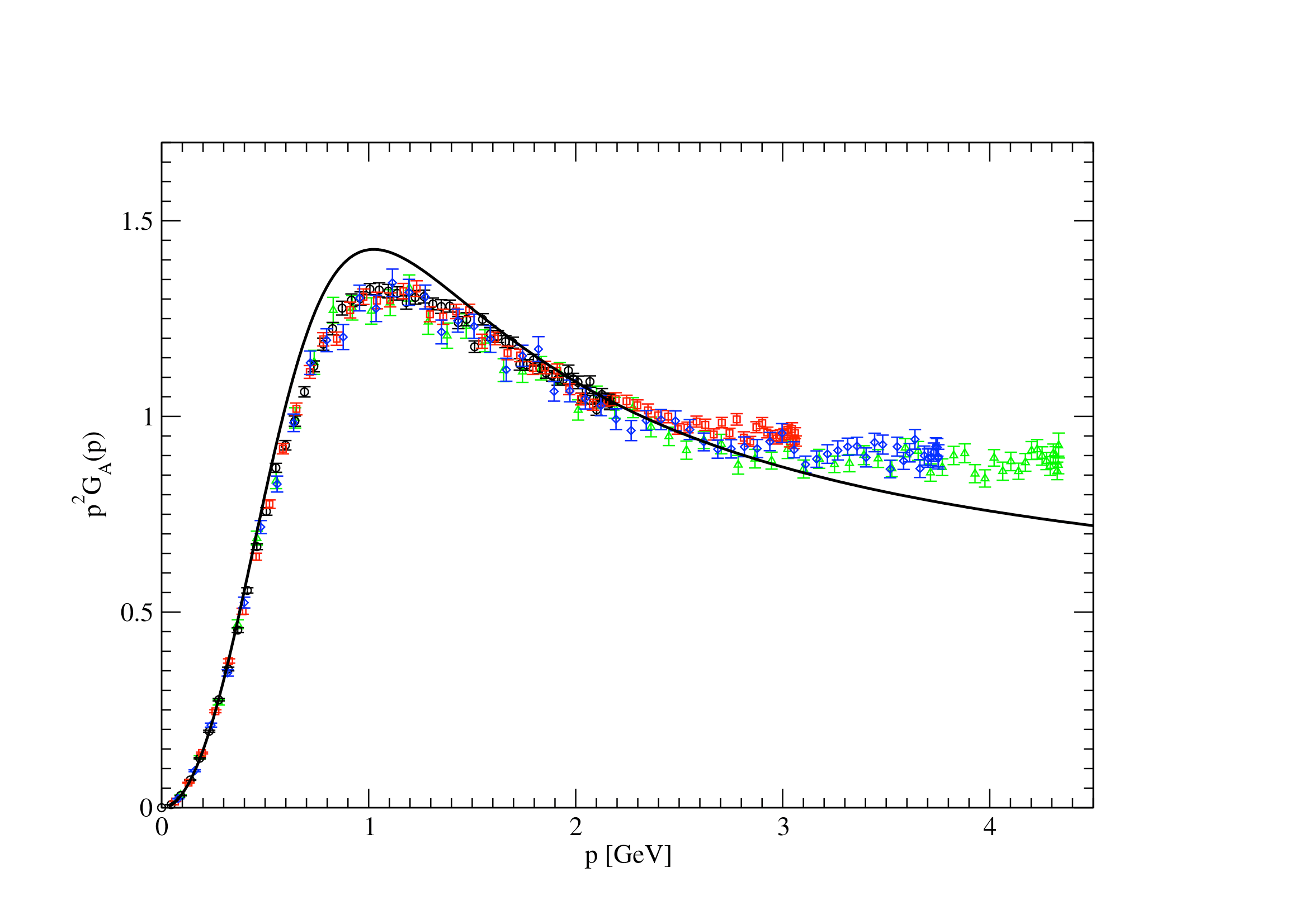} 
\end{center} 
\vspace{-1cm}
\caption{The gluon dressing function $F_A (p^2) = p^2 \, G_A (p^2)$ as a 
  function of momentum in the scale-dependent derivative scheme with the 
  scale-dependent definition of the renormalized coupling 
  constant and using the 
  same initial conditions as in Fig.\ \ref{figgluongdyn}, compared to the 
  lattice data of Ref.\ \cite{CM08a}.} 
\label{figgluondressgdyn}
\end{figure}

Our results are presented in Figs.\ \ref{figghosttw}--\ref{figgluondressgdyn} 
for the different renormalization schemes. In particular, Figs.\ 
\ref{figghosttw}--\ref{figgluondresstw} show the best fits for 
Tissier-Wschebor's original renormalization scheme, while Figs.\ 
\ref{figghostsymm}--\ref{figgluondresssymm} correspond to the 
renormalization scheme with the same 
normalization conditions \eqref{TWs1}--\eqref{TWs3} for the two-point 
functions but the renormalized coupling constant defined from the 
ghost-gluon vertex at the symmetry point as in Eq.\ \eqref{symmg}, 
rather than in the Taylor limit. While the fit to the ghost dressing
function (and also to the ghost propagator) is perfect for momenta up to
$3{.}7$ GeV in the Tissier-Wschebor scheme, there are no initial values
$g (\mu_0^2)$ and $m^2 (\mu_0^2)$ for which the renormalization group
improved ghost dressing function in the symmetry point scheme would stay
within the error bars of the lattice data for momenta below $3{.}7$ GeV.
The latter fact demonstrates that it is by no means obvious that a perfect
fit to the ghost propagator and the ghost dressing function can be achieved
(in this momentum range) by tuning the two initial values and the overall
multiplicative constant. The fit to the gluon propagator and the gluon
dressing function is also poorer in the symmetry point scheme than in the
Tissier-Wschebor (Taylor) scheme. We conclude that the original 
Tissier-Wschebor scheme leads to better results for the gluon and ghost 
propagators than the symmetry point scheme.

Note that our fits to the lattice data differ somewhat from the ones
obtained by Pel\'aez, Tissier and Wschebor in Ref.\
\cite{TW13}, in particular for the ghost dressing function, even though
the differential equations we have used are exactly the same. The reason
for this difference appears to be that Pel\'aez, Tissier and Wschebor
have fitted the renormalization
group improved results for the ghost dressing function to the lattice
data for momenta in the directions $(0, 0, 1, 1)$ and $(1, 1, 1, 1)$ on
the lattice (red and blue data points in the online version), while we have
found better overall fits by adjusting our curves to the data for
momentum directions $(0, 0, 0, 1)$ and $(0, 1, 1, 1)$, in the momentum
regime where the breaking of rotational invariance on the lattice is visible.

We present the results for all other renormalization schemes, with the
corresponding normalization conditions described in Subsection \ref{altRG},
in Figs.\ \ref{figgluonder}--\ref{figgluondressgdyn}. All these plots are
for the case where the renormalized coupling constant
is defined from an evaluation of the
ghost-gluon vertex in the Taylor limit, Eq.\ \eqref{defg}, with the
exception of the scale-dependent derivative scheme where we have also 
considered the scale-dependent definition \eqref{dyng} of the coupling 
constant. Just as in the
case of the normalization conditions \eqref{TWs1}--\eqref{TWs3} before,
the definition of the coupling constant from the ghost-gluon vertex at
the symmetry point leads to less satisfactory fits to the lattice data for
the ghost and gluon propagators (and dressing functions) than the definition
of the coupling constant that uses the Taylor limit of the same vertex,
with the same normalization conditions for the two-point functions.
In the paragraph that contains Eq.\ \eqref{dyng}, we had already
presented an intuitive argument in favor of the Taylor limit for the
description of the IR regime.

We do not show the fits to the ghost dressing functions in Figs.\
\ref{figgluonder}--\ref{figgluondressgdyn} since they are very similar 
to Fig.\ \ref{figghosttw} for the Tissier-Wschebor scheme in all cases 
(that use the Taylor scheme for the coupling constant), i.e., they 
are perfect fits for momenta below $3{.}7$ GeV, and the same is true for the
fits to the ghost propagator function. In general, our results compare fairly
well to the lattice simulations, in particular when one considers that we
have calculated the flow functions only to one-loop order. There is inevitably
an element of subjectivity left in our fitting procedure, but from the figures
we have produced it appears that, among all the renormalization 
schemes that we have considered, the best fits to the lattice data 
can be achieved in the critical ($\zeta = 1/3$) and the
scale-dependent derivative schemes (with the renormalized coupling
constant defined from the ghost-gluon vertex in the Taylor limit). In 
particular, these derivative schemes reproduce the gluon and ghost 
propagators obtained in the lattice simulations more closely than 
Tissier-Wschebor's original renormalization scheme.

In all renormalization schemes we have found a decrease of the gluon
propagator $G_A (p^2)$ towards smaller momenta in the extreme IR regime.
This decrease is not (clearly) seen in the lattice data, although at least
a slight decrease cannot be excluded from the present data, either. We shall
analytically confirm in the following subsection, through an expansion for
small momenta, that the decrease is present in all renormalization schemes
described in the previous section, in particular for all values $\zeta \ge 1/3$
in the derivative schemes, and it can thus be considered a general
  feature of the gluon propagator in the
renormalization group improved perturbative approach to Yang-Mills theory.
In principle, the effect could be overturned by higher-loop contributions
to the flow functions. However, this seems unlikely given that the decrease
is already visible in the plain perturbative one-loop expression \eqref{GAtir}
for $\Gamma^\perp_{AA} (p^2)$, where the term proportional to
$(p^2 \ln (p^2/m^2))$ dominates the $p^2$-dependence in the limit $p^2 \to 0$.

The decrease of the gluon propagator function towards small momenta
  implies a non-monotonous momentum dependence of the propagator that
  is clearly incompatible with the existence of a spectral representation
  with a non-negative spectral function $\rho (M^2)$,
  \be
G_A (p^2) = \int_0^\infty \frac{d M^2}{2 \pi} \, \frac{\rho(M^2)}{p^2 + M^2} \,,
  \ee
  as recognized before in Refs.\ \cite{CFM16} and
    \cite{RSTW17}. In other words, the non-monotony of the gluon propagator
    is sufficient for the violation of positivity.

From a comparison
of the extreme IR behavior of the numerical solutions in the derivative schemes
for different values of the parameter $\zeta$, one concludes that the decrease
of the gluon propagator towards smaller momenta gets more pronounced 
for larger values of $\zeta$ (compare Figs.\ \ref{figgluonder} and
\ref{figgluoncrit} for $\zeta = 1$ and $\zeta = 1/3$, respectively; 
this tendency can be seen to extend to larger values of $\zeta$). This 
$\zeta$-dependence of the IR decrease is at least one of the reasons
why, among the scale-independent derivative schemes, the 
scheme with the smallest possible (critical) value $\zeta = 1/3$ 
gives the best fit to the lattice data.

Even in the critical and the scale-dependent derivative schemes, 
the renormalization group improved curves for the gluon dressing 
function in Figs.\ \ref{figgluondresscrit} and \ref{figgluondressdyn} seem 
to systematically deviate from the lattice data for momenta above (roughly)
$3{.}5$ GeV. The same deviation of the gluon dressing function from the 
lattice data in the UV regime is seen in all the other renormalization
schemes as well, typically already at slightly lower momentum scales.
Our interpretation is that in this case it is the lattice data that do not
reproduce the correct UV behavior with sufficient precision.
In fact, the value of the lattice spacing used in the numerical calculations
of Ref.\ \cite{CM08a} is $a = 1{.}066$ $\mbox{GeV}^{-1}$ in physical units,
so that momenta around $3{.}5$ GeV are already of the order of the lattice
cutoff $(\pi/a)$ and the corresponding results cannot be compared
to calculations in the continuum (remember that the values that we cite for
the momenta correspond to improved lattice momenta \cite{Ma00}).

We can, however, compare the UV behavior of our renormalization group improved
results to the lattice simulations on finer (but smaller) lattices in Ref.\
\cite{BCL04}, where the physical scale was fixed via the string tension in
exactly the same way as in Ref.\ \cite{CM08a}. In Ref.\ \cite{BCL04}, the
results of lattice calculations of the gluon
and ghost dressing functions in the UV regime were compared to the usual
(renormalization group improved) perturbative two-loop propagators.
It was found that the perturbative propagators with a value of $1{.}2(1)$ GeV
for $\Lambda_{\text{QCD}}$, together with very small coefficients of the
renormalization scheme dependent two-loop terms (as in the
$\overline{\text{MS}}$ scheme), correctly reproduced the lattice results for
the dressing functions in the UV. Due to the strong renormalization scheme
dependence of $\Lambda_{\text{QCD}}$, we cannot relate the latter result for
$\Lambda_{\text{QCD}}$ [via Eq.\ \eqref{runningg}] directly to the values we have
used for $g (\mu_0)$ at $\mu_0 = 3$ GeV in our fits, but we can compare the
perturbative two-loop dressing functions to our numerical results for these
dressing functions in the UV. We obtain a value for $\Lambda_{\text{QCD}}$ of
around $1{.}0$ GeV for the fit of the perturbative two-loop propagators to
the numerical curves obtained from the renormalization group improvement in
the scale-dependent derivative scheme, and a value of about $0{.}95$ GeV for
the critical derivative scheme, where we have set the coefficients
of the renormalization scheme dependent two-loop terms to zero. These
values are quite remarkably close to and, within the error bars,
consistent with $\Lambda_{\text{QCD}} = 1{.}2$ GeV found from the comparison
of the perturbative two-loop propagators with the lattice data of
Ref.\ \cite{BCL04}. In contrast, if one tries to fit the perturbative
two-loop results to the lattice data of Refs.\ \cite{CM08a,CM08b} in the UV,
one finds a value of approximately $0{.}56$ GeV for $\Lambda_{\text{QCD}}$.

The last issue we shall discuss in this subsection 
is the violation of the Slavnov-Taylor identity \eqref{STIcomb} 
in all derivative schemes,
including the simple ($\zeta = 1$) scheme. As we have previously
explained in detail, Eq.\ \eqref{STIcomb} is a consequence of the massive
extension \eqref{BRST} of the BRST symmetry and predicts the combination
$(\Gamma_{c\bar{c}} (p^2) \Gamma_{AA}^\parallel (p^2)/p^2)$ to be 
$p^2$-independent to all orders of perturbation theory, 
even if the renormalization scheme should not properly 
preserve the extended BRST symmetry. We have already mentioned in the
previous subsection that the renormalization group improvement in such a
renormalization scheme (with the flow functions determined to a
given finite loop order) could lead to a $p^2$-dependence of the
combination $(\Gamma_{c\bar{c}} (p^2) \Gamma_{AA}^\parallel (p^2)/p^2)$, 
thus contradicting Eq.\ \eqref{STIcomb}. 

\begin{figure}
\begin{center}
\includegraphics[width=0.8\textwidth]{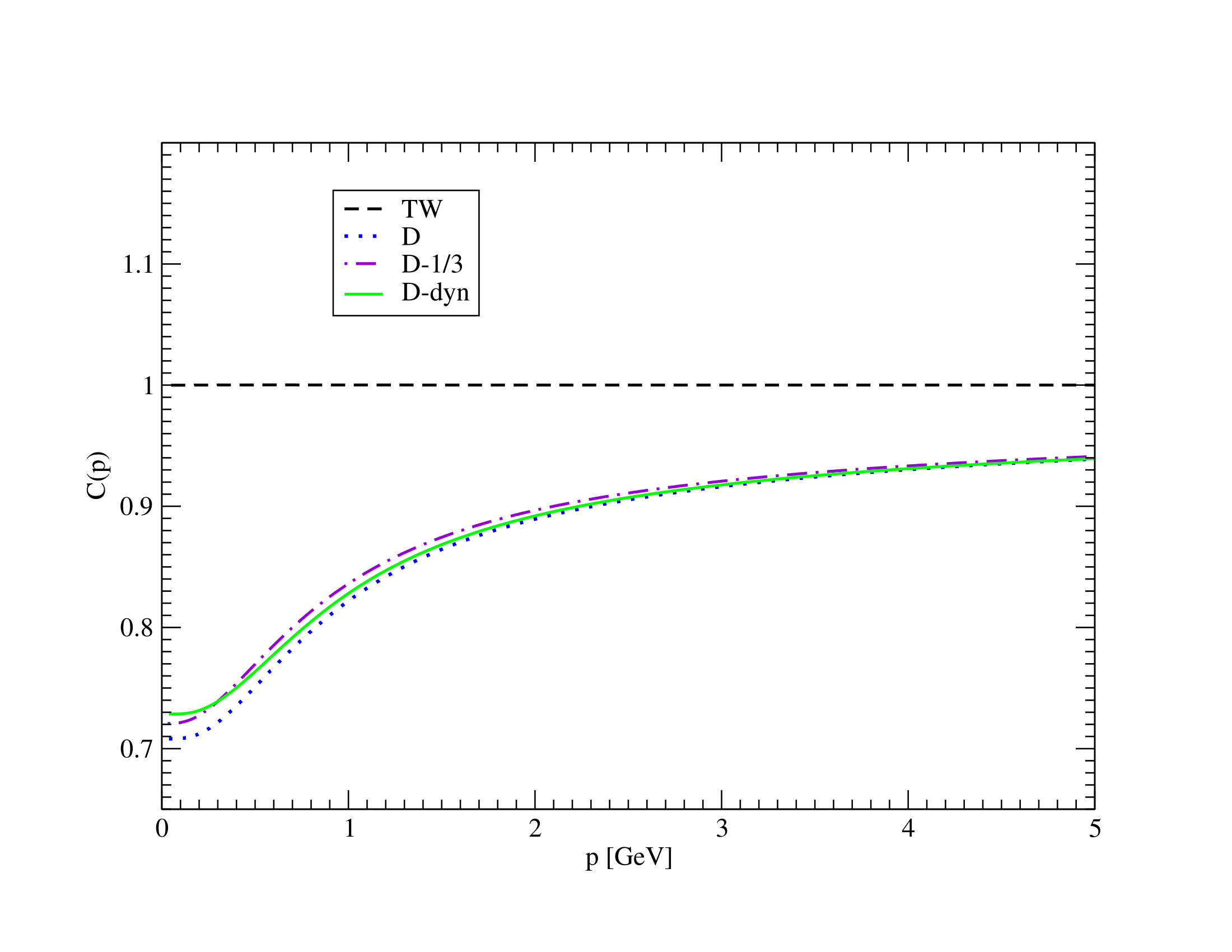} 
\end{center} 
\vspace{-1cm}
\caption{The normalized combination 
$C (p^2) = \Gamma_{c\bar{c}} (p^2) \Gamma_{AA}^\parallel (p^2)/p^2 m^2(\mu^2)$ 
as a function of momentum for the renormalization group improved
proper two-point functions
in the Tissier-Wschebor scheme (TW), the simple derivative scheme with 
$\zeta = 1$ (D), the critical ($\zeta = 1/3$) derivative scheme (D-1/3) and
the scale-dependent derivative scheme (D-dyn).}
\label{figSTI}
\end{figure}

In Fig.\ \ref{figSTI} we have plotted this
combination, normalized by dividing through the value of $m^2$
at the renormalization scale $\mu = 3$ GeV,
against $p = (p^2)^{1/2}$ for the simple ($\zeta = 1$) derivative scheme,
the critical ($\zeta = 1/3$) and the scale-dependent derivative schemes, using
our numerical results for the proper two-point functions in each of these
renormalization schemes. One clearly sees a nontrivial $p^2$-dependence 
in all these schemes. We shall confirm this finding, furthermore,
in the next subsection with the help of an analytical calculation. To
compare with, in Tissier-Wschebor's original renormalization scheme we find 
the same normalized combination, both numerically and analytically, 
to give a $p^2$-independent constant, which is also represented in Fig.\ 
\ref{figSTI}. Incidentally, at the renormalization scale 
$p = \mu = 3$ GeV, the normalized combination 
$(\Gamma_{c\bar{c}} (p^2) \Gamma_{AA}^\parallel (p^2)/p^2 m^2 (\mu^2))$
reduces to $\Gamma_{c\bar{c}} (\mu^2)/\mu^2$ as a consequence
of the normalization condition \eqref{TWs2} or \eqref{TWd2}. The
different value of the normalized combination at this scale in the original 
Tissier-Wschebor scheme as opposed to the derivative schemes is 
due to the different normalization conditions \eqref{TWs3} and \eqref{TWd3} 
for the ghost two-point function.

One may then argue that the derivative schemes do not constitute 
proper renormalizations of the theory, inasmuch as they do not respect (in 
the absence of anomalies) all the symmetries of the ``classical'' action.
As already discussed in the previous section, this is not our point 
of view here because we do not consider the massive extension of 
the BRST symmetry to be
a fundamental symmetry of the theory, i.e., of a formulation of Yang-Mills 
theory in the Landau gauge that takes care of the existence of 
gauge copies in the standard Faddeev-Popov quantization 
by restricting the integral over the gauge fields to the Gribov 
region. Rather, we look at the extended BRST symmetry as an 
``accidental'' symmetry of the simplest renormalizable effective theory 
that gives an accurate description of the crossover from 
the UV to the IR fixed point of the full theory, at least as far as 
the momentum dependence of the propagators is concerned, which is the 
Curci-Ferrari model \eqref{CF}.

\subsection{Analytical results \label{analytical}}

It is reassuring to back up the more interesting findings of the
numerical integration of the Callan-Symanzik equations with analytical
calculations, for instance to confirm that certain properties hold for
all values of the parameter $\zeta$ in the derivative schemes. Furthermore,
some of the analytical expressions, although approximate, are of
interest by themselves.

We will consider the description of the UV regime first. 
In the limit of large values of the renormalization scale,
$\mu^2 \gg m^2 (\mu^2)$, the one-loop expressions
for the flow functions, to leading order in an expansion in powers of
$(m^2/\mu^2)$, are the same in all (successful) renormalization schemes
considered in the previous section, namely
\be
\gamma_A  = -\frac{13}{6} \frac{N g^2 }{(4\pi)^2} \,, \quad
\gamma_c  = -\frac{3}{4} \frac{N g^2 }{(4\pi)^2}
\label{anomdimUV}
\ee
and
\be
\beta_g  = -\frac{11}{6} \frac{N g^2 }{(4\pi)^2} \, g   \label{betagUV}
\,.
\ee
These are, of course, the very well-known perturbative expressions in the
theory without a gluon mass term. The beta function of the mass,
\be
\beta_{m^2}  = -\frac{35}{12} \frac{N g^2 }{(4\pi)^2} \, 
m^2  \,,
\ee
is also the same in all renormalization schemes to leading order in 
$(m^2/\mu^2)$.

The integration of the differential equation for the renormalized
coupling constant to this order leads to the well-known result
\be \label{runningg}
\frac{N g^2 (\mu^2)}{(4\pi)^2} = \left( \frac{11}{3} \,
\ln \frac{\mu^2}{\Lambda_{\text{UV}}^2} \right)^{-1} \,,
\ee
where the scale $\Lambda_{\text{UV}}$ is defined through the same relation 
\eqref{runningg} in terms of the renormalized coupling constant $g (\mu_0^2)$ 
at some reference scale $\mu_0$. The scale $\Lambda_{\text{UV}}$ is
the same as the characteristic scale $\Lambda_{\text{QCD}}$ of perturbative 
Yang-Mills theory (or, generally, of perturbative QCD), but here we prefer
the notation $\Lambda_{\text{UV}}$ to distinguish it from an analogous
(different) characteristic scale $\Lambda_{\text{IR}}$ that will be introduced
below for the description of the infrared regime of the theory.

With the formula \eqref{runningg} for the running coupling constant, one can 
integrate the differential equation for the renormalized mass 
parameter, with the ``universal'' result
\be
m^2 (\mu^2) = m^2 (\mu_0^2) \left( 
\frac{\ln \, (\mu_0^2/\Lambda_{\text{UV}}^2)}
{\ln \, (\mu^2/\Lambda_{\text{UV}}^2)} \right)^{35/44} \label{runningm}
\ee
to leading order in $(m^2 (\mu_0^2)/\mu_0^2)$ and $(m^2 (\mu^2)/\mu^2)$. 
Equation \eqref{2pAparRG}, upon substituting Eq.\ \eqref{runningm} for 
$m^2 (\mu^2)$, and Eq.\ \eqref{runningg} for $g^2 (\mu^2)$ in the formula
\eqref{anomdimUV} for $\gamma_A (\mu^2)$, then leads to the equally
universal result for the longitudinal part of the proper gluonic 
two-point function
\be
\Gamma^\parallel_{AA} (p^2, \mu^2) = m^2 (\mu^2) \left(
\frac{\ln \, (\mu^2/\Lambda_{\text{UV}}^2)}
{\ln \, (p^2/\Lambda_{\text{UV}}^2)} \right)^{9/44} \,, \label{GAAparUV}
\ee
where we have put $\mu$ equal to the reference scale $\mu_0$
[alternatively, we may use Eq.\ \eqref{runningm} to write $m^2 (\mu_0^2)$ 
in terms of $m^2 (\mu^2)$]. Thus both the mass parameter and the longitudinal 
part of the gluonic two-point function tend logarithmically to zero in the 
UV limit $p^2 \to \infty$ ($\mu^2 \to \infty$ in the case of the mass
parameter), in particular, the proper two-point function becomes 
transverse in this limit. Note that the transverse part of the gluonic 
two-point function increases with $p^2$ in the UV and dominates over
the longitudinal part by a factor of $p^2/m^2 (\mu^2)$ [and a power of 
$\ln (p^2/\Lambda_{\text{UV}}^2)$, see below].

As for the ghost two-point function, using Eq.\ \eqref{runningg} for
$g^2 (\mu^2)$ in the formula \eqref{anomdimUV} for $\gamma_c (\mu^2)$
gives
\be
\exp \left( -\int_{\mu^2}^{p^2} \frac{d \mu^{\prime \, 2}}{\mu^{\prime \, 2}}
\, \gamma_c (\mu^{\prime \, 2}) \right) 
= \left( \frac{\ln \, (p^2/\Lambda_{\text{UV}}^2)}
{\ln \, (\mu^2/\Lambda_{\text{UV}}^2)} \right)^{9/44} \label{expgcUV}
\ee
in all renormalization schemes. Then Eq.\ \eqref{2pcRG} immediately yields
\be
\Gamma_{c\bar{c}}(p^2, \mu^2)
= p^2 \left( \frac{\ln \, (p^2/\Lambda_{\text{UV}}^2)}
{\ln \, (\mu^2/\Lambda_{\text{UV}}^2)} \right)^{9/44} \label{GccTW}
\ee
in Tissier-Wschebor's original scheme. For all the other (derivative) schemes,
we have from Eq.\ \eqref{2pcRGz} that
\be
\Gamma_{c\bar{c}}(p^2, \mu^2) = \int_0^{p^2} dp^{\prime \, 2}
\left( \frac{\ln \, (p^{\prime \, 2}/\Lambda_{\text{UV}}^2)}
{\ln \, (\mu^2/\Lambda_{\text{UV}}^2)} \right)^{9/44} \,. \label{intexpgcUV}
\ee

Note that Eq.\ \eqref{expgcUV} is expected to be valid only in the UV regime,
but the contribution to the integral over $p^{\prime \, 2}$ from this regime is
proportional to $p^2$ [times some power of $\ln (p^2/\Lambda_{\text{UV}}^2)$,
  see below] and thus dominates the integral, the contributions from the
other momentum regimes being suppressed by a relative factor of $1/p^2$.

The integral in Eq.\ \eqref{intexpgcUV}
can be expressed in closed form in terms of a
confluent hypergeometric or Kummer function of
$\ln (p^2/\Lambda_{\text{UV}}^2)$. However, a much more intuitive
representation for our purposes is obtained by writing the integral in
the form of a power series in $(\ln (p^2/\Lambda_{\text{UV}}^2))^{-1}$,
\bal
\lefteqn{\int_0^{p^2} dp^{\prime \, 2}
\left( \frac{\ln \, (p^{\prime \, 2}/\Lambda_{\text{UV}}^2)}
{\ln \, (\mu^2/\Lambda_{\text{UV}}^2)} \right)^{9/44}} \hspace{2cm} \n \\
&= p^2 \left( \frac{\ln \, (p^2/\Lambda_{\text{UV}}^2)}
{\ln \, (\mu^2/\Lambda_{\text{UV}}^2)} \right)^{9/44} 
\left[ 1 + \frac{A_1}{\ln \, (p^2/\Lambda_{\text{UV}}^2)}
  + \frac{A_2}{\big( \ln \, (p^2/\Lambda_{\text{UV}}^2) \big)^2}
  + \ldots \right] \,. \label{invlogexpand}
\eal
A recursion formula for the coefficients $A_i$, $i = 1, 2, \ldots$, can
easily be derived by differentiating both sides of this equation
with respect to $\ln (p^2/\Lambda_{\text{UV}}^2)$. The result coincides
with the use of the known asymptotic expansion of the Kummer function 
[for large values of $\ln (p^2/\Lambda_{\text{UV}}^2)$]. To first order in
$(\ln (p^2/\Lambda_{\text{UV}}^2))^{-1}$, one obtains
\be
\Gamma_{c\bar{c}}(p^2, \mu^2) =
p^2 \left( \frac{\ln \, (p^2/\Lambda_{\text{UV}}^2)}
{\ln \, (\mu^2/\Lambda_{\text{UV}}^2)} \right)^{9/44}
\left[ 1 - \frac{9/44}{\ln \, (p^2/\Lambda_{\text{UV}}^2)}
  + \cal{O} \left( \big( \ln \, (p^2/\Lambda_{\text{UV}}^2) \big)^{-2}
  \right) \right] \label{Gcczeta}
\ee
in all derivative schemes, including the simple ($\zeta = 1$) scheme,
the critical ($\zeta = 1/3$) and the scale-dependent derivative schemes.

Looking back at the derivation of the results \eqref{GccTW} and
\eqref{Gcczeta} which are obtained from the leading-order expressions
for the flow functions in an expansion in powers of $(m^2/\mu^2)$, one
realizes that only the Eqs.\ \eqref{anomdimUV} and \eqref{runningg} and the
formal integrals \eqref{2pcRG} and \eqref{2pcRGz} 
of the Callan-Symanzik
equations are involved, hence the difference between the results
\eqref{GccTW} in the Tissier-Wschebor scheme and \eqref{Gcczeta} in the
derivative schemes (in particular, in the simple derivative scheme with
$\zeta = 1$) is entirely due to the difference between the normalization
conditions \eqref{TWs3} and \eqref{TWd3} and, in particular, is
independent of the presence or not of a gluon mass term. We have also
carefully checked that including the contributions of the order of
$(m^2/\mu^2)$ in the flow functions generates contributions to
$\Gamma_{c\bar{c}}(p^2, \mu^2)$ that are suppressed in the UV limit
by relative factors $(m^2/p^2)$ and $(m^2/\mu^2)$, and thus are negligible
in comparison with the terms that arise from the renormalization
scheme dependence (even in a theory without a gluon mass term), i.e.,
these contributions are negligible compared to the difference between 
Eqs.\ \eqref{GccTW} and \eqref{Gcczeta}.

The results \eqref{GAAparUV}, \eqref{GccTW} and \eqref{Gcczeta} also
give a quantitative expression for the UV limit of the combination
\be
C (p^2) = \frac{\Gamma_{AA}^\parallel(p^2, \mu^2) \, 
\Gamma_{c\bar{c}}(p^2, \mu^2)}{p^2 \, m^2 (\mu^2)} \label{defC}
\ee
that is related to the Slavnov-Taylor identity \eqref{STIcomb} [the massive
  extension of the BRST symmetry implies that $C(p^2)$ should be
  $p^2$-independent] and has been plotted for the different renormalization
schemes in Fig.\ \ref{figSTI}. From Eqs.\ \eqref{GAAparUV} and \eqref{GccTW}
we find $C(p^2) = 1$ in the Tissier-Wschebor scheme, which is numerically
confirmed in Fig.\ \ref{figSTI}. On the other hand, Eqs.\ \eqref{GAAparUV}
and \eqref{Gcczeta} imply that
\be
C(p^2) = 1 - \frac{9/44}{\ln \, (p^2/\Lambda_{\text{UV}}^2)}
  + \cal{O} \left( \big( \ln \, (p^2/\Lambda_{\text{UV}}^2) \big)^{-2}
  \right) \label{STIcombUV}
\ee
in all derivative schemes, thus violating the Slavnov-Taylor identity. The
violation of the identity is also numerically confirmed in 
Fig.\ \ref{figSTI}. For a quantitative comparison, one may express
the term proportional to $(\ln (p^2/\Lambda_{\text{UV}}^2))^{-1}$
on the right-hand side of Eq.\ \eqref{STIcombUV} in terms of the
initial condition $g(\mu^2)$ for the solution of the Callan-Symanzik
equations. With
\be
\ln \frac{p^2}{\Lambda_{\text{UV}}^2}
= \ln \frac{p^2}{\mu^2} + \ln \frac{\mu^2}{\Lambda_{\text{UV}}^2}
= \ln \frac{p^2}{\mu^2} + \left( \frac{11}{3} \,
\frac{N g^2 (\mu^2)}{(4 \pi)^2} \right)^{-1} \label{compare1}
\ee
from Eq.\ \eqref{runningg} one has, to the present order,
\be
C (p^2) = 1 - \frac{3}{4} \, \frac{N g^2 (\mu^2)}{(4 \pi)^2}
\left( 1 + \frac{11}{3} \, \frac{N g^2 (\mu^2)}{(4 \pi)^2} \,
\ln \frac{p^2}{\mu^2} \right)^{-1} \,. \label{STIcombUVg}
\ee
Comparing this formula with the numerical curves shows that Eq.\
\eqref{STIcombUV} does correctly describe the qualitative behavior observed
in the UV, however, in a quantitative sense the analytical formula 
matches the numerical curves
only asymptotically for \emph{very} large momenta [which has nothing to do
with the expansion in inverse powers of logarithms: the exact evaluation
of the integral \eqref{intexpgcUV} leads to the same result]. We can also 
explicitly see the effect of the renormalization group improvement in the 
expression \eqref{STIcombUVg}: from an expansion in powers of $g^2 (\mu^2)$
it is clear that, to one-loop order, $C (p^2)$ is a
$p^2$-independent constant different from one, as expected, while the
$p^2$-dependence generated by the renormalization group improvement
sets in already at two-loop order.

For the Tissier-Wschebor scheme, it is easy to show that $C (p^2) = 1$
holds to any perturbative order in the renormalization group improved
theory: Eqs.\ \eqref{2pcRG} and \eqref{2pAparRG} in this renormalization
scheme imply that
\bal
\lefteqn{p^2 \frac{\d}{\d p^2} \left( \frac{\Gamma_{AA}^\parallel(p^2, \mu^2)
    \, \Gamma_{c\bar{c}}(p^2, \mu^2)}{p^2} \right)} \hspace{1cm} \n \\
&= p^2 \frac{\d}{\d p^2} \left[ m^2 (p^2) \,
  \exp \left( -\int_{\mu^2}^{p^2} \frac{d \mu^{\prime \, 2}}{\mu^{\prime \, 2}}
  \big[ \gamma_A (\mu^{\prime \, 2}) + \gamma_c (\mu^{\prime \, 2}) \big] 
  \right) \right] \n \\
&= \left( p^2 \frac{d}{d p^2} \, m^2 (p^2) - m^2 (p^2) \big[
  \gamma_A (p^2) + \gamma_c (p^2) \big] \right)
\exp \left( -\int_{\mu^2}^{p^2} \frac{d \mu^{\prime \, 2}}{\mu^{\prime \, 2}}
\big[ \gamma_A (\mu^{\prime \, 2}) + \gamma_c (\mu^{\prime \, 2}) \big] \right)
\n \\
&= 0
\eal
to all orders. The last equality is a consequence of
\be
Z_A (\mu^2) Z_c (\mu^2) Z_{m^2} (\mu^2) = 1
\ee
for all $\mu^2$ [cf.\ Eq.\ \eqref{STIGl}], since it follows from
the latter relation that
\bal
\mu^2 \frac{d}{d \mu^2} \, m^2 (\mu^2)
&= \mu^2 \frac{d}{d \mu^2} Z_{m^2}^{-1} (\mu^2) \, m_B^2 \n \\
&= -m^2 (\mu^2) \, \mu^2 \frac{d}{d \mu^2} \ln Z_{m^2} (\mu^2) \n \\
&= m^2 (\mu^2) \, \mu^2 \frac{d}{d \mu^2} \big[ \ln Z_A (\mu^2)
  + \ln Z_c (\mu^2) \big] \n \\
&= m^2 (\mu^2) \big[ \gamma_A (\mu^2) + \gamma_c (\mu^2) \big] \,,
\eal
i.e., Eq.\ \eqref{betam} is valid to all perturbative orders. For
$p^2 = \mu^2$, we have
\be
\frac{\Gamma_{AA}^\parallel(\mu^2, \mu^2) \, 
\Gamma_{c\bar{c}}(\mu^2, \mu^2)}{\mu^2} = m^2 (\mu^2)
\ee
from the normalization conditions \eqref{TWs2} and \eqref{TWs3}, and the
$p^2$-independence of $(\Gamma_{AA}^\parallel(p^2, \mu^2) \,
\Gamma_{c\bar{c}}(p^2, \mu^2)/p^2)$ then implies $C(p^2)=1$ to all 
perturbative orders.

We will also have a look at the UV behavior of the transverse part 
of the proper gluonic two-point function or, equivalently, at the 
UV behavior of the gluon propagator. To leading order in the UV limit, we 
have from Eqs.\ \eqref{anomdimUV} and \eqref{runningg} that
\be
\exp \left( -\int_{\mu^2}^{p^2} \frac{d \mu^{\prime \, 2}}{\mu^{\prime \, 2}}
\, \gamma_A (\mu^{\prime \, 2}) \right) 
= \left( \frac{\ln \, (p^2/\Lambda_{\text{UV}}^2)}
{\ln \, (\mu^2/\Lambda_{\text{UV}}^2)} \right)^{13/22} \,.
\ee
Then Eq.\ \eqref{2pARG} implies
\be
\Gamma^\perp_{AA} (p^2, \mu^2) =
p^2 \left( \frac{\ln \, (p^2/\Lambda_{\text{UV}}^2)}
{\ln \, (\mu^2/\Lambda_{\text{UV}}^2)} \right)^{13/22} \label{GperpUVtw}
\ee
in the Tissier-Wschebor scheme. In Eq.\ \eqref{GperpUVtw},
we have systematically suppressed
contributions of the relative order $(m^2 (\mu^2)/p^2)$ and
$(m^2 (\mu^2)/\mu^2)$ [times some power of $\ln (p^2/\Lambda^2_{\text{UV}})$
  and $\ln (\mu^2/\Lambda^2_{\text{UV}})$], in particular the contribution
from $m^2 (p^2)$ in Eq.\ \eqref{2pARG} (see also below). To the same order,
we find in all derivative schemes that
\be
\Gamma^\perp_{AA}(p^2, \mu^2) =
p^2 \left( \frac{\ln \, (p^2/\Lambda_{\text{UV}}^2)}
{\ln \, (\mu^2/\Lambda_{\text{UV}}^2)} \right)^{13/22}
\left[ 1 - \frac{13/22}{\ln \, (p^2/\Lambda_{\text{UV}}^2)}
  + \cal{O} \left( \big( \ln \, (p^2/\Lambda_{\text{UV}}^2) \big)^{-2}
  \right) \right] \label{GperpUVz}
\ee
by performing the $p^{\prime \, 2}$-integrals in Eqs.\ \eqref{2pAperpRGz}
or \eqref{2pAperpRGdyn}, respectively, and using an expansion analogous to
Eq.\ \eqref{invlogexpand}. The reason for the independence of the result
with respect to the derivative scheme employed for the renormalization [cf.,
  e.g., Eq.\ \eqref{2pAperpRGz}] is the relative suppression of
$\Gamma^\parallel_{AA}(p^2, \mu^2)$ by a factor of $(m^2 (\mu^2)/p^2)$,
see Eq.\ \eqref{GAAparUV}, and also the suppression of all
contributions from the low and intermediate momentum regimes, like
$\Gamma^\parallel_{AA}(0, \mu^2)$, by a relative factor of $1/p^2$,
cf.\ the argument after Eq.\ \eqref{intexpgcUV}.

The derivation of the results \eqref{GperpUVtw} and \eqref{GperpUVz}
only involves Eqs.\ \eqref{anomdimUV} and \eqref{runningg} and the
integrals \eqref{2pARG} and \eqref{2pAperpRGz} or \eqref{2pAperpRGdyn}
of the Callan-Symanzik equations, where all terms involving
$m^2 (p^2)$ or $\Gamma^\parallel_{AA}(p^{\prime \, 2}, \mu^2)$ can be
neglected to this order. The difference between Eqs.\ \eqref{GperpUVtw}
and \eqref{GperpUVz} then merely reflects the difference between the
normalization conditions \eqref{GAdiff} and \eqref{TWz1} [or \eqref{TWd1}],
where the longitudinal part $\Gamma^\parallel_{AA}(p^2, \mu^2)$ appearing
in these conditions is again irrelevant in the UV limit. In particular,
the presence or not of a gluonic mass term is irrelevant in the UV limit;
its contribution is strongly suppressed relative to the renormalization
scheme dependence which equally appears in the usual perturbative
formulation of Yang-Mills theory without a gluon mass term.

For a quantitative evaluation of the corrections due to the presence of
a gluon mass term in our present formulation, we have calculated the gluon
propagator in the UV limit to next-to-leading order in an expansion in
powers of $(m^2 (\mu^2)/p^2)$ and $(m^2 (\mu^2)/\mu^2)$ for the case of
Tissier-Wschebor's original scheme, where the corrections of the order of
inverse powers of $\ln (p^2/\Lambda^2_{\text{UV}})$ that appear in Eq.\
\eqref{GperpUVz} are absent. For a calculation to this order, one needs
the next-to-leading order expressions for the anomalous dimensions,
\bal
\gamma_A &= -\frac{13}{6} \frac{N g^2}{(4\pi)^2}
+ \frac{3}{4} \frac{N g^2}{(4\pi)^2} \, \frac{m^2}{\mu^2}
\left( \ln \frac{\mu^2}{m^2} + \frac{65}{6} \right) \,, \n \\
\gamma_c &= -\frac{3}{4} \frac{N g^2}{(4\pi)^2}
+ \frac{3}{4} \frac{N g^2}{(4\pi)^2} \, \frac{m^2}{\mu^2}
\left( \ln \frac{\mu^2}{m^2} - \frac{1}{2} \right) \,,
\label{anomdimUVnlo}
\eal
which are obtained from the explicit expressions \eqref{GAtuv}--\eqref{Gcuv}
upon implementing the normalization conditions \eqref{TWs3} and \eqref{GAdiff}.

The beta function for the coupling constant to next-to-leading order is
then determined from the relation \eqref{betag}. Even though 
in principle the beta functions of the coupling constant and the 
mass parameter couple the differential equations \eqref{2coupdiffeq}
for $g (\mu^2)$ and $m^2 (\mu^2)$, the equation for the coupling constant
can be integrated independently to next-to-leading order by replacing
$m^2 (\mu^2)$ with the leading-order expression \eqref{runningm},
\be
m^2 (\mu^2) = m^2 (\mu_0^2) \left( 
\frac{\ln \, (\mu_0^2/\Lambda_{\text{UV}}^2)}
{\ln \, (\mu^2/\Lambda_{\text{UV}}^2)} \right)^{35/44} \label{runningm2}
\ee
(with an arbitrary UV reference scale $\mu_0$), given that $m^2 (\mu^2)$
appears only in the next-to-leading order term in the expression for
$\beta_g (\mu^2)$. The result for $g (\mu^2)$ including terms of the order
of $(m^2 (\mu_0^2)/\mu^2)$ and $(m^2 (\mu_0^2)/\mu_0^2)$ is
\bal
\lefteqn{\frac{N g^2 (\mu^2)}{(4\pi)^2} = \left( \frac{11}{3} \,
\ln \frac{\mu^2}{\Lambda_{\text{UV}}^2} \right)^{-1}} \hspace{1.5cm} \n \\
&{}\times \left[ 1 - \frac{27}{44} \frac{m^2 (\mu_0^2)}{\mu^2}
  \left( \frac{\ln \, (\mu_0^2/\Lambda_{\text{UV}}^2)}
       {\ln \, (\mu^2/\Lambda_{\text{UV}}^2)} \right)^{35/44}
       + \frac{27}{44} \frac{m^2 (\mu_0^2)}{\mu_0^2} \,
       \frac{\ln \, (\mu_0^2/\Lambda_{\text{UV}}^2)}
       {\ln \, (\mu^2/\Lambda_{\text{UV}}^2)} \right] \,. \label{runninggnlo}
\eal
In the course of the calculation, an integral of the type \eqref{intexpgcUV}
is encountered, which we have again expanded in a series in powers of
$(\ln (\mu^2/\Lambda_{\text{UV}}^2))^{-1}$. Incidentally, we still define
$\Lambda_{\text{UV}}$ through the relation \eqref{runningg} at some reference
scale $\mu_0$ [although now there are corrections of the order
  $(m^2 (\mu_0^2)/\mu^2)$ and $(m^2 (\mu_0^2)/\mu_0^2)$ to this relation
  for $\mu \neq \mu_0$].

The result \eqref{runninggnlo} is now inserted in (the leading-order term
in) the expression \eqref{anomdimUVnlo} for $\gamma_A (\mu^2)$ and then,
together with the formula \eqref{runningm2} for $m^2 (\mu^2)$, used in
Eq.\ \eqref{2pARG} to calculate $\Gamma^\perp_{AA}(p^2, \mu^2)$. The
result is
\bal
\Gamma_{AA}^\perp (p^2, \mu^2) &= p^2 \left(
\frac{\ln \, (p^2/\Lambda_{\text{UV}}^2)}{\ln \, (\mu^2/\Lambda_{\text{UV}}^2)}
\right)^{13/22} \Bigg[ 1 + \frac{153}{968} \frac{m^2 (\mu^2)}{\mu^2}
  - \frac{351}{968} \frac{m^2 (\mu^2)}{\mu^2} \, \frac{\ln \,
    (\mu^2/\Lambda_{\text{UV}}^2)}{\ln \, (p^2/\Lambda_{\text{UV}}^2)} \n \\
&\hspace{4.5cm} {}+ \frac{53}{44} \frac{m^2 (\mu^2)}{p^2}
  \left( \frac{\ln \, (\mu^2/\Lambda_{\text{UV}}^2)}
       {\ln \, (p^2/\Lambda_{\text{UV}}^2)}\right)^{35/44} \Bigg]
\label{GperpUVnlo}
\eal
to next-to-leading order in $(m^2 (\mu^2)/p^2)$ and $(m^2 (\mu^2)/\mu^2)$,
and to leading order in an expansion in powers of
$(\ln (p^2/\Lambda_{\text{UV}}^2))^{-1}$ and
$(\ln (\mu^2/\Lambda_{\text{UV}}^2))^{-1}$. We have again employed the latter
expansion to evaluate the integral involved in Eq.\ \eqref{2pARG}. In order
to simplify the result \eqref{GperpUVnlo}, we have identified the scale
$\mu$ with the reference scale $\mu_0$.

We have also compared the expression \eqref{GperpUVnlo} to our numerical
result for the gluon dressing function $p^2/\Gamma_{AA}^\perp (p^2, \mu^2)$ in
the Tissier-Wschebor scheme. For the comparison, we have used that
\be
\frac{\ln \, (p^2/\Lambda_{\text{UV}}^2)}{\ln \, (\mu^2/\Lambda_{\text{UV}}^2)}
= 1 + \frac{11}{3} \frac{N g^2 (\mu^2)}{(4 \pi)^2} \,
\ln \frac{p^2}{\mu^2} \,, \label{logratiotog}
\ee
see Eq.\ \eqref{compare1}, which is valid at the reference scale $\mu$
where Eq.\ \eqref{runningg} holds. Equation \eqref{GperpUVnlo} provides
an excellent fit to the numerical curve down to momenta
of about $2{.}6$ GeV. This fit is clearly better than the one provided by
the leading-order expression \eqref{GperpUVtw}. We emphasize again that the
renormalization scheme dependence leads to much larger corrections to Eq.\
\eqref{GperpUVtw} than the next-to-leading order terms in Eq.\
\eqref{GperpUVnlo}, see Eq.\ \eqref{GperpUVz}. We have also extended the
analytical calculation in the Tissier-Wschebor scheme to the
next-to-leading order in $(\ln (p^2/\Lambda_{\text{UV}}^2))^{-1}$ and
$(\ln (\mu^2/\Lambda_{\text{UV}}^2))^{-1}$ [at next-to-leading order in
  $(m^2 (\mu^2)/p^2)$ and $(m^2 (\mu^2)/\mu^2)$], which results in an even
slightly better fit to our numerical result for the gluon dressing function
in this scheme.

As the final topic in the discussion of the UV behavior of 
the renormalization
group improved propagators, we show the inconsistency of the renormalization
scheme defined by Eqs.\ \eqref{tt2}--\eqref{TWd3} in the sense that the
longitudinal part of the proper gluonic two-point function does not tend
to zero in the limit of large momenta, but actually grows without bound.
In particular, one does not recover the usual (not extended) BRST symmetry
in this limit.

In this renormalization scheme, the anomalous dimensions $\gamma_A$ and
$\gamma_c$ are the same as in all derivative schemes since they are obtained
from the normalization conditions \eqref{TWd1} and \eqref{TWd3}. Hence,
to leading order in the UV limit, Eqs.\ \eqref{anomdimUV} and also Eq.\
\eqref{betagUV} for the beta function of the coupling constant hold, while
we have already obtained
\be
\beta_{m^2} = -\frac{13}{6} \frac{N g^2}{(4\pi)^2} \, \mu^2
\ee
to leading order in Eq.\ \eqref{betamincons} in the previous section. From
the integration of the differential equations for $g (\mu^2)$ and
$m^2 (\mu^2)$ one obtains Eq.\ \eqref{runningg} for the running coupling
constant and
\be
m^2 (\mu^2) = m^2 (\mu_0^2) + \frac{13}{22} \,
\frac{\mu_0^2}{\ln \, (\mu_0^2/\Lambda_{\text{UV}}^2)} - \frac{13}{22} \,
\frac{\mu^2}{\ln \, (\mu^2/\Lambda_{\text{UV}}^2)} \label{runmincons}
\ee
for the mass parameter (with an arbitrary UV reference scale $\mu_0$).
For the integration of the differential equation for $m^2 (\mu^2)$, 
we have used
an expansion in inverse powers of $(\ln (\mu^2/\Lambda_{\text{UV}}^2))$ and
$(\ln (\mu_0^2/\Lambda_{\text{UV}}^2))$ just as in Eq.\ \eqref{invlogexpand},
while the exact integral is given in this case by an exponential integral
function. The first corrections to Eq.\ \eqref{runmincons} are
thus suppressed by relative factors of
$(\ln (\mu^2/\Lambda_{\text{UV}}^2))^{-1}$ or
$(\ln (\mu_0^2/\Lambda_{\text{UV}}^2))^{-1}$.

We shall still have to improve on the approximations of the flow functions
by incorporating the first corrections in an expansion in powers of
$(m^2/\mu^2)$. From Eq.\ \eqref{runmincons} it is clear that these
corrections are only suppressed by one power of
$(\ln (\mu^2/\Lambda_{\text{UV}}^2))^{-1}$ or
$(\ln (\mu_0^2/\Lambda_{\text{UV}}^2))^{-1}$ [all terms in Eq.\ 
\eqref{runmincons} are formally considered to be of the same order 
in the expansion in powers of inverse logarithms]. The expansions of the 
flow functions to this order are
\bal
\gamma_A &= - \frac{13}{6} \frac{N g^2}{(4\pi)^2}
\left( 1 - \frac{9}{26} \frac{m^2}{\mu^2} \right) \,, \n \\
\gamma_c &= - \frac{3}{4} \frac{N g^2}{(4\pi)^2}
\left( 1 - \frac{m^2}{\mu^2} \right) \,, \n \\
\beta_{m^2} &= -\frac{13}{6} \frac{N g^2}{(4\pi)^2} \, \mu^2
\left( 1 + \frac{m^2}{\mu^2} \right) \,. \label{flowfincons}
\eal
We will first
use these expressions to improve on the leading-order results for
$g (\mu^2)$ and $m^2 (\mu^2)$. To this end, we will iteratively
solve the corresponding system of differential equations.

In each iteration, we first solve the differential equation for $g (\mu^2)$
and then the one for $m^2 (\mu^2)$, using the solution of the former
equation in the latter. To leading order we obtain, of course, Eqs.\
\eqref{runningg} and \eqref{runmincons}. We insert the leading-order result
\eqref{runmincons} for $m^2 (\mu^2)$ in the next-to-leading order flow
functions \eqref{flowfincons} [using relation \eqref{betag} for
the beta function $\beta_g$] to start the second iteration and retain
all contributions up to next-to-leading order in the integration. The
results are
\be
\frac{N g^2 (\mu^2)}{(4\pi)^2} = \left( \frac{11}{3} \, \ln \frac{\mu^2}
  {\Lambda_{\text{UV}}^2} \right)^{-1} \left[ 1 - 
  \frac{351}{968} \, \frac{\ds \ln
  \left( \frac{\ln \, (\mu^2/\Lambda_{\text{UV}}^2)}
       {\ln \, (\mu_0^2/\Lambda_{\text{UV}}^2)} \right)}
  {\ds \ln \frac{\mu^2}{\Lambda_{\text{UV}}^2}} \right]
 \label{runginconsnlo}
\ee
and
\bal
m^2 (\mu^2) &= m^2 (\mu_0^2) \left[ 1 -
\frac{13}{22} \, \ln
  \left( \frac{\ln \, (\mu^2/\Lambda_{\text{UV}}^2)}
       {\ln \, (\mu_0^2/\Lambda_{\text{UV}}^2)} \right) \right] \n \\[2mm]
&\phantom{=} {}+ \frac{13}{22} \,
\frac{\mu_0^2}{\ln \, (\mu_0^2/\Lambda_{\text{UV}}^2)}
\left[ 1 - \frac{13}{22} \, \ln
  \left( \frac{\ln \, (\mu^2/\Lambda_{\text{UV}}^2)}
       {\ln \, (\mu_0^2/\Lambda_{\text{UV}}^2)} \right) 
+ \frac{9/22}{\ln \, (\mu_0^2/\Lambda_{\text{UV}}^2)} \right] \n \\[1mm]
&\phantom{=} {}- \frac{13}{22} \,
\frac{\mu^2}{\ln \, (\mu^2/\Lambda_{\text{UV}}^2)}
\left[ 1 - 
  \frac{351}{968} \, \frac{\ds \ln
  \left( \frac{\ln \, (\mu^2/\Lambda_{\text{UV}}^2)}
       {\ln \, (\mu_0^2/\Lambda_{\text{UV}}^2)} \right)}
  {\ds \ln \frac{\mu^2}{\Lambda_{\text{UV}}^2}}
+ \frac{9/22}{\ln \, (\mu^2/\Lambda_{\text{UV}}^2)} \right] \,.
\label{runminconsnlo}
\eal

Unexpectedly, the iteration has not only produced terms to next-to-leading
order, but also new contributions to leading order in the expansion in
powers of inverse logarithms, in the result for $m^2 (\mu^2)$. If one
keeps iterating the equations to the same next-to-leading order in the
expansion, the expression for $g (\mu^2)$ in Eq.\ \eqref{runginconsnlo}
remains unchanged, but new leading-order terms are generated in each
iteration for $m^2 (\mu^2)$. After an infinite number of iterations, the
complete next-to-leading order results are Eq.\ \eqref{runginconsnlo} for
$g (\mu^2)$, and
\bal
m^2 (\mu^2) &= m^2 (\mu_0^2) 
  \left( \frac{\ln \, (\mu_0^2/\Lambda_{\text{UV}}^2)}
       {\ln \, (\mu^2/\Lambda_{\text{UV}}^2)} \right)^{13/22} \n \\[2mm]
&\phantom{=} {}+ \frac{13}{22} \,
       \frac{\mu_0^2}{\ln \, (\mu_0^2/\Lambda_{\text{UV}}^2)}
\left[ \left( \frac{\ln \, (\mu_0^2/\Lambda_{\text{UV}}^2)}
       {\ln \, (\mu^2/\Lambda_{\text{UV}}^2)} \right)^{13/22} 
+ \frac{9/22}{\ln \, (\mu_0^2/\Lambda_{\text{UV}}^2)} \right] \n \\[1mm]
&\phantom{=} {}- \frac{13}{22} \,
\frac{\mu^2}{\ln \, (\mu^2/\Lambda_{\text{UV}}^2)}
\left[ 1 - 
  \frac{351}{968} \, \frac{\ds \ln
  \left( \frac{\ln \, (\mu^2/\Lambda_{\text{UV}}^2)}
       {\ln \, (\mu_0^2/\Lambda_{\text{UV}}^2)} \right)}
  {\ds \ln \frac{\mu^2}{\Lambda_{\text{UV}}^2}}
+ \frac{9/22}{\ln \, (\mu^2/\Lambda_{\text{UV}}^2)} \right] \,.
\label{runminconsinf}
\eal
Reassuringly, the exact solution of the same equation for $m^2 (\mu^2)$,
with the beta function $\beta_{m^2}$ from Eq.\ \eqref{flowfincons} and
taking Eq.\ \eqref{runginconsnlo} for $g (\mu^2)$, leads to the same
result for the leading-order terms as in Eq.\ \eqref{runminconsinf}.
Compared to Eq.\ \eqref{runminconsinf}, the exact solution generates
additional next-to-leading order terms, which is not surprising, nor
does it signal any inconsistency of the iterative procedure, as we shall
see.

We have carried through the iterative procedure to the
next-to-next-to-leading order (the corresponding calculations are
\emph{extremely} tedious). Again, we find that an infinite number of
iterations is necessary to stabilize the result to this order, and the
final expression for $m^2 (\mu^2)$ is found to contain new
next-to-leading order terms as compared to Eq.\ \eqref{runminconsinf},
however, the leading-order terms are the same as in Eq.\ \eqref{runminconsinf}.
In the following considerations, we will only use the (complete)
leading-order results, Eq.\ \eqref{runningg} for $g (\mu^2)$ and
\be
m^2 (\mu^2) = \left( m^2 (\mu_0^2) + \frac{13}{22} \,
\frac{\mu_0^2}{\ln \, (\mu_0^2/\Lambda_{\text{UV}}^2)} \right)
\left( \frac{\ln \, (\mu_0^2/\Lambda_{\text{UV}}^2)}
{\ln \, (\mu^2/\Lambda_{\text{UV}}^2)} \right)^{13/22}
- \frac{13}{22} \, \frac{\mu^2}{\ln \, (\mu^2/\Lambda_{\text{UV}}^2)} \,.
\label{runminconsflo}
\ee
The most important qualitative feature of the latter result is
that, for sufficiently large momentum scales $\mu$, $m^2 (\mu^2)$ becomes 
negative and its absolute value increases with $\mu$, while 
$m^2 (\mu^2)/\mu^2$ tends logarithmically
to zero. This behavior should be compared with the ``consistent''
leading-order result \eqref{runningm} in all other renormalization 
schemes.

Let us now determine the UV behavior of $\Gamma_{AA}^\parallel (p^2, \mu^2)$
in this scheme. As a consequence of the normalization condition \eqref{tt2},
we have
\be
\Gamma^\perp_{AA}(p^2, \mu^2) = \big( m^2(p^2) + p^2 \big)
\exp \left( -\int_{\mu^2}^{p^2} \frac{d \mu^{\prime \, 2}}{\mu^{\prime \, 2}} 
\, \gamma_A (\mu^{\prime \, 2}) \right) \,, \label{GAperpincons}
\ee
the same as in Tissier-Wschebor's scheme. Then
\be
p^2 \frac{\d}{\d p^2} \, \Gamma^\perp_{AA}(p^2, \mu^2) 
= \big[ \beta_{m^2} (p^2) + p^2 - \big( m^2(p^2) + p^2 \big) \gamma_A (p^2)
\big] \exp \left( -\int_{\mu^2}^{p^2} 
\frac{d \mu^{\prime \, 2}}{\mu^{\prime \, 2}} \, 
\gamma_A (\mu^{\prime \, 2}) \right) \,.
\ee
On the other hand, the normalization condition \eqref{TWd1} implies
the relation \eqref{2pARGzdif} (with $\zeta = 1$), so that
\bal
p^2 \frac{\d}{\d p^2} \, \Gamma^\parallel_{AA}(p^2, \mu^2)
&= p^2 \frac{\d}{\d p^2} \, \Gamma^\perp_{AA}(p^2, \mu^2) 
- p^2 \frac{\d}{\d p^2} \left( \Gamma^\perp_{AA}(p^2, \mu^2) 
- \Gamma^\parallel_{AA}(p^2, \mu^2) \right) \n \\
&= \big[ \beta_{m^2} (p^2) - \big( m^2(p^2) + p^2 \big) \gamma_A (p^2) \big] 
\exp \left( -\int_{\mu^2}^{p^2} \frac{d \mu^{\prime \, 2}}{\mu^{\prime \, 2}} 
\, \gamma_A (\mu^{\prime \, 2}) \right) \n \\
&= \left[ -\frac{13}{6} \frac{N g^2 (p^2)}{(4\pi)^2} \, p^2
\left( 1 + \frac{m^2 (p^2)}{p^2} \right)
+ \frac{13}{6} \frac{N g^2 (p^2)}{(4\pi)^2} \, p^2
\left( 1 + \frac{17}{26} \frac{m^2 (p^2)}{p^2} \right) \right] \n \\
&\phantom{=\bigg[} {}\times
\exp \left( -\int_{\mu^2}^{p^2} \frac{d \mu^{\prime \, 2}}{\mu^{\prime \, 2}} 
\, \gamma_A (\mu^{\prime \, 2}) \right) \n \\
&= -\frac{3}{4} \frac{N g^2 (p^2)}{(4\pi)^2} \, m^2 (p^2) \,
\exp \left( -\int_{\mu^2}^{p^2} \frac{d \mu^{\prime \, 2}}{\mu^{\prime \, 2}} 
\, \gamma_A (\mu^{\prime \, 2}) \right) \,, \label{derGAparincons}
\eal
where we have expanded the term in square brackets in powers of
$(m^2 (p^2)/p^2)$ using Eq.\ \eqref{flowfincons}, and retained only
the dominant term after the cancellation of the leading-order terms in
this expansion.

Finally, we substitute the leading-order expressions for $g (\mu^2)$ and
$\gamma_A (\mu^2)$ and Eq.\ \eqref{runminconsflo} for $m^2 (\mu^2)$ in
Eq.\ \eqref{derGAparincons} and integrate over $p^2$ to obtain
\bal
\Gamma^\parallel_{AA}(p^2, \mu^2) 
&= \Gamma^\parallel_{AA}(\mu^2, \mu^2) 
- \frac{9}{44} \left( m^2 (\mu^2) + \frac{13}{22} \,
\frac{\mu^2}{\ln \, (\mu^2/\Lambda_{\text{UV}}^2)} \right)
\ln \left( \frac{\ln \, (p^2/\Lambda_{\text{UV}}^2)}
{\ln \, (\mu^2/\Lambda_{\text{UV}}^2)} \right) \n \\
&\phantom{=} {}+ \frac{117}{968} \, \frac{p^2}
{\big( \ln \, (p^2/\Lambda_{\text{UV}}^2) \big)^2}
\left( \frac{\ln \, (p^2/\Lambda_{\text{UV}}^2)}
{\ln \, (\mu^2/\Lambda_{\text{UV}}^2)} \right)^{13/22}
- \frac{117}{968} \, \frac{\mu^2}
{\big( \ln \, (\mu^2/\Lambda_{\text{UV}}^2) \big)^2} \,, \label{GAparincons}
\eal
where we have identified the scale $\mu$ with the reference scale $\mu_0$
for simplicity. Note that our leading-order calculation has produced a
leading-order term [first line in Eq.\ \eqref{GAparincons}] and a
next-to-leading order term (second line), in terms of the powers
of $(\ln (p^2/\Lambda_{\text{UV}}^2))^{-1}$ and
$(\ln (\mu^2/\Lambda_{\text{UV}}^2))^{-1}$. We have also carried through a
complete next-to-leading order calculation, including the
next-to-leading order term in Eq.\ \eqref{derGAparincons} (in the
expansion in square brackets) and using the (complete)
next-to-leading order expressions for $g (\mu^2)$ and $m^2 (\mu^2)$,
which we have obtained before from an infinite number of iterations of
the corresponding system of differential equations at
next-to-next-to-leading order. As a result, many additional
next-to-leading order terms are generated for
$\Gamma^\parallel_{AA}(p^2, \mu^2)$, however, their respective
$p^2$-dependences are much weaker than the one of the next-to-leading
order term in Eq.\ \eqref{GAparincons}, in fact, they are at most as
strong as the one of the leading-order term in Eq.\ \eqref{GAparincons}.

We conclude that the
longitudinal part of the proper gluonic two-point function \emph{grows}
with increasing momentum, for sufficiently large values of $p$. It is
positive (as it should be), contrary to $m^2 (p^2)$ itself, but the
usual, not extended, BRST symmetry is not restored in the UV limit.
It can be deduced from Eq.\ \eqref{GAperpincons} that the longitudinal part
is suppressed with respect to the transverse part by only a factor of
$(\ln (p^2/\Lambda_{\text{UV}}^2))^{-2}$ in the large-$p$ limit.

For the rest of this subsection, we shall analyze the IR limit of the
two-point functions with very similar techniques. We will show some of the
details of the calculations for the case of the (non-critical and
scale-independent) derivative schemes, and at the end 
comment on the corresponding results for the other schemes. 

Let us begin with the ghost propagator. The calculation is particularly
simple in this case since we need the flow functions, 
{as we shall see, only
to leading order in the IR expansion in powers of $(\mu^2/m^2)$ (remember
that we have $\mu^2 \ll m^2$ in the IR regime even though the mass
parameter tends to zero with the renormalization scale because it does
so only logarithmically; see below for the precise formula). Explicitly,
from Eqs.\ \eqref{GAtir}--\eqref{Gcir} and the normalization conditions
\eqref{TWd3}, \eqref{TWd2} and \eqref{TWz1},
\bal
\gamma_A &= \frac{3 \zeta - 1}{12} \, 
\frac{N g^2}{(4\pi)^2} \,, \n \\
\gamma_c &= -\frac{1}{2} \frac{N g^2}{(4\pi)^2} \, \frac{\mu^2}{m^2}
\left( \ln \frac{m^2}{\mu^2} + \frac{1}{3} \right) \,, \n \\
\beta_{m^2} &= \frac{3 \zeta - 1}{12} \, 
\frac{N g^2}{(4\pi)^2} \, m^2 \,. \label{flowfir}
\eal
The anomalous dimension $\gamma_c$ is already of first order in
$(\mu^2/m^2)$, which is precisely what makes the calculation of the ghost
propagator simple. It also implies that the beta function of the
coupling constant is simply
\be
\beta_g = \frac{g}{2} \, \gamma_A = \frac{3 \zeta - 1}{24} \, 
\frac{N g^2}{(4\pi)^2} \, g 
\ee
to leading order in $(\mu^2/m^2)$. The integration of the corresponding 
differential equation for the coupling constant yields
\be
\frac{N g^2 (\mu^2)}{(4\pi)^2} =
\left( \frac{3 \zeta - 1}{12} \, \ln \frac{\Lambda_{\text{IR}}^2}{\mu^2} 
\right)^{-1} \,, \label{runninggir}
\ee
with a characteristic scale $\Lambda_{\text{IR}}$ for the IR regime. 
Its relation to the UV scale $\Lambda_{\text{UV}} 
= \Lambda_{\text{QCD}}$ can only be established by integrating the
renormalization group equations from the UV to the IR. Equation
\eqref{runninggir} is supposed to be valid only for renormalization scales
$\mu^2 \ll \Lambda_{\text{IR}}^2$, and the value of $\Lambda_{\text{IR}}$
is fixed through the same relation \eqref{runninggir} in terms of the value 
of the coupling constant at one such scale $\mu = \mu_0$. Of course 
we assume that $\zeta > 1/3$ (so that the beta function is
positive in the IR).

When we use the leading-order result for the coupling constant in the
beta function for the mass, we can integrate the corresponding differential 
equation and obtain
\be
m^2 (\mu^2) = m^2 (\mu_0^2) \, \frac{\ln \, (\Lambda_{\text{IR}}^2/\mu_0^2)}
{\ln \, (\Lambda_{\text{IR}}^2/\mu^2)} \,, \label{runningmir}
\ee
in terms of the value of $m^2$ at an arbitrary IR reference scale $\mu_0$.
As anticipated, the mass parameter vanishes logarithmically in the limit
$\mu \to 0$.

We use the leading-order formulas for both the coupling constant and the
mass parameter in the expression \eqref{flowfir} for $\gamma_c$ to
calculate
\be
\exp \left( -\int_{\mu^2}^{p^{\prime 2}} 
\frac{d \mu^{\prime \, 2}}{\mu^{\prime \, 2}} \, 
\gamma_c (\mu^{\prime \, 2}) \right) \label{expintgammacir}
\ee
to leading order. The $\mu^{\prime \, 2}$-integral over $\gamma_c$ can be 
evaluated exactly to the present order in terms of the exponential
integral function, but as before in the discussion of the UV regime we 
find it more intuitive to write the result in an expansion in powers of
$(\ln (\Lambda_{\text{IR}}^2/p^{\prime 2}))^{-1}$ and
$(\ln (\Lambda_{\text{IR}}^2/\mu^2))^{-1}$. For simplicity, we will
identify the scale $\mu$ appearing as the lower limit of the integral
with the reference scale $\mu_0$ in Eq.\ \eqref{runningmir}.

Since $\gamma_c$ and thus
the integral are already of first order in $(\mu^{\prime \, 2}/m^2)$,
the exponential in Eq.\ \eqref{expintgammacir} can be 
approximated simply by
\be
1 - \int_{\mu^2}^{p^{\prime 2}} 
\frac{d \mu^{\prime \, 2}}{\mu^{\prime \, 2}} \, 
\gamma_c (\mu^{\prime \, 2})
\ee
to the present order. Then the $p^{\prime 2}$-integration in Eq.\
\eqref{2pcRGz} can be performed, where we again use an expansion in
powers of $(\ln (\Lambda_{\text{IR}}^2/p^2))^{-1}$ instead of the
exponential integral function. The final result for the proper ghost
two-point function is
\bal
\Gamma_{c\bar{c}} (p^2, \mu^2) &= p^2 \Bigg\{ 1 + \frac{3}{3\zeta - 1} \,
\frac{p^2}{m^2 (\mu^2) \, \ln \, (\Lambda_{\text{IR}}^2/\mu^2)}
\bigg[ \ln \frac{m^2 (\mu^2)}{p^2} 
- \ln \bigg( \frac{\ln \, (\Lambda_{\text{IR}}^2/p^2)}
{\ln \, (\Lambda_{\text{IR}}^2/\mu^2)} \bigg) \n \\
&\hspace{7cm} {}+ \frac{11}{6} -
\frac{3/2}{\ln \, (\Lambda_{\text{IR}}^2/p^2)} \bigg] \n \\
&\phantom{=p^2\;} {}- \frac{6}{3\zeta - 1} \,
\frac{\mu^2}{m^2 (\mu^2) \, \ln \, (\Lambda_{\text{IR}}^2/\mu^2)}
\left[ \ln \frac{m^2 (\mu^2)}{\mu^2} + \frac{4}{3} 
- \frac{1}{\ln \, (\Lambda_{\text{IR}}^2/\mu^2)} \right] \Bigg\} \,.
\label{Gcbarcir}
\eal
The corrections to this expression are suppressed by at least one power of
$(\ln (\Lambda_{\text{IR}}^2/p^2))^{-1}$ or 
$(\ln (\Lambda_{\text{IR}}^2/\mu^2))^{-1}$.
Note the sign of the quantum correction in the limit $p^2 \to 0$, which
implies that the ghost dressing function decreases with increasing $p^2$
in this limit, in agreement with the numerical findings.

For a detailed comparison to the numerical solution, we express the
characteristic scale $\Lambda_{\text{IR}}$ via Eq.\ \eqref{runninggir}
through the value of the coupling constant at the reference scale 
(now) $\mu$. By use of the $\zeta$-dependent analogue 
of Eq.\ \eqref{compare1} for the infrared scale $\Lambda_{\text{IR}}$, 
\be
\left( \ln \frac{\Lambda_{\text{IR}}^2}{p^2} \right)^{-1}
= \frac{3 \zeta - 1}{12} \, \frac{N g^2 (\mu^2)}{(4\pi)^2}
\left( 1 + \frac{3 \zeta - 1}{12} \, \frac{N g^2 (\mu^2)}{(4\pi)^2}
\ln \frac{\mu^2}{p^2} \right)^{-1} \,, \label{compare1ir}
\ee
Eq.\ \eqref{Gcbarcir} can be written in the form
\bal
\Gamma_{c\bar{c}} (p^2, \mu^2) &= p^2 \Bigg\{ 1 + \frac{1}{4} \,
\frac{N g^2 (\mu^2)}{(4\pi)^2} \, \frac{p^2}{m^2 (\mu^2)}
\bigg[ \ln \frac{m^2 (\mu^2)}{p^2} - \ln \left( 1 + 
\frac{3 \zeta - 1}{12} \, \frac{N g^2 (\mu^2)}{(4\pi)^2}
\ln \frac{\mu^2}{p^2} \right) \n \\
&\hspace{2cm} {}+ \frac{11}{6} - \frac{3}{2} \, 
\frac{3 \zeta - 1}{12} \, \frac{N g^2 (\mu^2)}{(4\pi)^2}
\left( 1 + \frac{3 \zeta - 1}{12} \, \frac{N g^2 (\mu^2)}{(4\pi)^2}
\ln \frac{\mu^2}{p^2} \right)^{-1} \bigg] \n \\
&\phantom{p^2 \Bigg\{ 1 +} {}- \frac{1}{2} \, \frac{N g^2 (\mu^2)}{(4\pi)^2} \,
\frac{\mu^2}{m^2 (\mu^2)} \bigg[ \ln \frac{m^2 (\mu^2)}{\mu^2} 
  + \frac{4}{3} - \frac{3 \zeta - 1}{12} \,
  \frac{N g^2 (\mu^2)}{(4\pi)^2} \bigg] \Bigg\} \,. \label{Gcbarcirg}
\eal
This equation provides an excellent fit to the 
corresponding numerical curves, but only for momenta up to approximately
$0{.}05$ GeV (and a decent fit for $p$ up to roughly $0{.}1$ GeV), for
different values of the parameter $\zeta$.

We now come to the IR behavior of the proper gluonic two-point function.
We begin with its longitudinal part which we will have to determine to
next-to-leading order in powers of $(p^2/m^2)$ and $(\mu^2/m^2)$. To this
end, we need the flow functions to this order,
\bal
\gamma_A &= \frac{N g^2}{(4\pi)^2} \left[ \frac{3 \zeta - 1}{12} 
- \frac{45 \zeta + 389}{360} \, \frac{\mu^2}{m^2} \right] \,, \n \\
\beta_g &= \frac{N g^2}{(4\pi)^2} \, g \left[ \frac{3 \zeta - 1}{24}
- \frac{1}{2} \frac{\mu^2}{m^2} \left( \ln \frac{m^2}{\mu^2} 
+ \frac{45 \zeta + 509}{360} \right) \right] \,, \n \\
\beta_{m^2} &= \frac{N g^2}{(4\pi)^2} \, m^2 \left[ \frac{3 \zeta - 1}{12} 
- \frac{1}{4} \frac{\mu^2}{m^2} \left( \ln \frac{m^2}{\mu^2} 
+ \frac{45 \zeta + 464}{90} \right) \right] \,, \label{flowfirnlo}
\eal
where we have used the expression \eqref{flowfir} for $\gamma_c$.
After inserting the leading-order expression \eqref{runningmir} for the mass 
parameter in the formula for $\beta_g$, the differential equation for the
coupling constant can be integrated. The result to next-to-leading order, 
but restricted to the leading order in powers of
$(\ln (\Lambda_{\text{IR}}^2/\mu^2))^{-1}$ and
$(\ln (\Lambda_{\text{IR}}^2/\mu_0^2))^{-1}$, is
\bal
\frac{N g^2 (\mu^2)}{(4\pi)^2} &=
\frac{12}{3 \zeta - 1} \, \frac{1}{\ln \, (\Lambda_{\text{IR}}^2/\mu^2)}
\Bigg\{ 1 - \frac{12}{3\zeta - 1} \,
\frac{\mu^2}{m^2 (\mu_0^2) \, \ln \, (\Lambda_{\text{IR}}^2/\mu_0^2)}
\bigg[ \ln \frac{m^2 (\mu_0^2)}{\mu^2} \n \\
&\hspace{6.3cm} {}- \ln \bigg( \frac{\ln \, (\Lambda_{\text{IR}}^2/\mu^2)}
  {\ln \, (\Lambda_{\text{IR}}^2/\mu_0^2)} \bigg)
  + \frac{45 \zeta + 1129}{360} \bigg] \n \\
&\hspace{2.8cm} {}+ \frac{12}{3\zeta - 1} \,
\frac{\mu_0^2}{m^2 (\mu_0^2) \, \ln \, (\Lambda_{\text{IR}}^2/\mu^2)}
\left[ \ln \frac{m^2 (\mu_0^2)}{\mu_0^2} + \frac{45 \zeta + 1129}{360} 
 \right] \Bigg\} \,. \label{rungirnlo}
\eal
At the reference scale $\mu_0$, the characteristic scale 
$\Lambda_{\text{IR}}$ is related to the coupling constant via Eq.\
\eqref{runninggir}. For the results to be presented below, it is actually
necessary to explicitly determine all terms up to the next-to-leading order 
in powers of $(\ln (\Lambda_{\text{IR}}^2/\mu^2))^{-1}$ and
$(\ln (\Lambda_{\text{IR}}^2/\mu_0^2))^{-1}$ [at next-to-leading order
in $(\mu^2/m^2)$ and $(\mu_0^2/m^2)$]. These terms are not displayed in Eq.\
  \eqref{rungirnlo}.
  
In the formula for $\beta_{m^2}$, the complete next-to-leading order 
expression is inserted for $g (\mu)$, and $m^2 (\mu^2)$ inside the square
bracket is replaced with Eq.\ \eqref{runningmir}. The differential
equation for the mass parameter can then be integrated to next-to-leading
order. The result is lengthy and will not be shown here. In a similar
fashion, the next-to-leading order formula for $g (\mu)$ and the
leading-order expression for $m^2 (\mu^2)$ are substituted in the formula
for $\gamma_A$ in Eq.\ \eqref{flowfirnlo}, and the integral
\be
\int_{\mu^2}^{p^2} \frac{d \mu^{\prime \, 2}}{\mu^{\prime \, 2}} \, 
\gamma_A (\mu^{\prime \, 2}) \label{intgamAnor}
\ee
is determined to next-to-leading order. Here we will only present the
leading-order result for the exponential,
\be
\exp \left( -\int_{\mu^2}^{p^2} \frac{d \mu^{\prime \, 2}}{\mu^{\prime \, 2}} 
\, \gamma_A (\mu^{\prime \, 2}) \right)
= \frac{\ln \, (\Lambda_{\text{IR}}^2/p^2)}
{\ln \, (\Lambda_{\text{IR}}^2/\mu^2)} \,, \label{intgamAirLO}
\ee
which is valid in all derivative schemes with $\zeta > 1/3$,
including the scale-dependent scheme, and also in
Tissier-Wschebor's original IR safe renormalization scheme. It is an
important result because, together with Eqs.\ \eqref{2pAparRG} and 
\eqref{runningmir}, it shows that $\Gamma_{AA}^\parallel (p^2, \mu^2)$
tends towards a nonzero constant in the limit $p^2 \to 0$, despite the
vanishing of $m^2 (p^2)$ in this limit. Locality implies that the same
should be true for $\Gamma_{AA}^\perp (p^2, \mu^2)$, which is indeed the
case as can be confirmed from Eq.\ \eqref{2pARG} for the Tissier-Wschebor 
scheme and Eq.\ \eqref{2pAperpRGz} for the derivative schemes.

To next-to-leading order, the result for the integral 
\eqref{intgamAnor} to this order, together with the 
next-to-leading order formula for the 
mass parameter, leads via Eq.\ \eqref{2pAparRG} to the following relatively 
compact expression for the longitudinal part of the proper gluonic two-point 
function to next-to-leading order in the derivative schemes:
\bal
\Gamma_{AA}^\parallel (p^2, \mu^2)
&= m^2 (\mu^2) - \frac{3}{3\zeta - 1} \,
\frac{p^2}{\ln \, (\Lambda_{\text{IR}}^2/\mu^2)}
\left[ \ln \frac{m^2 (\mu^2)}{p^2} -
\ln \left( \frac{\ln \, (\Lambda_{\text{IR}}^2/p^2)}
{\ln \, (\Lambda_{\text{IR}}^2/\mu^2)} \right) \right. \n \\
&\hspace{5.8cm} \left. {}+ \frac{11}{6}
- \frac{1}{\ln \, (\Lambda_{\text{IR}}^2/p^2)} \right] \n \\
&\phantom{=\;} {}+ \frac{3}{3\zeta - 1} \,
\frac{\mu^2}{\ln \, (\Lambda_{\text{IR}}^2/\mu^2)}
\left[ \ln \frac{m^2 (\mu^2)}{\mu^2} + \frac{11}{6}
- \frac{1}{\ln \, (\Lambda_{\text{IR}}^2/\mu^2)} \right] \,, \label{GAAparir}
\eal
with corrections suppressed by relative factors of
$(\ln (\Lambda_{\text{IR}}^2/p^2))^{-1}$ or 
$(\ln (\Lambda_{\text{IR}}^2/\mu^2))^{-1}$. The scale $\mu$ has been
identified with the reference scale $\mu_0$ for simplicity.
For future reference, we will also write Eq.\ \eqref{GAAparir} in 
terms of the coupling constant $g (\mu)$ instead of the scale 
$\Lambda_{\text{IR}}$. The result is
\bal
\Gamma_{AA}^\parallel (p^2, \mu^2) &= m^2 (\mu^2) - \frac{1}{4} \,
\frac{N g^2 (\mu^2)}{(4\pi)^2} \, p^2
\bigg[ \ln \frac{m^2 (\mu^2)}{p^2} - \ln \left( 1 + 
\frac{3 \zeta - 1}{12} \, \frac{N g^2 (\mu^2)}{(4\pi)^2}
\ln \frac{\mu^2}{p^2} \right) \n \\
&\hspace{3cm} {}+ \frac{11}{6} - 
\frac{3 \zeta - 1}{12} \, \frac{N g^2 (\mu^2)}{(4\pi)^2}
\left( 1 + \frac{3 \zeta - 1}{12} \, \frac{N g^2 (\mu^2)}{(4\pi)^2}
\ln \frac{\mu^2}{p^2} \right)^{-1} \bigg] \n \\
&\phantom{p^2 \Bigg\{ 1 +} {}+ \frac{1}{4} \, \frac{N g^2 (\mu^2)}{(4\pi)^2} 
\, \mu^2 \bigg[ \ln \frac{m^2 (\mu^2)}{\mu^2} 
  + \frac{11}{6} - \frac{3 \zeta - 1}{12} \, \frac{N g^2 (\mu^2)}{(4\pi)^2}
  \bigg] \,. \label{GAAparirg}
\eal

We shall postpone the discussion of the Slavnov-Taylor identity and turn
to the determination of the transverse part of the gluonic two-point 
function in the IR limit first. It is given by Eq.\ \eqref{2pAperpRGz},
for which we use the result \eqref{GAAparir} for the longitudinal 
part and
\be
\int_0^{p^2} dp^{\prime \, 2} \, \exp \left( 
-\int_{\mu^2}^{p^{\prime 2}} \frac{d \mu^{\prime \, 2}}{\mu^{\prime \, 2}} \, 
\gamma_A (\mu^{\prime \, 2}) \right) 
= p^2 + \frac{p^2}{\ln \, (\Lambda_{\text{IR}}^2/\mu^2)}
\left( \ln \frac{m^2 (\mu^2)}{p^2} - \ln \frac{m^2 (\mu^2)}{\mu^2}
+ 1 \right) \,, \label{intgamAir}
\ee
which is only needed to leading order in the expansion in powers of 
$(p^2/m^2)$. Then we get for the transverse part
\bal
\Gamma_{AA}^\perp (p^2, \mu^2)
&= m^2 (\mu^2) + p^2 + \frac{3}{3\zeta - 1} \,
\frac{\mu^2}{\ln \, (\Lambda_{\text{IR}}^2/\mu^2)}
\left[ \ln \frac{m^2 (\mu^2)}{\mu^2} + \frac{11}{6}
- \frac{1}{\ln \, (\Lambda_{\text{IR}}^2/\mu^2)} \right] \n \\
&\phantom{= m^2} {}- \frac{1}{3\zeta - 1} \,
\frac{p^2}{\ln \, (\Lambda_{\text{IR}}^2/\mu^2)}
\left[ \ln \frac{m^2 (\mu^2)}{p^2} 
+ (3 \zeta - 1) \ln \frac{m^2 (\mu^2)}{\mu^2} \right. \n \\
&\hspace{4cm} \left. {}-3 \zeta \ln
\left( \frac{\ln \, (\Lambda_{\text{IR}}^2/p^2)}
{\ln \, (\Lambda_{\text{IR}}^2/\mu^2)} \right) + \frac{5 \zeta + 2}{2}
- \frac{3 \zeta}{\ln \, (\Lambda_{\text{IR}}^2/p^2)} \right] \,. 
\label{GAAperpir}
\eal
Again, corrections to this formula are suppressed by relative 
factors of $(\ln (\Lambda_{\text{IR}}^2/p^2))^{-1}$ or 
$(\ln (\Lambda_{\text{IR}}^2/\mu^2))^{-1}$, and the scale $\mu$ has been
put equal to the reference scale $\mu_0$.

The most important feature of Eq.\ \eqref{GAAperpir} is the \emph{decrease}
of $\Gamma_{AA}^\perp (p^2, \mu^2)$ with increasing $p$ for the
smallest momenta, for any value $\zeta > 1/3$. This corresponds to the 
increase of the gluon propagator function in the extreme IR regime, or its 
decrease as the momentum approaches zero, which was observed in the
numerical solutions of the Callan-Symanizk equations in the previous
subsection and related to the violation of positivity.
It is now clear that this behavior persists for all values
of $\zeta$ (greater than $1/3$, such that there is no Landau pole; 
the same behavior is found analytically in the critical and 
scale-dependent derivative schemes and also in the original
Tissier-Wschebor scheme, see below). Note that the term that dominates 
the momentum behavior of the gluon propagator in the extreme IR is 
contained in the next-to-leading order term in the expansion 
\eqref{GAAparir} of the longitudinal part of the proper gluonic two-point 
function, while the corresponding contribution from Eq.\ 
\eqref{intgamAir} has the opposite sign.

For comparison to the numerical solutions, we will express the
characteristic scale $\Lambda_{\text{IR}}$
through the value of the coupling constant at the reference scale 
$\mu$. Then Eq.\ \eqref{GAAperpir} takes the form
\bal
\Gamma_{AA}^\perp (p^2, \mu^2)
&= m^2 (\mu^2) + p^2 + \frac{1}{4} \frac{N g^2 (\mu^2)}{(4\pi)^2} \, \mu^2
\left[ \ln \frac{m^2 (\mu^2)}{\mu^2} + \frac{11}{6}
- \frac{3 \zeta - 1}{12} \, \frac{N g^2 (\mu^2)}{(4\pi)^2} \right] \n \\
&\phantom{= m^2} {}- \frac{1}{12} \frac{N g^2 (\mu^2)}{(4\pi)^2} \,
p^2 \left[ \ln \frac{m^2 (\mu^2)}{p^2}
  + (3 \zeta - 1) \ln \frac{m^2 (\mu^2)}{\mu^2} \right. \n \\
  &\hspace{3.6cm} {}- 3 \zeta \ln \left( 1 + \frac{3 \zeta - 1}{12} \,
  \frac{N g^2 (\mu^2)}{(4\pi)^2} \ln \frac{\mu^2}{p^2} \right) 
+ \frac{5 \zeta + 2}{2} \n \\
&\phantom{= -\frac{1}{12} \frac{N g^2 (\mu^2)}{(4\pi)^2}} \left.
        {}- 3 \zeta \, \frac{3 \zeta - 1}{12} \, \frac{N g^2 (\mu^2)}{(4\pi)^2}
\left( 1 + \frac{3 \zeta - 1}{12} \, \frac{N g^2 (\mu^2)}{(4\pi)^2}
\ln \frac{\mu^2}{p^2} \right)^{-1} \right] \,. \label{GAAperpirg}
\eal
Similarly to the case of the proper ghost two-point function 
before, Eq.\ \eqref{GAAperpir} provides an excellent approximation 
to the numerical solutions, but only for very small momenta up to 
approximately $p = 0{.}05$ GeV. 
Incidentally, as already mentioned, the expansion in
powers of $(\ln (\Lambda_{\text{IR}}^2/p^2))^{-1}$ and 
$(\ln (\Lambda_{\text{IR}}^2/\mu^2))^{-1}$ in Eq.\ \eqref{GAAperpir} can be
avoided and the complete next-to-leading order contribution in 
the expansion in powers of $(p^2/m^2)$ and $(\mu^2/m^2)$ be 
expressed in terms of the exponential integral
function (the corresponding calculation is, actually, not more
complicated), but the result is essentially the same as far as the
comparison to the numerical solutions is concerned.

We have remarked in the previous subsection that the decrease of the
gluon propagator function as the momentum tends to zero is more
pronounced for larger values of $\zeta$. This tendency is clearly seen
in the numerical solutions, but it is not visible in Eqs.\ 
\eqref{GAAperpir} or \eqref{GAAperpirg}. As it turns out, the extreme IR
behavior of the gluon propagator as parameterized by the latter of these 
equations depends very sensibly on the value of the coupling constant
at the IR reference scale $\mu$, and this value is obtained through the
numerical integration of the renormalization group equations from the
UV to the IR regime and our fitting procedure, for every fixed value
of the parameter $\zeta$. It is hence impossible to determine the
$\zeta$-dependence of $g (\mu)$ analytically.
In addition, the higher-order terms in the expansion in powers of $(p^2/m^2)$
are expected to be important for the description of the decrease of the
gluon propagator towards small momenta, given that the 
decrease of the numerical solutions extends to momenta far 
beyond $0{.}05$ GeV.

Turning now to the discussion of the Slavnov-Taylor identity 
\eqref{STIcomb}, from Eq.\ \eqref{Gcbarcir}
together with Eq.\ \eqref{GAAparir} one may determine the next-to-leading 
order expression in the IR regime for the quantity $C (p^2)$ defined in Eq.\ 
\eqref{defC}. The result is
\bal
C (p^2) &= \frac{\Gamma_{AA}^\parallel (p^2, \mu^2) \, 
\Gamma_{c\bar{c}} (p^2, \mu^2)}{p^2 \, m^2 (\mu^2)} \n \\
&= 1 - \frac{3}{3\zeta - 1} \, 
\frac{\mu^2}{m^2 (\mu^2) \, \ln \, (\Lambda_{\text{IR}}^2/\mu^2)}
\left[ \ln \frac{m^2 (\mu^2)}{\mu^2} + \frac{5}{6} 
- \frac{1}{\ln \, (\Lambda_{\text{IR}}^2/\mu^2)} \right] \n \\
&\phantom{= 1 -} {}- \frac{3/2}{3\zeta - 1} \, 
\frac{p^2}{m^2 (\mu^2) \, \ln \, (\Lambda_{\text{IR}}^2/\mu^2)}
\, \frac{1}{\ln \, (\Lambda_{\text{IR}}^2/p^2)} \,. \label{BRSTviolir}
\eal
As far as the $p^2$-dependence is concerned, in the next-to-leading
order term in powers of $(p^2/m^2)$, both the contributions of the order
$(p^2/m^2) (\ln (\Lambda_{\text{IR}}^2/p^2) )^1$ [note that
$\ln (m^2/p^2) = \ln (\Lambda_{\text{IR}}^2/p^2) - 
\ln (\Lambda_{\text{IR}}^2/m^2)$] and
$(p^2/m^2) (\ln (\Lambda_{\text{IR}}^2/p^2) )^0$ have cancelled, and the
dominant contribution [in the expansion in powers of
$(\ln (\Lambda_{\text{IR}}^2/p^2) )^{-1}$] is of the order of
$(p^2/m^2) (\ln (\Lambda_{\text{IR}}^2/p^2) )^{-1}$. Hence the Slavnov-Taylor
identity \eqref{STIcomb} is violated in the IR, but only relatively
weakly.

Equation \eqref{BRSTviolir} predicts that the combination $C (p^2)$
decreases with increasing momentum $p$ in the extreme IR (for any
$\zeta > 1/3$), a behavior that is not visible in the numerical curve
for $\zeta = 1$ in Fig.\ \ref{figSTI}. A very precise numerical
evaluation in the extreme IR regime reveals that the curve actually
does very slightly decrease for momenta close to zero. Formula
\eqref{BRSTviolir} provides an accurate description, but only for momenta
up to, approximately, $p = 0{.}01$ GeV. For only slightly larger momenta, 
higher-order terms in the expansion in powers of $(p^2/m^2)$ dominate and 
lead to a strong deviation from the prediction \eqref{BRSTviolir}. A
more precise next-to-leading order evaluation of $C (p^2)$ that avoids
the expansion in powers of $(\ln (\Lambda_{\text{IR}}^2/p^2) )^{-1}$ 
[and $(\ln (\Lambda_{\text{IR}}^2/\mu^2) )^{-1}$] gives an even
(slightly) better fit to the numerical curve, but does not appreciably extend 
the range of momenta where the analytical prediction coincides with the
numerical solution. Of course, the reason for the very limited validity
of formula \eqref{BRSTviolir} is the weak momentum dependence it predicts,
so that higher-order corrections already dominate for comparatively
small momenta.

We shall now briefly comment on the other renormalization schemes, starting
with Tissier-Wschebor's original IR safe scheme. The flow functions in this
scheme coincide with the ones of the simple derivative scheme
with $\zeta = 1$ to leading order in the expansion in powers of
$(\mu^2/m^2)$, but differ from the latter to next-to-leading order.
In particular, the dominant contribution to $\gamma_c$ which is already
of the order of $(\mu^2/m^2)^1$, is different from the one in Eq.\
\eqref{flowfir}. As a consequence, the result for the coupling
constant, for example, while being similar to Eq.\ \eqref{rungirnlo} with
$\zeta = 1$, differs from it in the values of the coefficients in the
  next-to-leading order contributions.
Nevertheless, the complete next-to-leading order result for the
longitudinal part of the proper gluonic two-point function is identical
to Eq.\ \eqref{GAAparir} with $\zeta = 1$. On the other hand, the result
for the ghost two-point function differs from Eq.\ \eqref{Gcbarcir}
(for $\zeta = 1$) in just such a way that the normalized combination
$C (p^2)$ [see Eq.\ \eqref{BRSTviolir}] is equal to one. The latter equality 
holds to all orders in the Tissier-Wschebor scheme, as we have shown before.
The transverse part of the proper gluonic two-point function is, again,
surprisingly similar to its counterpart \eqref{GAAperpir} for $\zeta = 1$,
only the constant $(5 \zeta + 2)/2$ in the next-to-leading order
contributions (in square brackets) is replaced by $11/2$.

As for the scale-dependent derivative scheme, the expressions for the flow
functions in this scheme differ substantially from the ones of the $\zeta = 1$
derivative scheme to next-to-leading order, and the same is true for
the coupling constant and the mass parameter. Nevertheless, the final complete
next-to-leading order results for the longitudinal and the transverse
part of the proper gluonic two-point function and for the
ghost two-point function are identical to the corresponding results in the
$\zeta = 1$ derivative scheme, to which the scale-dependent scheme reduces in
the limit $\mu \to 0$.

In the critical derivative scheme, the flow functions are given by Eq.\
\eqref{flowfirnlo}, substituting $\zeta = 1/3$. Since this substitution
eliminates the leading-order contributions in these flow functions, in the
critical scheme one has
\be
g (\mu^2) = g (\mu_0^2) \quad \text{and} \quad m^2 (\mu^2) = m^2 (\mu_0^2)
\ee
to leading order, which makes the iterative solution of the Callan-Symanzik
equations much simpler in this scheme. In particular, the choice $\mu_0^2 = 0$
for the reference scale is possible here. 

Proceeding in the same (iterative) way as for 
all the other schemes, the integration of the differential equation for
the coupling constant with $\beta_g$ from Eq.\ \eqref{flowfirnlo} 
gives, with the help of the leading-order approximation
$m^2 (\mu^2) = m^2 (\mu_0^2)$,
\bal
\frac{N g^2 (\mu^2)}{(4\pi)^2} &= \frac{N g^2 (\mu_0^2)}{(4\pi)^2}
\Bigg\{ 1 - \frac{N g^2 (\mu_0^2)}{(4\pi)^2} \, \frac{\mu^2}{m^2 (\mu_0^2)}
\left[ \ln \frac{m^2 (\mu_0^2)}{\mu^2} + \frac{221}{90} \right] \n \\
&\phantom{\frac{N g^2 (\mu_0^2)}{(4\pi)^2} \Bigg\{ 1 -}
{}+ \frac{N g^2 (\mu_0^2)}{(4\pi)^2} \, \frac{\mu_0^2}{m^2 (\mu_0^2)}
\left[ \ln \frac{m^2 (\mu_0^2)}{\mu_0^2} + \frac{221}{90} \right]
\Bigg\} \,. \label{rungircrit}
\eal
This is the complete result to next-to-leading order in $(\mu^2/m^2)$
and $(\mu_0^2/m^2)$, no expansion in inverse powers of logarithms is
necessary here. Equation \eqref{rungircrit} is certainly very different
in appearance from Eq.\ \eqref{rungirnlo} in the general derivative
scheme, basically because there is no characteristic infrared scale 
$\Lambda_{\text{IR}}$ in the critical derivative scheme, in particular,
it is unclear how to define the limit $\zeta \to 1/3$ in the form
\eqref{rungirnlo}.

However, the result \eqref{rungirnlo} can be rewritten by eliminating
the scale $\Lambda_{\text{IR}}$ in favor of the value $g (\mu_0)$ of
the coupling constant at the reference scale, which leads to the 
next-to-leading order expression
\bal
\lefteqn{\frac{N g^2 (\mu^2)}{(4\pi)^2} = \frac{N g^2 (\mu_0^2)}{(4\pi)^2}
\left( 1 + \frac{3 \zeta - 1}{12} \, \frac{N g^2 (\mu_0^2)}{(4\pi)^2}
\ln \frac{\mu_0^2}{\mu^2} \right)^{-1}} \hspace{1cm} \n \\
&{}\times \Bigg\{ 1 - \frac{N g^2 (\mu_0^2)}{(4\pi)^2} \,
\frac{\mu^2}{m^2 (\mu_0^2)} \left[ \ln \frac{m^2 (\mu_0^2)}{\mu^2}
- \ln \left( 1 + \frac{3 \zeta - 1}{12} \, \frac{N g^2 (\mu_0^2)}{(4\pi)^2}
\ln \frac{\mu_0^2}{\mu^2} \right) + \frac{45 \zeta + 1229}{360} \right] \n \\
&\phantom{{}\times \Bigg\{}
{}+ \frac{N g^2 (\mu_0^2)}{(4\pi)^2} \, \frac{\mu_0^2}{m^2 (\mu_0^2)}
\left( 1 + \frac{3 \zeta - 1}{12} \, \frac{N g^2 (\mu_0^2)}{(4\pi)^2}
\ln \frac{\mu_0^2}{\mu^2} \right)^{-1} \left[ 
\ln \frac{m^2 (\mu_0^2)}{\mu_0^2} + \frac{45 \zeta + 1229}{360} \right]
\Bigg\}
\eal
for the running coupling constant in the general derivative scheme. In this
form, the limit $\zeta \to 1/3$ can be taken and leads to Eq.\
\eqref{rungircrit} with the exception of the constant
$(45 \zeta + 1229)/360$, the limit of which differs by one from
$221/90$. Note that the complete result for the running coupling constant
in the general derivative schemes to next-to-leading order in 
$(\mu^2/m^2)$ and $(\mu_0^2/m^2)$ contains a series in powers of
$(\ln (\Lambda_{\text{IR}}^2/\mu^2))^{-1}$ and
$(\ln (\Lambda_{\text{IR}}^2/\mu_0^2))^{-1}$, and we have only written
down the leading term of this series in Eq.\ \eqref{rungirnlo}. However,
as a consequence of Eqs.\ \eqref{runninggir} and \eqref{compare1ir}
(replacing $\mu^2$ with $\mu_0^2$ and $p^2$ with $\mu^2$ there),
all the other terms of the series vanish in the limit $\zeta \to 1/3$.

The results for the proper ghost two-point function and the longitudinal
and the transverse part of the proper gluonic two-point function
in the critical derivative scheme are precisely the limits $\zeta \to 1/3$
of Eqs.\ \eqref{Gcbarcirg}, \eqref{GAAparirg} and \eqref{GAAperpirg} above.
As a consequence, the same is true for Eq.\ \eqref{BRSTviolir} when rewritten
in terms of $g (\mu)$. In view
of Eq.\ \eqref{compare1ir}, the $p^2$-dependent term in Eq.\ \eqref{BRSTviolir}
vanishes in the critical derivative scheme, and the same goes for all
higher terms in the expansion in powers of
$(\ln (\Lambda_{\text{IR}}^2/p^2))^{-1}$. In particular, the Slavnov-Taylor
identity \eqref{STIcomb} is \emph{not} violated in the critical scheme to
next-to-leading order in $(\mu^2/m^2)$ and $(p^2/m^2)$. However, beyond
the extreme IR limit [to the present order in $(\mu^2/m^2)$ and $(p^2/m^2)$]
the identity is obviously not fulfilled, as is evident from Eq.\
\eqref{STIcombUV} and the numerical results represented in Fig.\
\ref{figSTI}.

\section{Conclusions}

In this paper, we have considered a variety of renormalization schemes for
four-dimensional massive Yang-Mills theory in the Landau gauge, comparing 
the resulting one-loop renormalization group improved ghost and gluon 
propagators to the data produced by gauge fixed lattice calculations. Our aim 
was to identify a formulation of the theory in the continuum (including the
renormalization scheme) that would, after the necessary renormalization 
group improvement, give the best fit to the completely nonperturbative 
results in the minimal Landau gauge on the lattice. In particular, we have
explored how much of the effect of restricting the functional integral 
to the first Gribov region could be captured by 
merely adding a gluon mass term to the standard Faddeev-Popov action.

The main result is that surprisingly accurate (almost perfect) fits to
the ghost and gluon propagators and their dressing functions as measured on
the lattice can be achieved in certain renormalization schemes after adjusting
the two parameters of the theory, the values of the coupling constant and
the mass parameter at an arbitrary fixed scale. We are aware of the
fact that, physically speaking, the two parameters are related; however,
by construction their relation cannot be determined in our effective 
approach. We stress that the only input we use to generate our results
for the propagators are the perturbative one-loop 
corrections to the propagators and, for some of the renormalization schemes, 
to the ghost-gluon vertex (for the calculation of the flow functions in the
renormalization group equations) and, 
of course, the definition of the respective renormalization 
scheme and the two parameters as initial conditions for the integration of 
the differential equations.

All the renormalization schemes we have considered are defined through 
normalization conditions for the propagators and the ghost-gluon vertex. 
In addition to the original Tissier-Wschebor scheme \cite{TW011}, 
which we have reformulated in a more
standard fashion, we have introduced a family of derivative schemes (where
the normalization conditions are imposed on the derivatives of the proper
two-point functions with respect to momentum) which are parameterized by
the value of $\zeta$, a constant restricted to the range $\zeta \ge 1/3$
to ensure the IR safety of the schemes. 
The ``critical'' value $\zeta = 1/3$
corresponds to a particularly interesting case, the only case among the
schemes we have considered where the running coupling constant does not tend 
to zero in the extreme IR limit. Finally, we have also looked at a scheme 
where the parameter $\zeta$ varies with the renormalization scale.
In all renormalization schemes, we have found that the gluon propagator
functions, as well as the ghost dressing functions, tend to finite values at 
zero momentum (and thus correspond to the decoupling solutions of the 
Dyson-Schwinger equations).

It is the two derivative schemes with the critical value $\zeta = 1/3$ and
with a scale-dependent $\zeta$-parameter that give the best fits to the
lattice data \cite{CM08a, CM08b} for the propagators 
and their dressing functions (see Figs.\ 
\ref{figghosttw} and \ref{figgluoncrit}--\ref{figgluondressdyn})
when, at the same time, the coupling constant is defined from the ghost-gluon 
vertex in the Taylor limit where the momentum of the incoming ghost is set 
to zero (this limit has the additional virtue of being particularly simple, 
because the loop corrections to the vertex vanish in this limit). 
One could, furthermore, consider different forms of the scale dependence 
of the $\zeta$-parameter to try to obtain an even better fit to 
the lattice propagators, but here we have restricted our attention to one
very simple form of the scale dependence which appeared particularly natural
to us.

On the more theoretical side, we have analyzed the necessary ingredients for
a renormalization scheme to be IR safe (the coupling constant does not run
into a Landau pole) and UV consistent (the nilpotent BRST symmetry is 
restored in the UV limit). 
We have found it necessary to involve the longitudinal
part $\Gamma_{AA}^\parallel (p^2)$ of the proper gluonic two-point
function in the normalization conditions both to obtain a positive
beta function for the coupling constant in the IR limit and to recover
the nilpotent BRST symmetry in the UV limit [with respect
to the latter, see the discussion that includes Eqs.\ 
\eqref{dZAUVtrouble}--\eqref{TWd2} and the tedious iterative solution 
of the renormalization group equations for this case in Subsection 
\ref{analytical}]. Given that, in the Landau gauge, all 
connected n-point functions as well as
the transverse parts of all proper n-point functions are determined
in terms of only the transverse part of the proper gluonic two-point
function and the transverse parts of the proper vertices, it is quite
surprising that the longitudinal part $\Gamma_{AA}^\parallel (p^2)$ should
play such an important role in the renormalization group improvement.
Note, however, that the appearance of a longitudinal part in the term in the
``classical'' action \eqref{CF} that is quadratic in the gluon field
becomes inevitable as soon as one introduces a mass parameter in the 
transverse part of this term, in order to satisfy the requirement of locality
which is one of the backbones of the renormalization program.

There are several properties that hold in all consistent renormalization 
schemes: in the UV limit, the mass parameter goes to zero like
\be
m^2 (\mu^2) \propto \big( \ln(\mu^2/\Lambda_{\text{QCD}}^2) \big)^{-35/44}
\ee
and the longitudinal part of the proper gluonic two-point function behaves
like
\be
\Gamma_{AA}^\parallel (p^2, \mu^2) = m^2 (\mu^2) \left( 
\frac{\ln(\mu^2/\Lambda_{\text{QCD}}^2)}{\ln(p^2/\Lambda_{\text{QCD}}^2)}
\right)^{9/44}
\ee
(note that we have used the notation $\Lambda_{\text{UV}}$ for
$\Lambda_{\text{QCD}}$ in Subsection \ref{analytical}, to distinguish it 
from an analogous scale for the IR regime of the theory).
The dominant UV behavior of the transverse part, on the other hand,
is the same as for the
usual Faddeev-Popov action and slightly varies with the renormalization 
scheme, while the presence of a gluon mass term is a
subdominant effect compared to
the renormalization scheme dependence. As for the IR behavior, we have
found that the gluon propagator function increases with momentum in the
extreme IR region in all renormalization schemes (the increase being more
pronounced for larger values of $\zeta$ in the derivative schemes),
whereas this increase is not clearly visible in the lattice data.
The non-monotonous momentum dependence of the gluon propagator
  resulting from the increase already implies the violation of positivity.

Another interesting result of the present analysis is the fact that 
the renormalization schemes that lead to the best agreement with the lattice 
data, belong to a class of schemes for which the Slavnov-Taylor identity 
connecting the longitudinal part of the proper gluonic two-point 
function with the ghost two-point function, which was a key element in 
identifying an IR safe scheme in the first place \cite{TW011}, gets violated 
by the renormalization group improvement. This fact confirms our 
understanding that the extended BRST symmetry involving the gluon mass 
parameter (and lacking the nilpotency property), from which this 
Slavnov-Taylor identity is derived, is not a fundamental symmetry of 
quantized Yang-Mills theory. The Faddeev-Popov action with the massive 
extension has to be considered as an effective theory that is capable 
of producing accurate quantitative results for the gluon and ghost
propagators.

The general approach to quantum Yang-Mills theory that we have been
following here may be described as follows: generalize the Faddeev-Popov
action to an action that violates the common (nilpotent) BRST 
symmetry --- at present, we have only added a gluon mass term which, as a 
properly relevant parameter in the sense of the renormalization group, is 
naturally expected to have the greatest impact ---, calculate the perturbative 
loop corrections to the order desired, derive the corresponding flow functions
and use the Callan-Symanzik equations for renormalization group improvement.
Make sure, if necessary via fine-tuning, that the coupling constant does not
run into a Landau pole (IR safety) and that BRST symmetry in the usual sense
is recovered in the UV limit, so that the well-known (renormalization
group improved) perturbative UV behavior is reproduced.

The present approach has important advantages over other nonperturbative
continuum approaches like Dyson-Schwinger equations or the functional
renormalization group: it is comparatively simple and straightforward
(although the evaluation of the radiative corrections at higher loop orders
can be technically demanding), it is completely systematical once a 
renormalization scheme is chosen, and all predictions, in particular
\emph{all} n-point functions, are entirely determined by just fixing 
the two parameters of the theory --- at least in our present version
(note that the calculation of the vertex functions does not introduce
  any additional parameters in this approach). This should be compared to the
  enormous effort required to arrive at a self-consistent determination of the
  Yang-Mills propagators and (some of the) vertex functions from first
  principles in the functional formalisms, which was achieved in Ref.\
  \cite{CFM16} for the functional renormalization group and (very recently)
  in Ref.\ \cite{Hub20} for the Dyson-Schwinger equations (and the equations
  of motion for higher n-particle irreducible effective actions). We remark
  that the solutions of the decoupling type that are usually refered to as
  ``parameter-free'' in the functional approaches, actually do depend on two
  parameters (where one of the parameters just fixes the physical scale).

The fact that the input to the flow functions in the Callan-Symanzik equations 
is perturbative comes with additional benefits: at least in the
IR and UV limits, the behavior of the renormalization group improved
propagators or, in general, of the n-point functions, can be determined
analytically; furthermore, 
it is possible to use dimensional regularization which is,
actually, technically the simplest choice. For the renormalization of the
Dyson-Schwinger equations and the formulation of the functional renormalization
group, on the other hand, one will typically have to introduce a (UV or IR) 
momentum cutoff that by itself necessarily breaks manifest (nilpotent) BRST 
invariance. As a consequence, a gluon mass counterterm has to be
  introduced in order to restore the usual BRST invariance in the UV regime.
  It can be numerically very subtle to determine the counterterm to such
  precision that it does not alter
  the correct IR behavior, a difficulty known as the problem of quadratic
  divergencies in the case of the Dyson-Schwinger equations. It has only very
  recently (apparently) been solved \cite{Hub20}.

\begin{acknowledgments}
The authors would like to thank 
David Dudal, Jos\'e Rodr\'iguez-Quintero, Lorenz von Smekal,
Christian Fischer, M. Q. Huber, A. K. Cyrol and Jan M. Pawlowski for discussions during the various
stages of this work.
They also thank Attilio Cucchieri and Tereza Mendes for
allowing them to use their lattice data.
A.W. is grateful to the Institute for Theoretical Physics at the
University of Heidelberg for the warm hospitality extended to him
on a sabbatical stay during which the present work was completed.
Support by Conacyt project no.\ CB-2013/222812 and CIC-UMSNH is gratefully 
acknowledged.
P.D. is supported by Direcci\'on General de Asuntos del Personal Acad\'emico (DGAPA)
grant – Universidad Nacional Autónoma de M\'exico (UNAM).
\end{acknowledgments}

\appendix

\section{Transversality of the gluon propagator \label{transverse}}
We show here how, even in the presence of a gluon mass term that
breaks the usual nilpotent BRST symmetry, the gluon propagator remains
transverse in the Landau gauge. This is easily demonstrated 
using the Dyson-Schwinger equation for the auxiliary field. 

Adding a term to the action \eqref{CF} with sources
coupled to the fields and their nonlinear nilpotent transformations,
\be
S_{\text{sources}}=\int d^D x \left[ J^a_\mu A^a_\mu+ \bar{\eta}^a c^a
  +\bar{c}^a\eta^a+R^aB^a +K^a_\mu(s A_\mu)^a+L^a(s c)^a\right] \,,
\ee
the generating functional $Z$ is defined as the functional
integral
\be
Z[J_\mu,\eta, \bar{\eta},R,K_\mu,L] = e^{G[J_\mu,\eta, \bar{\eta},R,K_\mu,L]}
=\int \mathcal{D}[A_\mu, c ,\bar{c},B]\, e^{-S+S_{\text{sources}}} \,.
\ee
The vanishing of the functional integral of a functional
derivative gives, for the auxiliary field,
\be
0=\int D[A_\mu, c ,\bar{c},B]\frac{\delta}{\delta B^a(x)}
\,e^{-S+ S_{\text{sources}}}
=\int D[A_\mu, c ,\bar{c},B] \left(-i\partial_\mu A^a_\mu+R^a \right)
e^{-S+ S_{\text{sources}}} \,.
\ee
This relation translates into
\be
-i\partial_\mu \frac{\delta G}{\delta J^a_\mu(x)}+R^a(x)=-i\partial_\mu A^a_\mu(x)
+\frac{\delta \Gamma}{\delta B^a(x)}=0 \,, \label{bdse}
\ee
where we have introduced the effective action
$\Gamma$, 
\be
\Gamma[A_\mu,c,\bar{c},B,K_\mu,L]=-G[J_\mu,\eta, \bar{\eta},R,K_\mu,L]
+\int d^D x \left( J_\mu^a A_\mu^a+\bar{\eta}^a c^a +\bar{c}^a\eta^a+
R^aB^a \right) \,,
\ee
the Legendre transform of the logarithm of the generating functional
with respect to the classical fields, which are defined as the vacuum
expectation values of the corresponding quantum fields.

By taking another derivative of the left-hand side of Eq.\
\eqref{bdse} with respect to $J$, the transversality of the connected gluon
propagator is
immediately obtained. Incidentally, by differentiating the same equation
written in terms of the effective action, it is also proved that the 
proper two-point functions involving the auxiliary field do not
receive any quantum corrections, i.e., following the notation
\eqref{notateGamma} (although here it would not be necessary to
put the external sources equal to zero),
 \be
 \Gamma_{BB}(p )=0 \,, \qquad \Gamma_{A_\mu B}(p )=p_\mu \,. \label{abdse}
 \ee

 We can also extract information about the transverse and longitudinal parts
 of the proper gluonic two-point function (its longitudinal
 part, unlike the longitudinal part of the propagator, does not vanish in
 the presence of a mass term) by looking at the identity
\be
\begin{pmatrix} {\ds \Gamma_{A_\mu A_\nu}(p)} & {\ds\Gamma_{A_\mu B}(p)}\\
  {\ds  \Gamma_{ B  A_\nu}(p)} & {\ds \Gamma_{B B}(p)} \end{pmatrix}
\cdot \begin{pmatrix} {\ds G_{J_\nu  J_\rho}(p)} & {\ds G_{ J_\nu  R}(p)}\\
  {\ds  G_{ R  J_\rho}(p)} & {\ds  G_{ R  R}(p)} \end{pmatrix}
= \begin{pmatrix}{\ds \delta_{\mu \rho}} & {\ds 0} \\
  {\ds 0} & {\ds 1}  \end{pmatrix} \,,
\ee
which gives a nontrivial matrix relation due to the nonvanishing mixed terms
involving both gauge and auxiliary fields.
With the notations $\Gamma_{A_\mu A_\nu}(p)=\Gamma^{\perp}_{AA}(p)
(\delta_{\mu \nu} - p_\mu p_\nu/p^2) + \Gamma^{\parallel}_{AA}(p)\, 
p_\mu p_\nu/p^2$ and
$G_{J_\mu J_\nu}(p)= G_{JJ}(p) (\delta_{\mu \nu} - p_\mu p_\nu/p^2)$, 
the identity implies
\be
\Gamma^{\perp}_{A A}(p )=\big( G_{JJ}(p) \big)^{-1} \,, \qquad
\Gamma^{\parallel}_{A A}(p )=p^2 G_{RR}(p ) \,.
\ee

\section{Slavnov-Taylor identity \label{derSTI}}
Exploiting the invariance of the action and of the integration measure under
the BRST variation \eqref{BRST}, it is easy to derive the Slavnov-Taylor
identity for the effective action,
\be
\int d^D x \left[\frac{\delta \Gamma}{\delta A^a_\mu(x)}
  \frac{\delta \Gamma}{\delta K^a_\mu(x)}+\frac{\delta \Gamma}{\delta c^a(x)}
  \frac{\delta\Gamma}{\delta L^a(x)}-iB^a(x)
  \frac{\delta \Gamma}{\delta \bar{c^a}(x)}+im^2c^a(x)
  \frac{\delta \Gamma}{\delta B^a(x)} \right]=0 \,. \label{ST}
\ee 
We will also make use of the antighost equation (the Dyson-Schwinger
equation for the antighost field)
\be
\partial_\mu\frac{\delta G}{\delta K^a (x)}+\eta^a(x)=-\partial_\mu
\frac{\delta \Gamma}{\delta K^a(x)}+\frac{\delta \Gamma}{\bar{c}^a(x)}=0 \,,
\ee
which in momentum space reads
\be
ip_\mu \frac{\delta \Gamma}{\delta K^a(p)}+
\frac{\delta \Gamma}{\delta \bar{c}^a(p)}=0 \,.
\ee
Taking another derivative with respect to $c^b(-q)$ we obtain
\be
\Gamma_{c \bar{c}}(p)=-ip_\mu \Gamma_{c K_\mu}(p) \,, \label{ckdse}
\ee
which, by Lorentz covariance, implies that $\Gamma_{c K_\mu}(p)=i
\Gamma_{c \bar{c}}(p)p_\mu/p^2$.

Differentiating Eq.\
\eqref{ST} with respect to $A^b_\nu(y)$ and $c^c(z)$ and switching
to momentum space, we get (at vanishing sources), considering ghost
number conservation at the quantum level,
\be
\Gamma_{A_\mu A_\nu}(p)\Gamma_{c K_\mu}(p) = im^2 \Gamma_{A_\nu B}(p) \,.
\label{aackdse}
\ee
Using Eqs.\ \eqref{abdse} and \eqref{ckdse} in
\eqref{aackdse}, it follows that
\be
p_\mu\Gamma_{A_\mu A_\nu}(p) \Gamma_{c \bar{c}}(p)=m^2p^2p_\nu \,.
\ee
Finally, decomposing the proper gluonic two-point function in its
transverse and longitudinal parts, Eq.\ \eqref{STI} is obtained.

\section{Two-point functions to one-loop order \label{twop}}

Below we give explicit expressions for the Feynman diagrams that
contribute to one-loop order to the ghost and gluon self-energies, in
$D=4-\epsilon$ Euclidean space-time dimensions, with the corresponding
IR and UV limits. 

The one-loop ghost self-energy is
\be
\begin{split}
  \Sigma^B(p)&=\imineq{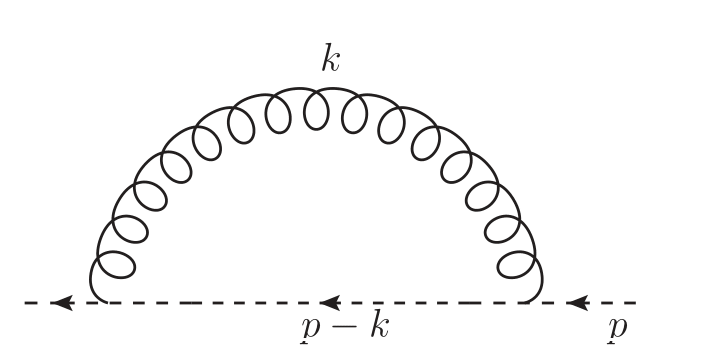}{10} = \frac{3}{4} \, p^2
  \frac{Ng^2}{(4\pi)^2}\left(\overline{\frac{2}{\epsilon}}
  -\ln \frac{m^2}{\kappa^2} \right)+\frac{1}{4} \, p^2 \frac{Ng^2}{(4\pi)^2}
  \Big[s^{-1}+5+s\ln s \\
&\phantom{=}{}-s^{-2}(1+s)^3\ln(1+s) \Big] \,,
\end{split}
\ee
where $s=p^2/m^2$, and $\overline{(2/\epsilon)}$ was defined in Eq.\
\eqref{overeps}. In the IR limit ($s \ll 1$), the first terms of the
diagram's expansion in powers of $s$ are
\be
\Sigma^B(p)/p^2 =  \frac{3}{4} \,  \frac{N g^2}{(4\pi)^2}
\left( \overline{\frac{2}{\epsilon}} - \ln \frac{m^2}{\kappa^2} 
+ \frac{5}{6} \right)+ \frac{1}{4} \, s \,
\frac{N g^2}{(4\pi)^2} \left( \ln s - \frac{11}{6} \right)+ \cal{O}(s^2) \,,
\ee
while the UV limit ($s \gg 1$) is given by
\be
\Sigma^B(p)/p^2 =  \frac{3}{4} \,  \frac{N g^2}{(4\pi)^2}
\left( \overline{\frac{2}{\epsilon}} - \ln \frac{p^2}{\kappa^2} 
+ \frac{4}{3} \right)-\frac{3}{4} \,  \frac{1}{s}
\frac{N g^2}{(4\pi)^2} \left( \ln s + \frac{1}{2} \right)+ \cal{O}(1/s^2) \,.
\ee

The three diagrams contributing to the gluon self-energy are written
in the simpler decomposition $(\delta_{\mu \nu}-p_{\mu}p_{\nu}/p^2)[\Pi^{\perp}(p^2)
  -\Pi^{\parallel}(p^2)]+\delta_{\mu \nu}\Pi^{\parallel}(p^2)$. The only gluonic
two-point diagram with a nontrivial expansion for $p^2\ll m^2$ and $p^2\gg m^2$
is the one with the gluon loop.
\be
\begin{split}
\hspace{0pt}
\Pi^{B\,gh}_{\mu \nu}(p)&=\imineq{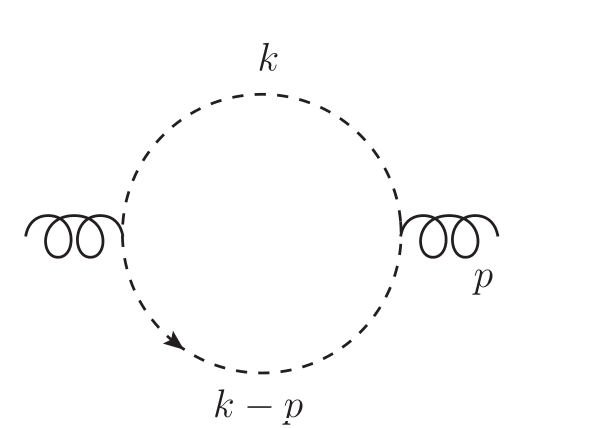}{14}
=-\left(\delta_{\mu \nu}-\frac{p_\mu p_\nu}{p^2} \right) \frac{1}{6} \, p^2
\frac{Ng^2}{(4\pi)^2}\left(\overline{\frac{2}{\epsilon}}
-\ln \frac{m^2}{\kappa^2}-\ln s+\frac{5}{3}\right) \\
&\phantom{=}{}+\delta_{\mu \nu}\frac{1}{4} \, p^2 \frac{Ng^2}{(4\pi)^2}
\left(\overline{\frac{2}{\epsilon}}-\ln \frac{m^2}{\kappa^2}-\ln s+2\right) \,,
\hspace{\linewidth minus\linewidth}
\end{split} \label{glselfgh}
\ee

\be
\hspace{0pt} \Pi^{B\,tad}_{\mu \nu}(p)
=\imineq{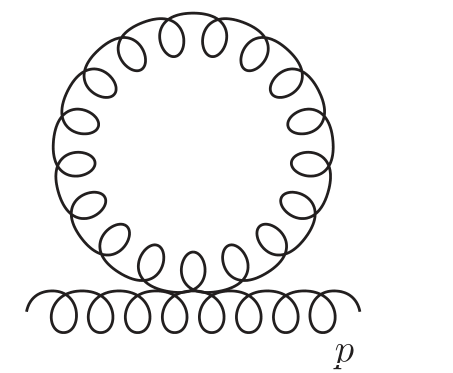}{12}
=\delta_{\mu \nu}\, \frac{9}{4} \, m^2 \frac{Ng^2}{(4\pi)^2}
\left(\overline{\frac{2}{\epsilon}}-\ln \frac{m^2}{\kappa^2}+\frac{1}{6}\right)
\,, \hspace{\linewidth minus\linewidth}
\ee

\be
\begin{split}
  \hspace{0pt}\Pi^{B\,gl}_{\mu \nu}(p)
  &=\imineq{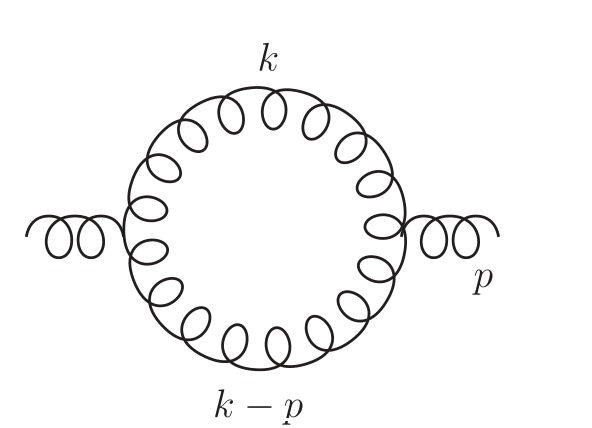}{14}
  =\left(\delta_{\mu \nu}-\frac{p_\mu p_\nu}{p^2} \right) p^2
  \frac{Ng^2}{(4\pi)^2} \left\{\frac{7}{3}
    \left( \overline{\frac{2}{\epsilon}}-\ln \frac{m^2}{\kappa^2} \right)
    \right.\\
    &\phantom{=}{}+\frac{1}{72 s^2} \bigg[2(12-144s+131s^2)
    -6s^{-1}(1+s)^3(4-10s+s^2)\ln(1+s) \\
    &\phantom{=}{}\left. +3s^4\ln s -3\sqrt{s}(4+s)^{3/2}(s^2-20s+12)\ln
    \left(\frac{\sqrt{4+s}-\sqrt{s}}{\sqrt{4+s}+\sqrt{s}} \right)\bigg]
    \right\}\\
  &\phantom{=}{}-\delta_{\mu \nu} \, p^2 \frac{Ng^2}{(4\pi)^2}\bigg\{\frac{1}{4}
    \left( \overline{\frac{2}{\epsilon}}-\ln \frac{m^2}{\kappa^2}\right)
    (12s^{-1}+1)+\frac{1}{8s^2}\Big[2+13s+4s^2 \\
&\phantom{=}{}-2s^{-1}(1+s)^3\ln(1+s) \Big]\bigg\} \,. 
\hspace{\linewidth minus\linewidth} \label{glselfgl}
\end{split}
\ee
The last diagram has the following expansion in the deep IR ($s\ll 1$):
\be
\begin{split}
  \Pi^{B\,gl}_{\mu \nu}(p)/p^2&=\left(\delta_{\mu \nu}-\frac{p_\mu p_\nu}{p^2} \right)
  \frac{Ng^2}{(4\pi)^2}\left[ \frac{7}{3}\left( \overline{\frac{2}{\epsilon}}
    -\ln \frac{m^2}{\kappa^2} \right)-\frac{11}{36}-\frac{217}{360}s
    +\cal{O}(s^2)\right] \\
  &\phantom{=}{}-\delta_{\mu \nu}\frac{Ng^2}{(4\pi)^2}\left[ \frac{1}{4}
    \left( \overline{\frac{2}{\epsilon}}-\ln \frac{m^2}{\kappa^2}\right)
    (12s^{-1}+1)+\frac{1}{24}\left(24s^{-1}+1-\frac{3}{2}s \right)+\cal{O}(s^2)
    \right] \,,
\end{split}
\ee
and in the UV ($s\gg 1$)
\be
\begin{split}
  \Pi^{B\,gl}_{\mu \nu}(p)/p^2&=\left(\delta_{\mu \nu}-\frac{p_\mu p_\nu}{p^2} \right)
  \frac{Ng^2}{(4\pi)^2}\left[ \frac{7}{3}\left( \overline{\frac{2}{\epsilon}}
    -\ln \frac{p^2}{\kappa^2} \right)+\frac{107}{36}-\frac{1}{8s}(6\ln s+71)
    +\cal{O}(1/s^2) \right] \\
  &\phantom{=}{}-\delta_{\mu \nu}\frac{Ng^2}{(4\pi)^2}\left[ \frac{1}{4}
    \left( \overline{\frac{2}{\epsilon}}-\ln \frac{p^2}{\kappa^2}\right)
    (12s^{-1}+1)+\frac{1}{2}+\frac{1}{8s}\left(18\ln s+11 \right)+\cal{O}(1/s^2)
    \right] \,.
\end{split}
\ee

In any renormalization scheme, the dominant contribution
in the deep IR to the anomalous dimension $\gamma_A$, and hence also
to the beta functions $\beta_{m^2}$ and $\beta_g$, comes from the
ghost loop diagram \eqref{glselfgh}, a phenomenon that is known as
  ghost dominance. The intuitive reason is that diagrams with internal
massless (ghost) propagators dominate over diagrams with internal
massive (gluon) propagators in the IR. However, note that
  there is a UV divergent $p^2$-dependent contribution to the longitudinal
  part in Eq.\ \eqref{glselfgh} that would spoil the renormalizability of the
  theory if it was not cancelled by an identical contribution from the gluon
  loop diagram \eqref{glselfgl}.

\section{Three-point integrals \label{threep}}

In this appendix, we present explicit expressions for the two diagrams
contributing to the one-loop correction of the ghost-gluon vertex, in
$D=4$ dimensions (the diagrams are UV convergent in Landau gauge). The
diagrams, evaluated at the symmetry point $k^2=p^2=(p-k)^2=\mu^2$ in order to
define the renormalized coupling constant in alternative
renormalization schemes, have been calculated using the ordinary tensor
reduction to scalar integrals with two and three propagators. The three-point
scalar integrals have been left in symbolic notation in the following
expressions, and are given below in terms of one-dimensional integrals whose
form is suitable for both numerical evaluation and analytical expansion in the
IR and UV limits. 
Following the notation in Ref.\ \cite{Dav00}, we denote as
$\tilde{J_n}(t)=\mu^2 J_n(1,1,1)$ the three-point scalar integral at the
symmetry point \cite{footn4},
where the subscript denotes the number of massive propagators, and
$t=\mu^2/m^2$. Since we are interested in the transverse parts of the
corrections, these are to be thought as contracted with the transverse
projector corresponding to the external gluon momentum.

The first one-loop contribution to the ghost-gluon vertex function is
\be
\begin{split}
\hspace{0pt}
\imineq{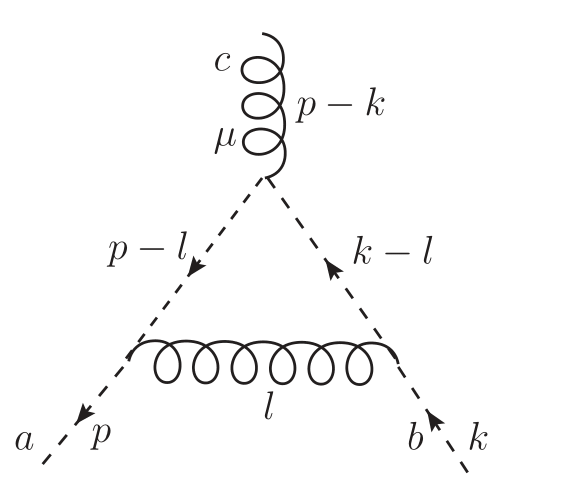}{18} &=ip_\mu
\frac{Ng^3 f^{abc}}{24(4\pi)^2t^2} \Big[ 3t +t(2-t^2)\ln t -(1+t)(5-t^2)\ln(1+t)
  \\
&\phantom{=}{}-t^3\tilde{J_0}+(2+t)(1+t^2)\tilde{J_1}(t) \Big] \,,
\hspace{\linewidth minus\linewidth} 
\end{split}
\ee
which in the IR ($t \ll 1$) behaves like
\be
ip_\mu \frac{Ng^3 f^{abc}}{288(4\pi)^2}\Big[t(-30\ln t+31-12\tilde{J_0} )
  +\cal{O}(t^2)\Big] \,,
\ee
and in the UV ($t \gg 1$) 
\be
ip_\mu \frac{Ng^3 f^{abc}}{96(4\pi)^2}\left[8\tilde{J_0}+\frac{1}{t}
  (-18\ln t+3+4\tilde{J_0})+\cal{O}(1/t^2)\right] \,.
\ee

The second one-loop contribution is
\be
\begin{split}
\hspace{0pt}
&\imineq{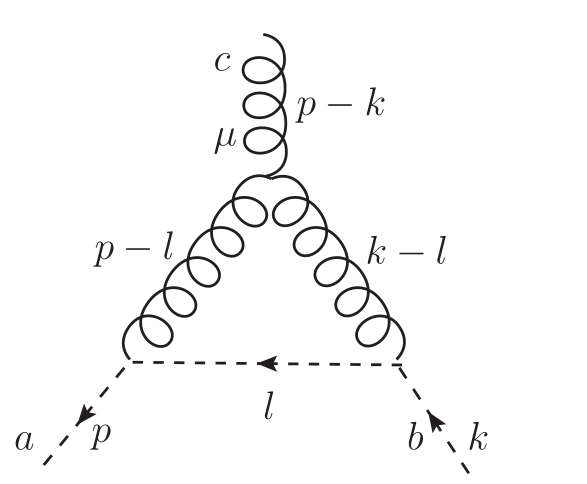}{18} =ip_\mu
\frac{Ng^3 f^{abc}}{48(4\pi)^2t^2} \bigg[2t(2+9t)-t^2(4+18t-3t^2)\ln t \\
  &{}-6(1+t)^2(2-t+t^2)\ln (1+t)-\sqrt{t(4+t)}(8-4t+18t^2+3t^3)\ln \left(
  \frac{\sqrt{4+t}-\sqrt{t}}{\sqrt{4+t}+\sqrt{t}} \right) \\
  &{}+2t^4\tilde{J_0}-4t(1+t)(1+4t+t^2)\tilde{J_1}(t)-2(1+t)^2(4-8t-t^2)
  \tilde{J_2}(t) \bigg] \,,
 \hspace{\linewidth minus\linewidth} 
\end{split}
\ee
whose limit in the IR is
\be
ip_\mu \frac{Ng^3 f^{abc}}{72(4\pi)^2}\Big[51t+t^2(-31+3\tilde{J_0} )+\cal{O}(t^3)
  \Big] \,,
\ee
and in the UV
\be
ip_\mu \frac{Ng^3 f^{abc}}{24(4\pi)^2}\, \left [18+3\tilde{J_0}-\frac{2}{t}
  (6+9\ln t +\tilde{J_0})+\cal{O}(1/t^2) \right] \,.
\ee

The three-point scalar integrals $\tilde J_n(t)$ in $D=4$ dimensions
have been evaluated by the use of Feynman parameters
$x_1, x_2, x_3$, releasing two of them from the constraint
$\delta(x_1+x_2+x_3-1)$ according to the Cheng-Wu theorem \cite{CW87}:
\be
\int_0^1 \!\!\!dx_1\!\! \int_0^1 \!\!\! dx_2 \!\! \int_0^1 \!\!\! dx_3\,
\delta(x_1+x_2+x_3-1) F(x_i) = \int_0^\infty \!\!\! dx_1 \!\! \int_0^\infty \!\!\!
dx_2 \!\! \int_0^1 \!\!\! dx_3 \, \delta(x_3-1) F(x_i) \,,
 \ee
where
\be
F(x_i) = 2 \int d^4 l \, \frac{1}{\left [\sum_{i=1}^3 x_i \big((l-q_i)^2 +\chi_i
    m^2 \big) \right]^3}
\ee
with $q_1 =0$, $q_2=p$, $q_3=k$ and $\chi_i=0$ or $1$
depending on whether the corresponding propagator is massless or
massive. The Cheng-Wu theorem is a general result, extendable to an
arbitrary number of propagators, which states that, within the Feynman
parameterization, only a (non-empty) subset of the Feynman
parameters needs to be included in the sum constrained by the
delta function, while the rest of them remain unconstrained and are
integrated up to infinity.


For our purposes (albeit not
in its most general form), the Cheng-Wu theorem is easily demonstrated by
starting from the Schwinger parameterization of a product of arbitrary powers
of (scalar) propagators $1/A_i$,
\bal
\prod_{i=1}^n \frac{1}{A_i^{b_i}} 
= \left( \prod_{i=1}^n \frac{1}{\Gamma(b_i)} \right)
\int_0^\infty \!\!\!d\alpha_1 \cdots \!\int_0^\infty \!\!\!d\alpha_n \,
\alpha_1^{b_1 - 1} \cdots \alpha_n^{b_n - 1} e^{-\sum_{i=1}^n \alpha_i A_i} \,.
\eal
Now the integral over the rescaling factor $\lambda$ is introduced in the
usual way, but the sum over the Schwinger parameters constrained by the
delta function may be restricted to any non-empty subset of them, e.g.,
\be
\prod_{i=1}^n \frac{1}{A_i^{b_i}}
= \left( \prod_{i=1}^n \frac{1}{\Gamma(b_i)} \right) \int_0^\infty \!\!\!
d\lambda \!\int_0^\infty \!\!\!d\alpha_1 \cdots \!\int_0^\infty \!\!\!d\alpha_n
\, \delta \big( {\ts
  \lambda - \sum_{i=k}^n \alpha_i} \big) \left( \prod_{i=1}^n
\alpha_i^{b_i - 1} \right) e^{-\sum_{i=1}^n \alpha_i A_i}
\ee
with $1 \le k \le n$, where we have arbitrarily included the last $(n - k + 1)$
Schwinger parameters in the sum constrained by the delta function, while we
clearly could have chosen any $(n - k + 1)$ among them. Then, rescaling
\emph{all} Schwinger parameters as $\alpha_i = \lambda x_i$ and integrating
over $\lambda$, one obtains
\be
\prod_{i=1}^n \frac{1}{A_i^{b_i}} 
= \frac{ \Gamma (\sum_{i=1}^n b_i)}
{\prod_{i=1}^n \Gamma(b_i)} \int_0^\infty \!\!\!d x_1 \cdots \!
\int_0^\infty \!\!\!d x_{k-1} \!\int_0^1 \!\!\!d x_k \cdots \!\int_0^1 \!\!\!
d x_n \, \delta \big( {\ts 1 - \sum_{i=k}^n x_i} \big) \,
\frac{\prod_{i=1}^n x_i^{b_i - 1}}
{\left[ \sum_{i=1}^n x_i A_i \right]^{\sum_{i=1}^n b_i}}
\ee
for any $k$, $1 \le k \le n$. In particular, for $k = 1$ one recovers the
usual Feynman parameterization.

The three-point scalar integrals can therefore be written as
unconstrained integrals over two Feynman parameters, one of which
may be analytically evaluated and the remaining one rescaled to the 
compact domain $[0,1]$, leading to the following form which is best suited 
to both numerical and analytical calculations:
\be
\begin{split} \label{threeD4} 
  &\tilde{J_0}= \left. \mu^2 \int d^4 l \, \frac{1}{l^2(l-p)^2(l-k)^2}\right
  \vert_{s.p.}= -2 \int_0^1 \frac{dx}{1-x+x^2} \ln (x) \,, \\
  &\tilde{J_1}(t)= \left. \mu^2 \int d^4 l \, \frac{1}{(l^2+m^2)(l-p)^2(l-k)^2}
  \right\vert_{s.p.}=  \int_0^1 \frac{dx}{1-x+x^2} \, \ln \left (
  \frac{t+x}{x(1+t x)} \right) \,, \\
  &\tilde{J_2}(t)= \left. \mu^2 \int d^4 l \, \frac{1}
       {(l^2+m^2)\big((l-p)^2+m^2\big)(l-k)^2} \right \vert_{s.p.}=
       \int_0^1 \frac{dx}{1-x+x^2} \, \ln \left( \frac{1+t}{1+t x-t x^2} 
  \right) \,.
\end{split}
\ee


\begin{thebibliography}{99}
\bibitem{footn3}
It is well known today that the first Gribov region still contains
gauge copies and that one should, in principle, restrict the integration over
the gauge fields to the fundamental modular region defined as the set of the
absolute minima of the functional $\int_x A^a_\mu(x) A^a_\mu(x)$ for every 
gauge orbit \cite{SF86, DZ91}, but here we do not wish to enter the discussion 
about the possible effect that the restriction to the fundamental modular 
region, as opposed to the Gribov region, might have on the n-point functions 
\cite{Maa13, Maa17}.
\bibitem{footn2}
The corrections are strictly of the order
of $(m^2/\mu^2)^0$ in the UV limit, i.e., there are no corrections of the
order of $\ln (\mu^2/m^2)$ in the present case. Note that the normalization
condition \eqref{tt2}, on the other hand, would lead to corrections of the
order of $(\mu^2/m^2)$ in this limit.
\bibitem{footn4}
With the difference that here we consider Euclidean
space-time, so that a Wick rotation is implied compared to Ref.\ \cite{Dav00}.
\end{thebibliography}
\end{document}